\documentclass[12pt,onecolumn,a4paper,compsoc,twoside,openright]{IEEEtran}
\usepackage[english]{babel}
\usepackage[labelfont=bf,font=small]{caption}
\usepackage[printonlyused]{acronym}
\usepackage{graphicx}
\usepackage{tabularx}
\usepackage{enumerate}
\usepackage{color}
\usepackage{amsmath}
\usepackage{paralist}
\usepackage{threeparttable}
\usepackage{booktabs}
\usepackage{colortbl,multirow,hhline}
\usepackage{boldline}
\usepackage{colortbl}
\usepackage{array}
\usepackage{adjustbox}
\usepackage{multirow,tabu}
\usepackage{algorithm,algpseudocode}
\usepackage{tikz,pgfplots}
\usetikzlibrary{arrows}
\usepackage{pgfplotstable}
\usepgfplotslibrary{units,groupplots}
\usepackage{filecontents}
\usepackage{wrapfig}
\usepackage{varwidth,xcolor}
\usepackage{subfigure}
\usepackage{epsfig}
\usepackage{boldline}
\usepackage{stackengine}
\usepackage{color}
\usepackage{xspace}
\usepackage{listings}
\usepackage{fancyvrb}
\usepackage{verbatim}
\usepackage[hidelinks]{hyperref}
\usepackage{booktabs}
\usepackage{colortbl}
\usepackage{relsize}
\usepackage{paralist}
\usepackage{threeparttable}
\usepackage{amsmath}
\usepackage{amssymb}
\usepackage{booktabs}
\usepackage{ifthen}
\usepackage{url}
\usepackage{mathtools}
\usepackage{nccmath}
\usepackage{capt-of}
\usepackage{hhline}
\usepackage{setspace}
\usepackage{framed,varwidth}
\usepackage{lettrine}
\usepackage[margin=2.5cm,twoside]{geometry}

\newcommand{\subparagraph}{} 
\usepackage{titlesec}

\usepackage{fancyhdr}
\usepackage{lastpage}

\newcommand{\roberto}[1]{{\leavevmode\color{black}{#1}}}

\newcommand{\highlightnewtext}{0}

\ifnum\highlightnewtext=1
\newcommand{\revised}[1]{{\leavevmode\color{blue}{#1}}}
\else
\newcommand{\revised}[1]{{\leavevmode\color{black}{#1}}}
\fi

\newcommand{\mytinysize}{\fontsize{6}{7}\selectfont}
\pgfplotsset{
	compat = newest,
	tick label style={font=\sffamily\scriptsize},
	label style={font=\sffamily\scriptsize},
	legend style={font=\sffamily\mytinysize\raggedleft},
	legend cell align=left,
	grid style={dotted,gray}
}

\newcolumntype{?}{!{\vrule width 0.8pt}}

\definecolor{mygray}{RGB}{220,220,220}
\definecolor{skyblue}{RGB}{86,180,233}
\definecolor{bluish-green}{RGB}{0,158,115}
\definecolor{myblue}{RGB}{86,114,178}
\definecolor{vermilion}{RGB}{213,94,0}
\definecolor{reddish-purple}{RGB}{204,121,167}
\definecolor{persianorange}{rgb}{0.85, 0.56, 0.35}
\definecolor{saffron}{rgb}{0.96, 0.77, 0.19}
\definecolor{mediumorchid}{rgb}{0.73, 0.33, 0.83}
\definecolor{mediumseagreen}{rgb}{0.24, 0.7, 0.44}
\definecolor{mangotango}{rgb}{1.0, 0.51, 0.26}
\definecolor{mahogany}{rgb}{0.75, 0.25, 0.0}
\definecolor{hollywoodcerise}{rgb}{0.96, 0.0, 0.63}
\definecolor{deepcarmine}{rgb}{0.66, 0.13, 0.24}

\newlength{\Oldarrayrulewidth}
\newcommand{\Cline}[2]{%
	\noalign{\global\setlength{\Oldarrayrulewidth}{\arrayrulewidth}}%
	\noalign{\global\setlength{\arrayrulewidth}{#1}}\cline{#2}%
	\noalign{\global\setlength{\arrayrulewidth}{\Oldarrayrulewidth}}}

\def\nullpage{%
	\clearpage%
	\thispagestyle{empty}%
	\addtocounter{page}{-1}%
	\null%
	\clearpage}

\def\blankpage{%
	\clearpage%
	\thispagestyle{empty}%
	\null%
	\clearpage}

\titleformat{\section}{\normalfont\Huge\bfseries}{\thesection}{1em}{}[\vspace{1.8cm}]
\titleformat{\subsection}{\normalfont\Large\bfseries}{\thesubsection}{1em}{}[\vspace{2mm}]
\titleformat{\subsubsection}{\normalfont\large\bfseries}{\thesubsubsection}{1em}{}

\graphicspath{{./artworks/}}
\hypersetup{pdfinfo={
		Title={Methods and Techniques for Dynamic Deployability of Software-Defined Security Services},
		Author={Roberto Doriguzzi Corin},
		Keywords={Network Function Virtualisation, Network Service Chaining, Progressive Embedding, Application-Aware Network Security, DDoS, Deep Learning, Convolutional Neural Networks, XDP, eBPF}
}}

\makeatletter
\renewcommand*\cleardoublepage{\clearpage\if@twoside
	\ifeven\c@page \hbox{}\newpage\if@twocolumn\hbox{}%
	\newpage\fi\fi\fi}
\makeatother

\begin{document}

\begin{titlepage}
	\thispagestyle{empty}
	\begin{titlepage}
	\begin{center}
		{{\Large{\textsc{Alma Mater Studiorum $\cdot$ Universit\`a di
						Bologna}}}} \rule[0.1cm]{15.8cm}{0.1mm}
		\rule[0.5cm]{15.8cm}{0.6mm}
		{\Large{\bf Ph.D. in Electronics, Telecommunications and\\
				Information Technologies Engineering \\}}
		\vspace{3mm}
		
	\end{center}
	\vspace{5mm}


	\vspace{25mm}
	\begin{center}
		{\LARGE{\bf \textsc{Methods and Techniques for}}}\\
		\vspace{3mm}
		{\LARGE{\bf \textsc{Dynamic Deployability of}}}\\
		\vspace{3mm}
		{\LARGE{\bf \textsc{Software-Defined Security Services}}}\\
	\end{center}
	\vspace{40mm}
	\par
	\noindent
	\begin{minipage}[t]{0.55\textwidth}
		{\large{Author:\\
				\textbf{Roberto Doriguzzi-Corin}}}
			
	\vspace{20mm}
	{\large{Supervisor of the Doctoral Program:\\
			\textbf{Prof. Alessandra Costanzo}}}
		
	\end{minipage}
	\hfill
	\begin{minipage}[t]{0.45\textwidth}\raggedleft
		\vspace{26.5mm}
		{\large{Supervisors:\\
				\textbf{Prof. Franco Callegati}
		}}\\
		{\large{	
				\textbf{Dr. Domenico Siracusa}
		}}\\
		

	\end{minipage}
	\vspace{40mm}
	\begin{center}
	{\small{Dipartimento di Ingegneria dell'Energia Elettrica e
			dell'Informazione ``Guglielmo Marconi''}}
\rule[0.1cm]{15.8cm}{0.1mm}
		{\large{2020 -- Cycle XXXII}}
	\end{center}
\end{titlepage}
\end{titlepage}

\newgeometry{hmargin=3.3cm,top=2.5cm,bottom=3cm,outer=1.7cm}

\pagestyle{fancy}
\fancyhead[RO,LE]{\thepage}


\nullpage
\spacing{1.5}
\section*{Abstract}
\vspace{-1mm}
\lettrine[findent=2pt]{\textbf{W}}{ }ith the recent trend of ``network softwarisation'', enabled by emerging technologies such as \ac{sdn} and \ac{nfv}, system administrators of data centres and enterprise networks have started replacing dedicated hardware-based middleboxes with virtualised network functions running on servers and end hosts. 
This radical change has facilitated the provisioning of advanced and flexible network services, ultimately helping system administrators and network operators to cope with the rapid changes in service requirements and networking workloads.

This thesis investigates the challenges of provisioning network security services in ``softwarised'' networks, where the security of residential and business users can be provided by means of sets of software-based network functions running on high performance servers or on commodity compute devices. The study is approached from the perspective of the telecom operator, whose goal is to protect the customers from network threats and, at the same time, maximize the number of provisioned services, and thereby revenue. Specifically, the overall aim of the research presented in this thesis is proposing novel techniques for optimising the resource usage of software-based security services, hence for increasing the chances for the operator to accommodate more service requests while respecting the desired level of network security of its customers. In this direction, the contributions of this thesis are the following: (i) a solution for the dynamic provisioning of security services that minimises the utilisation of computing and network resources, and (ii) novel methods based on \acl{dl} and Linux kernel technologies for reducing the CPU usage of software-based security network functions, with specific focus on the defence against \ac{ddos} attacks.

The experimental results reported in this thesis demonstrate that the proposed solutions for service provisioning and \ac{ddos} defence require fewer computing resources, compared to similar approaches available in the scientific literature or adopted in production networks.

\newpage
\spacing{1.5}
\blankpage
\tableofcontents
\newpage

\blankpage
\section*{List of Acronyms}

\begin{acronym}[NFV MANO] 
	\acro{aaa}[AAA]{Authentication, Authorisation and Accounting}
	\acro{acl}[ACL]{Access Control List}
	\acro{aml}[AML]{Adversarial Machine Learning}
	\acro{ann}[ANN]{Artificial Neural Network}
	\acro{api}[API]{Application Programming Interface}
	\acro{bow}[BoW]{Bag-of-Words}
	\acro{cctv}[CCTV]{Closed Circuit Television}
	\acro{cnn}[CNN]{Convolutional Neural Network}
	\acro{cpe}[CPE]{Customer Premise Equipment}
	\acro{dl}[DL]{Deep Learning}
	\acro{dlp}[DLP]{Data Loss/Leakage Prevention}
	\acro{dpi}[DPI]{Deep Packet Inspection}
	\acro{dos}[DoS]{Denial of Service}
	\acro{ddos}[DDoS]{Distributed Denial of Service}
	\acro{ebpf}[eBPF]{extended Berkeley Packet Filter}
	\acro{ewma}[EWMA]{Exponential Weighted Moving Average}
	\acro{foss}[FOSS]{Free and Open-Source Software}
	\acro{fpr}[FPR]{False Positive Rate}
	\acro{gpu}[GPU]{Graphics Processing Unit}
	\acro{ha}[HA]{Hardware Appliance}
	\acro{ids}[IDS]{Intrusion Detection System}
	\acro{ilp}[ILP]{Integer Linear Programming}
	\acro{iot}[IoT]{Internet of Things}
	\acro{isp}[ISP]{Internet Service Provider}
	\acro{ips}[IPS]{Intrusion Prevention System}
	\acro{lstm}[LSTM]{Long Short-Term Memory}
	\acro{lucid}[\textsc{Lucid}]{Lightweight, Usable CNN in DDoS Detection}
	\acro{mips}[MIPS]{Millions of Instructions Per Second}
	\acro{ml}[ML]{Machine Learning}
	\acro{mlp}[MLP]{Multi-Layer Perceptron}
	\acro{nat}[NAT]{Network Address Translation}
	\acro{nic}[NIC]{Network Interface Controller}
	\acro{nids}[NIDS]{Network Intrusion Detection System}
	\acro{nf}[NF]{Network Function}
	\acro{nfv}[NFV]{Network Function Virtualisation}
	\acro{mano}[NFV MANO]{NFV Management and Orchestration}
	\acro{nsc}[NSC]{Network Service Chaining}
	\acro{of}[OF]{OpenFlow}
	\acro{os}[OS]{Operating System}
	\acro{pess}[PESS]{Progressive Embedding of Security Services}
	\acro{pop}[PoP]{Point of Presence}
	\acro{ppv}[PPV]{Positive Predictive Value}
	\acro{ps}[PS]{Port Scanner}
	\acro{qoe}[QoE]{Quality of Experience}
	\acro{qos}[QoS]{Quality of Service}
	\acro{qub}[QUB]{Queen's University Belfast}
	\acro{rnn}[RNN]{Recurrent Neural Network}
	\acro{sdbranch}[SD-Branch]{Software-Defined Branch}
	\acro{sdn}[SDN]{Software-Defined Networking}
	\acro{sla}[SLA]{Service Level Agreement}
	\acro{smartnic}[SmartNIC]{Smart Network Interface Card}
	\acro{snf}[SNF]{Security Network Function}
	\acro{svm}[SVM]{Support Vector Machine}
	\acro{tc}[TC]{Traffic Classifier}
	\acro{tor}[ToR]{Top of Rack}
	\acro{tpr}[TPR]{True Positive Rate}
	\acro{tsp}[TSP]{Telecommunication Service Provider}
	\acro{unb}[UNB]{University of New Brunswick}
	\acro{vm}[VM]{Virtual Machine}
	\acro{vne}[VNE]{Virtual Network Embedding}
	\acro{vnep}[VNEP]{Virtual Network Embedding Problem}
	\acro{vnf}[VNF]{Virtual Network Function}
	\acro{vsnf}[VSNF]{Virtual Security Network Function}
	\acro{vpn}[VPN]{Virtual Private Network}
	\acro{xdp}[XDP]{eXpress Data Path}
	\acro{wan}[WAN]{Wide Area Network}
	\acro{waf}[WAF]{Web Application Firewall}
\end{acronym}
\newpage





\spacing{1.5}
\section{Introduction}\label{sec:introduction}
\lettrine[findent=2pt]{\textbf{N}}{ }etwork security implemented by \acp{tsp} has traditionally been based on the deployment of specialised, closed, proprietary \acp{ha}. Such \acp{ha} are inflexible in terms of functionalities and placement in the network, which means that even slight changes in the security requirements generally necessitate manually intensive and time-consuming re-configuration tasks, the replacement of existing \acp{ha} or the deployment of additional \acp{ha}. 

The \ac{nfv}~\cite{7243304} initiative  has been proposed as a possible solution to address the operational challenges and high costs of managing proprietary \acp{ha}. The main idea behind \ac{nfv} is to transform network functions (e.g., firewalls, intrusion detection systems etc.) based on proprietary \acp{ha}, into software components (called \acp{vnf}) that can be deployed and executed in virtual machines on commodity, high-performance servers. By decoupling software from hardware, this approach allows any network function to be deployed in any server connected to the network. 
In this context, \ac{nsc} is a technique for selecting subsets of the network traffic and forcing them to traverse various \ac{vnf}s in sequence. For example, a firewall followed by an \ac{ips}, then a \ac{nat} service and so on. \ac{nsc} and \ac{nfv} enable flexible, dynamic service chain modifications to meet the real time network demands.

A promising area of application for \ac{nsc} and \ac{nfv} is in network security, where chains of \acfp{vsnf}, i.e., security-specific \acp{vnf} such as a firewall or an \ac{ips}, can be dynamically created and configured to inspect, filter or monitor the network traffic. The flexibility of the \ac{nsc} and \ac{nfv} paradigms brings many benefits, among others: (i) highly customizable security services based on the needs of the end-users, (ii) fast reaction to new security threats or variations of known attacks, and (iii) low Operating Expenditure (OpEx) and Capital Expenditure (CapEx) for the operator.

\subsection{Strategies for Security Service Provisioning}\label{sec:strategies-provisioning}
In an \ac{nfv}-enabled network, the \ac{tsp} decides which \acp{vsnf} should apply for a given service, where to place them and how to connect them. Such decisions are not only based on the requirements of the security service to be provisioned, but they are also influenced by the computing and memory requirements of each \ac{vsnf}, and by the computing, memory and network resources available in the \ac{nfv} infrastructure. Moreover, the specific position of a \ac{vsnf} might be dictated by specific \ac{tsp}'s security best practices and policies. For instance, an authentication/authorisation system should be placed inside the customer's premises to avoid transmitting sensitive data outside the local network and to reduce service latency. Another example is a \ac{dpi} system used for payload analysis, which requires powerful data centre servers to cope with the computational complexity of this type of \ac{vsnf}. In this case, the position of the data centre is important, as the \ac{tsp} might want to block malware or other malicious data as soon as it enters the network by placing the \ac{dpi} close to the border.

\begin{figure}[!h]
	\begin{center}
		\includegraphics[width=1\textwidth]{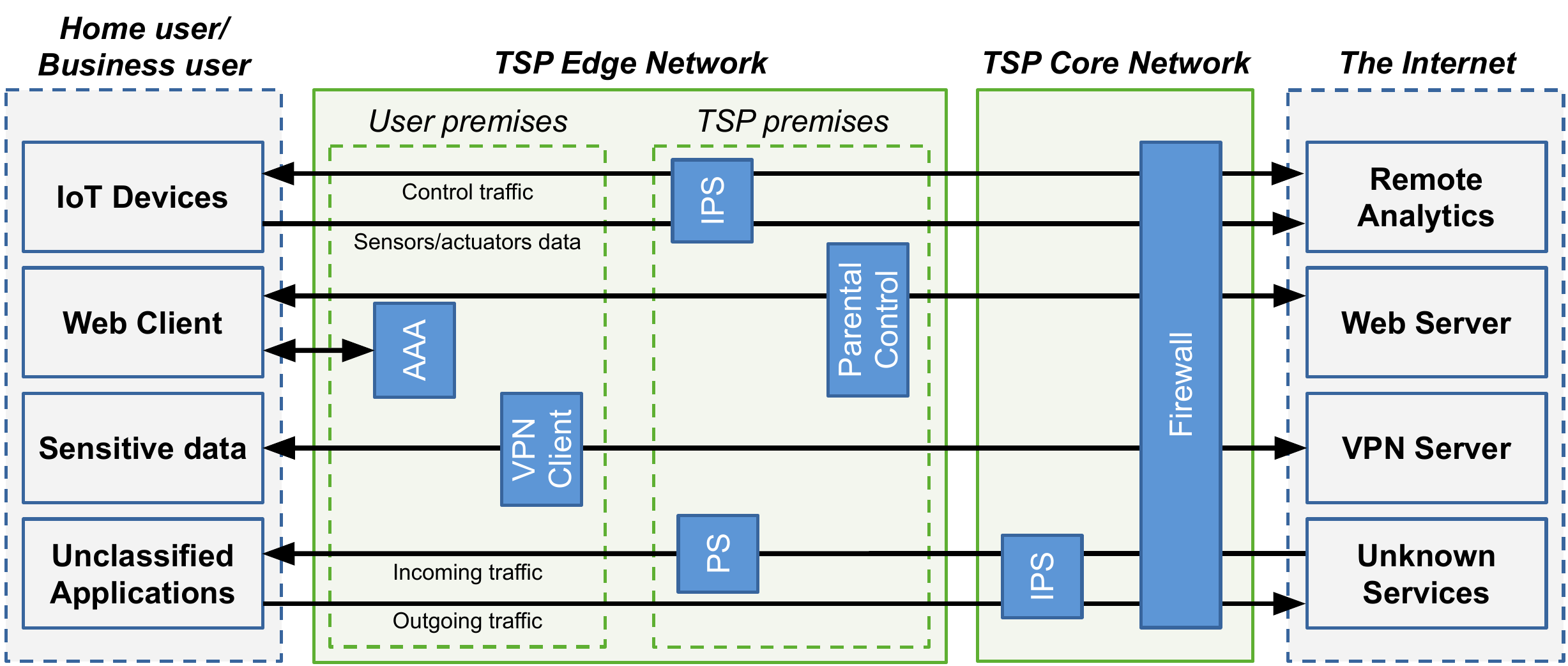}
		\caption{Examples of security service provisioning in a \ac{tsp} scenario.}
		\label{fig:user-edge-core}
	\end{center}
\end{figure}

Figure \ref{fig:user-edge-core} depicts a few examples of provisioning strategies, in which chains of \acp{vsnf} are configured across the \ac{tsp} infrastructure to secure the end-user traffic. For instance, the chain \textit{Firewall-\ac{ips}} is used to monitor and filter the traffic of the \ac{iot} devices. Other application-specific services represented in the figure include a Parental Control, an \ac{aaa} system, a \ac{vpn} client and a \ac{ps} tool used by the \ac{tsp} to detect potential security breaches in the customer's network.

\subsubsection{TSP Edge Network}
With the recent trend of \ac{iot}, \acp{tsp} have started moving part of the computing resources to the edge of their network with the aim of addressing concerns related to processing the data in remote data centres such as high latencies, bandwidth costs, security and others. As shown in Figure \ref{fig:user-edge-core}, devices at the \ac{tsp} network edge (often called \textit{edge nodes}) can be grouped in two categories: 
\begin{itemize}
	\item Nodes located at the end-user premises, including the \ac{cpe}, home gateways or any other network device leased by the user and under the control of the \ac{tsp}. Despite their limited computing and storage resources, services such as \ac{vpn} clients, parental controls, AAA and others can be provisioned at the customer's premises to minimize the service latency and to reduce privacy risks.
	\item Servers in small data centres close to the end-user but located at the \ac{tsp} premises. Compared to a \ac{cpe}, this class of nodes possesses higher storage and computing resources, although they can be shared among multiple users.    
\end{itemize}
The challenge of provisioning security services at the edge of the network is controlling the impact of CPU demanding \acp{vsnf} on the performance of other services running on the same edge node. For instance, a signature-based \ac{ips} might need a large amount of memory and CPU cycles for querying large databases, even under normal traffic conditions. Therefore, by overloading the edge node, the \ac{ips} would reduce the performance of latency-sensitive \acp{vsnf} running on the node, hence reducing part of the benefits obtained with this provisioning approach. On the other hand, given the increasing number of \ac{ddos} attacks involving compromised \ac{iot} devices, provisioning an \ac{ips} at the edge might be required for detecting and blocking the \ac{ddos} traffic as soon as it leaves the compromised devices. 

\subsubsection{TSP Core network}
The second option is placing part of the \acp{vsnf} in data centres located at the core of the \ac{tsp} network. Of course, the main benefit here is the massive computing and memory resources that can be devoted to security services. Moreover, data centres close to the border of the \ac{tsp} network can be used to mitigate network attacks coming from other administrative domains, hence before they compromise the rest of the network and user services. However, the location of the data centres in the network might lead to longer paths for the traffic, leading to higher latencies and bandwidth consumption.

\subsection{Research Challenges}\label{sec:challenges}
Compared to specialised \acp{ha}, \acp{vsnf} may have a significant impact on the performance of the network and on the \ac{qos} level experienced by the users. 
The virtualisation overhead, the utilisation level of the servers and the techniques adopted to implement the \acp{vsnf} are the most significant contributors to the QoS degradation.
Meaning that, the strategies for provisioning security services in softwarised networks must take into account not only the security requirements, but also the specific \ac{qos} needs of user applications. Omitting the latter may lead, for instance, to a provisioning mechanism that blindly forces all the user traffic to traverse the whole chain of \acp{vsnf}. As a result, computationally demanding \acp{vsnf} such as \acp{ips} may cause a noticeable performance degradation to latency-sensitive applications (e.g., online games~\cite{Claypool:2010:LKP:1730836.1730863}) or bandwidth sensitive applications (e.g., video streaming). From the \ac{tsp} perspective, the challenge is to ensure a consistent implementation of its security policies and best practices, while respecting the specific \ac{qos} requirements of customers' applications.

The latency introduced to the network traffic by softwarised functions is a major concern in \ac{nfv}-enabled networks. Indeed, unlike traditional network functions provided through dedicated \acp{ha}, \acp{vsnf} run on general purpose hardware, such as data centre servers, competing with other \acp{vsnf} for computing and memory resources. 
A notable example in this regard is the defence against \ac{ddos} attacks, which includes inspection and filtering of large volumes of malicious traffic (also in the order of Gigabits per second) generated by thousands of compromised devices. In such a scenario, a \ac{vsnf} devoted to the \ac{ddos} attack detection/mitigation might overload the CPU and exhaust the memory, resulting in serious delays for all the \acp{vsnf} running on the same server. Recently, technologies such as DPDK~\cite{dpdk:webpage} and Netmap~\cite{rizzo2012netmap}  have been proposed to improve the performance, reduce latency, and provide more predictable overall behaviour of generic \acp{vnf}. However, although \ac{nfv} performance is getting better, there is still a risk of bottlenecks, and \ac{vsnf} implementation techniques still need to be fine-tuned to guarantee maximisation of performance with minimum risk of latency-sensitive service disruption.

\subsection{Contributions and Outline of the Thesis}
In this thesis, we tackle the challenges outlined in Section \ref{sec:challenges} related to the provisioning of security services in softwarised network environments, where the availability of computing, memory and storage resources allows the dynamic deployment and customisation of chains of \acp{vsnf}. Specifically, we propose a novel approach, called \acs{pess} (\aclu{pess}), to provision security services by composing chains of VSNFs according to the specific QoS needs of user applications and the security policies defined by the \ac{tsp}. \ac{tsp}'s security policies (given as an input to \ac{pess}) include: the type of \acp{vsnf} (e.g., firewall, \ac{ips}, etc.) that should be deployed for a specific class of applications, their order (e.g., firewall first, then \ac{ips}, etc.), and their location in the \ac{tsp} infrastructure (e.g., a parental control should be installed at the edge of the \ac{tsp} network, hence close to the user's premises). 

To assess the performance issues of the \acp{vsnf}, we study the challenges of provisioning software-based security network functions in resource-constrained devices, such as the edge nodes. In this regard, one of the most complex and relevant problems in today's networks is the defence against volumetric \ac{ddos} attacks. As the malicious \ac{ddos} traffic often looks like normal network traffic, both detection and mitigation process are usually computationally expensive. Indeed, the detection requires complex algorithms and long training sessions to achieve the desired level of accuracy. The mitigation involves inspecting large volumes of traffic and comparing the traffic's characteristics with the information contained in large databases. Hence, the second part of this thesis focuses on the specific problem of the defence against \ac{ddos} attacks, by studying novel approaches for attack detection and mitigation that are suitable for devices with limited computing, memory and storage capabilities, such as the nodes located at the edge of the \ac{tsp} network. 

\noindent The remainder of this thesis is organised as follows:
\begin{itemize}
	\item In Chapter \ref{sec:application-aware}, we present \ac{pess}, a dynamic and application aware approach to provision security services by means of chains of \acp{vsnf}. \ac{pess} defines a mathematical formulation and a heuristic algorithm to tackle the provisioning problem in dynamic network scenarios, where the service requests are not known in advance. In contrast, advance knowledge of service requests is assumed by the majority of related works. \ac{pess} is evaluated in terms of quality of the solutions (deviation from optimality) and scalability performed on real-world and randomly generated topologies.
	\item In Chapter \ref{sec:detection-cnn}, we propose a lightweight \ac{dl}-based \ac{ddos} detection architecture suitable for online resource-constrained environments, which leverages \acfp{cnn} to learn the behaviour of \ac{ddos} and benign traffic flows with both low processing overhead and attack detection time. We call our model \acs{lucid} (\aclu{lucid}). \ac{lucid} is compared to state-of-the-art solutions and validated on a resource-constrained hardware platform to demonstrate the applicability of the approach in edge computing scenarios. 
	\item In Chapter \ref{sec:mitigation-xdp}, we study technological solutions to mitigate volumetric \ac{ddos} attacks in medium-sized servers, whose computing and memory resources are comparable to those of servers available in micro data centres at the edge of the \ac{tsp} network. We first analyse various approaches that can be used for an efficient and cost-effective \ac{ddos} mitigation. Then, we describe the design and the implementation of a \ac{ddos} mitigation pipeline that leverages the flexibility and efficiency of \ac{xdp} and the performance of the hardware-based filtering to handle large amounts of traffic and attackers.
	\item Chapter \ref{sec:conclusions} draws the conclusion of the thesis and highlights open issues and challenges to be further investigated.
\end{itemize}

\makeatletter
\algrenewcommand\ALG@beginalgorithmic{\small} 
\makeatother

\newcolumntype{L}[1]{>{\raggedright\let\newline\\\arraybackslash\hspace{0pt}}m{#1}}
\newcolumntype{C}[1]{>{\centering\let\newline\\\arraybackslash\hspace{0pt}}m{#1}}
\newcolumntype{R}[1]{>{\raggedleft\let\newline\\\arraybackslash\hspace{0pt}}m{#1}}

\graphicspath{{application-centric-security/artworks/}{application-centric-security/graphs/}}
\DeclareGraphicsExtensions{.pdf,.jpeg,.png}

\section{Application-Aware Security Services in Softwarised Networks}\label{sec:application-aware}

\lettrine[findent=2pt]{\textbf{I}}{}n this chapter, we present \acs{pess} (\aclu{pess}), a solution to efficiently deploy chains of \acp{vsnf} based on the \ac{qos} and security requirements of individual applications and operators' policies, while optimising resource utilisation. \ac{pess} defines an \ac{ilp} formulation for the progressive provisioning of security services (the \ac{pess} \ac{ilp} model), where the objective function requires minimisation of the usage of network and computing resources, and it is subject to routing, resource, \ac{qos} and security constraints. Moreover, \ac{pess} implements a heuristic algorithm, called \ac{pess} heuristic, to obtain near-optimal solutions of the provisioning problem in an acceptable time frame (in the order of a few milliseconds even in large network scenarios).
Although the \ac{pess} formulation and implementation presented in this chapter focus on security-specific services, the proposed approach is also suitable for more complex scenarios, where heterogeneous network services provided by means of generic \acp{vnf} coexist (e.g., security, video broadcasting, content caching, etc.).  

We prove that \ac{pess} can deploy more security services over the same infrastructure compared to an application-agnostic approach (the baseline), while still respecting the security policies and best practices defined by the \ac{tsp}.
With \ac{pess}, each traffic flow generated by each application can be served by the strict subset of \acp{vsnf} that are necessary to ensure its security, which means that no flow is burdened with any unnecessary security function that could affect its smooth execution, as would be the case with an application-agnostic approach. 
Of course, the capability of the \acp{vsnf} to properly contrast any security attack depends on the specific implementation of the \ac{vsnf} itself. This aspect is investigated in Chapters \ref{sec:detection-cnn} and \ref{sec:mitigation-xdp} in the context of \ac{ddos} attack detection and mitigation, respectively.

The study detailed in this chapter has been carried out in collaboration with Queen's University Belfast's Centre for Secure Information Technologies. Moreover, the results have been presented at the third IEEE International Workshop on Security in NFV-SDN \cite{pess-short-paper} and published in the IEEE Transactions on Network and Service Management \cite{pess-tnsm-paper}.  

The remainder of this chapter is structured as follows: Section \ref{sec:background} gives the relevant background information. Section \ref{sec:motivation} provides the motivation behind this work.
Section \ref{sec:model} details the mathematical formulation of the \ac{pess} \ac{ilp} model, while Section \ref{sec:implementation} describes the \ac{pess} heuristic algorithm that we implemented to solve the problem. In Section \ref{sec:pess-evaluation}, the heuristic algorithm is evaluated on real-world and random topologies. Section \ref{sec:sota} reviews and discusses the related work.
\subsection{Background}\label{sec:background}
The work presented in this chapter is underpinned by two emerging network technologies; \acf{nfv} and \ac{sdn} and their integration to provision network security solutions.


\subsubsection{\acl{nfv}}\label{sec:nfv}
Today's network functions such as firewalling, \acf{dpi}, \acp{ids}, etc. are provided by specialised proprietary hardware appliances (also called \textit{middleboxes}) strategically deployed in the network.
The \ac{nfv} paradigm separates the network functions from the underlying hardware by moving the functions from specialised devices to off-the-shelf commodity equipment such as industry standard servers or high-performance network devices. Therefore, network services can be decomposed into multiple \acp{vnf} running on physical or virtual machines, which could be located in data centres, network nodes or at the end-user premises. 

In contrast to middleboxes, the configuration of which requires intensive and time-consuming manual intervention, \ac{nfv} allows an automated and flexible deployment of network functions on any NFV-enabled device. The lifecycle management of \acp{vnf} and hardware resources is achieved through a centralised software component called the Orchestrator.

\subsubsection{\acl{sdn}}
\ac{sdn} is often referred to as a paradigm for network environments where the control plane is physically separated from the data plane and a logically centralised control plane controls several devices. This differs from traditional networks in which nodes are autonomous systems unaware of the overall state of the network. In \ac{sdn} deployments, the nodes are remotely controlled via standard protocols (e.g., OpenFlow~\cite{openflow}) by a logically centralised intelligent module called the \textit{\ac{sdn} controller}, which bases routing decisions on a global (domain) view of the network.\\
The controller is a software component which runs on commodity hardware appliances and provides an open \ac{api} to program the network for configuration, monitoring and troubleshooting purposes~\cite{sdn-survey}. Such programmability enables automated and dynamic network configurability and fine-grained control of the traffic based on the values of the packets' header fields (e.g., source/destination IP/MAC addresses, VLAN tags, TCP/UDP ports, etc.). 

\subsubsection{Service Function Chaining}
Service Function Chaining (also known as Network Service Chaining) is a technique for selecting and steering data traffic flows through network services. The network services can be traffic management applications such as load balancing, or security applications such as those detailed in Section~\ref{sec:vsnf}. Service function chaining combines the capabilities of \ac{sdn} and \ac{nfv} to connect a distributed set of \acp{vnf}.

\subsubsection{A Taxonomy of Security VNFs}\label{sec:vsnf}
As introduced in Section~\ref{sec:nfv}, a \ac{vnf} is a software implementation of a network function which is deployed on a virtual resource such as a Virtual Machine. 
Table \ref{tab:vnf_taxonomy} provides a list of the most common security functions. Traditionally, the majority of these functions would have been implemented on dedicated hardware (middleboxes) to process the network traffic along the data path. Today, these functions are deployed as \acp{vnf}. Table \ref{tab:vnf_taxonomy} includes a short description of each \ac{vnf} and some of the publicly available open-source implementations or commercial products.

\begin{table}[!t]
	\centering
	\scriptsize
	\renewcommand{\arraystretch}{1.5}
	\begin{threeparttable}
		\begin{tabular}{>{\bfseries}llll}
			\hlineB{2}
			\textbf{\small VNF} & \textbf{\small Description} & \textbf{\small Use in security} & \textbf{\small Implementations}  \\ \hlineB{2}
			\small Antispam  & Email filtering             &   \begin{tabular}{@{}l@{}}Malware detection,\\Phishing prevention\end{tabular}     & \begin{tabular}{@{}l@{}}SpamAssassin, rspamd,\\ASSP, Juniper vSRX\end{tabular} \\ \hline
			\small Antivirus  &\begin{tabular}{@{}l@{}}Email, Web scanning,\\Endpoint security\end{tabular}   &   \begin{tabular}{@{}l@{}}Virus, Trojan,\\Malware detection\end{tabular} & \begin{tabular}{@{}l@{}}ClamAV, ClamWin,\\Juniper vSRX\end{tabular} \\ \hline
			\small DLP  & \begin{tabular}{@{}l@{}}Data Loss,\\Leakage Prevention\end{tabular}    &   Data exfiltration detection    & myDLP, OpenDLP
			\\ \hline
			\small DPI  & Payload analysis      & \begin{tabular}{@{}l@{}}Spam Filtering, Intrusion detection,\\DDoS detection, Malware detection,\\Security Analytics\end{tabular}              & \begin{tabular}{@{}l@{}}OpenDPI, nDPI,\\L7-filter, Libprotoident,\\PACE, NBAR, Cisco ASAv\end{tabular}          \\ \hline
			\small Honeypot  & \begin{tabular}{@{}l@{}}Traffic redirection\\and inspection\end{tabular}             &   \begin{tabular}{@{}l@{}}Spam filtering, Malware detection,\\SQL database protection,\\Security Analytics \end{tabular}    & \begin{tabular}{@{}l@{}}HoneyD, SpamD, Kippo,\\Kojoney, Dionaea, Glastopf \end{tabular}        \\ \hline
			\small IDS  & \begin{tabular}{@{}l@{}}Traffic inspection \\ (header and payload)\end{tabular}            &   \begin{tabular}{@{}l@{}}Intrusion detection, Malware\\detection, 	DDoS detection,\\Security Analytics \end{tabular}   & \begin{tabular}{@{}l@{}}Snort, Bro, Suricata, AIDE,\\ACARM-ng, OSSEC, Samhain,\\Cuckoo, Cisco ASAv\end{tabular}         \\ \hline
			\small IPS  & \begin{tabular}{@{}l@{}}Traffic filtering based on\\ header and payload\end{tabular}             &   \begin{tabular}{@{}l@{}}Intrusion prevention,\\ DDoS prevention \end{tabular}    & \begin{tabular}{@{}l@{}}Snort, Suricata, ACARM-ng,\\Fail2Ban, Juniper vSRX \end{tabular}        \\ \hline
			\small NAT\tnote{1}  & IP address mapping               &   Intrusion prevention  & Netfilter, IPFilter, PF        \\ \hline
			\small \begin{tabular}{@{}l@{}}Packet Filter\\Firewall\end{tabular}  & \begin{tabular}{@{}l@{}}Header-based\\packet filtering\end{tabular}         &    Intrusion prevention      & \begin{tabular}{@{}l@{}}Netfilter, nftables, NuFW,\\IPFilter, Juniper vSRX,\\ipfw, PF, VMWare vShield,\\Fortigate FW\end{tabular}         \\ \hline
			\small \begin{tabular}{@{}l@{}}Parental\\Control\end{tabular}  & \begin{tabular}{@{}l@{}}Media content filtering\end{tabular}             &   \begin{tabular}{@{}l@{}}Blocking access to\\ inappropriate content\end{tabular}      & \begin{tabular}{@{}l@{}}OpenDNS, SquidGuard,\\ DansGuardian, pfsense\end{tabular}         \\ \hline
			\small VPN Gateway &  \begin{tabular}{@{}l@{}}Site-to-site VPN connection\\over unsecured networks\end{tabular}  & Data Tunneling/Encryption &   \begin{tabular}{@{}l@{}}OpenVPN, strongSwan,\\Juniper vSRX, Cisco ASAv,\\Fortigate VPN\end{tabular}    \\ \hline
			\small WAF  & \begin{tabular}{@{}l@{}}HTTP traffic monitoring,\\filtering, logging\end{tabular}             &   \begin{tabular}{@{}l@{}}Prevention of SQL injection,\\cross-site scripting\end{tabular}      & ModSecurity         \\ \hlineB{2}
		\end{tabular}
		\begin{tablenotes}
			\item[1] NAT is not a security function but inherently provides packet filtering similar to a firewall.
		\end{tablenotes}
	\end{threeparttable}
	\caption{Taxonomy of security \acp{vsnf}.}	
	\label{tab:vnf_taxonomy}
\end{table}
\subsection{Motivation}\label{sec:motivation}

We motivate our work by describing two use case scenarios, namely web browsing and online gaming, where the \ac{tsp} exploits the \ac{nsc} and \ac{nfv} technologies to provide security services tailored to specific users' application requirements.

\textbf{Web browsing.} Parental control is applied to Web traffic to block unwanted media and social-media content, while an \ac{ids} might be used to intercept malicious software (malware). Stateful \acp{vsnf} provide security functionality by tracking the state of network connections (e.g., Layer 4 firewall, \ac{nat}). In this case, the same \ac{vsnf} instance must be traversed by all traffic flows of a network conversation in order to maintain the correct state of the connection. More flexible provisioning schemes can be adopted for stateless \acp{vsnf}, where multiple instances of the same \ac{vsnf} might be deployed on different servers for load balancing. This example also illustrates the security best-practice that unwanted traffic should be blocked as soon as it enters the network by placing firewalls and \ac{ids}/\ac{ips} close to the border of the \ac{tsp} domain. Another generally accepted practice, is to place firewalls before \ac{ids}/\ac{ips} (from the point of view of incoming traffic). Firewalls are generally designed to drop unauthorised traffic very quickly, thus reducing the burden on \ac{ids}/\ac{ips}, which are more computationally expensive.

\textbf{Online gaming.} An \ac{ids} might also be used to detect possible threats due to the misuse of chat tools integrated within the gaming software (e.g., phishing~\cite{Gianvecchio:2011:HBI:2109150.2109174}, social engineering~\cite{Yan:2005:SCC:1103599.1103606}, etc.). As the communication between the client and the server relies on timely delivery of packets, \ac{ids} operations are not executed on the \textit{in-game} traffic. In this case the security is enforced by a faster \ac{vsnf} such as a Firewall, which checks the packet headers without any deep-payload analysis. It should be noted that web traffic and chat conversations are often encrypted by TLS/SSL cryptographic protocols. Although encryption preserves the confidentiality of the traffic, it also prevents \ac{ids}-based \acp{vsnf} such as Parental Control and \ac{ids} from inspecting the packets, thus allowing an attacker to obfuscate malicious data in encrypted payloads. However, the \ac{tsp} could overcome this limitation either using a Transparent Proxy \ac{vsnf} or by exploiting recent advances in network security~\cite{Sherry:2015:BDP:2785956.2787502,Canard:2017:BMP:3052973.3053013}.\\

\begin{table*}[h!]
	\centering
	\scriptsize
	\renewcommand{\arraystretch}{1.5}
	\begin{threeparttable}
		\begin{tabular}{>{\bfseries}lllll}
			\hlineB{2}
			\textbf{Application class} & \textbf{Description} & \textbf{Related threats}
			& \textbf{Relevant \ac{vsnf} \tnote{1}} & \textbf{\ac{qos} requirements}\\ \hlineB{2}
			CCTV systems  & \begin{tabular}{@{}l@{}}Closed Circuit TV\\for video surveillance\\accessible remotely \end{tabular}  &   \begin{tabular}{@{}l@{}}Port scanning, DDoS\\ password cracking \end{tabular}   & \begin{tabular}{@{}l@{}}Firewall, DPI,\\ IDS, IPS \end{tabular}     &  \begin{tabular}{@{}l@{}}Bandwidth: 10Mbps\\ (5 cameras, 720p, 15fps,\\ H.264, medium quality)\\ Latency: 200ms (PTZ\tnote{2} \\two-way latency~\cite{5384976})\end{tabular}  \\ \hline
			Email  & Electronic mail  &   \begin{tabular}{@{}l@{}}Malware, spam,\\phishing,\\data exfiltration  \end{tabular}   & \begin{tabular}{@{}l@{}}DPI, Antispam,\\IDS, DLP\end{tabular}   &   --   \\ \hline
			Instant messaging  & \begin{tabular}{@{}l@{}}Real-time text-based\\Internet chat\end{tabular}      & \begin{tabular}{@{}l@{}}Malware, DDoS,\\phishing (out-of-band)\end{tabular}              & \begin{tabular}{@{}l@{}}DPI, Antispam,\\IDS, IPS\end{tabular}     &  --   \\ \hline
			\begin{tabular}{@{}l@{}}Media streaming\end{tabular}  & \begin{tabular}{@{}l@{}}Audio/video content\\accessed over\\the Internet\end{tabular}  &   \begin{tabular}{@{}l@{}}Inappropriate content  \end{tabular}   & \begin{tabular}{@{}l@{}}Parental control\end{tabular}   &   \begin{tabular}{@{}l@{}} Bandwidth\tnote{3}: 5Mbps (HD)\\25Mbps (UHD)  \end{tabular}   \\ \hline
			\begin{tabular}{@{}l@{}}Remote storage\end{tabular}  & \begin{tabular}{@{}l@{}}File transfer over\\the network\end{tabular}  &   \begin{tabular}{@{}l@{}}Data exfiltration  \end{tabular}   & \begin{tabular}{@{}l@{}}VPN, Data\\Encryption\end{tabular}   &   \begin{tabular}{@{}l@{}} Bandwidth  \end{tabular}   \\ \hline
			\begin{tabular}{@{}l@{}}Network services\\ (DNS, VoD, file\\sharing, WWW)\end{tabular}  & \begin{tabular}{@{}l@{}}Server application\\ accessed by remote\\ client applications\end{tabular}  &   \begin{tabular}{@{}l@{}}DDoS, SQL injection,\\ remote code execution  \end{tabular}   & \begin{tabular}{@{}l@{}}Firewall, IDS,\\ WAF, Honeypot\end{tabular}     &  --  \\ \hline
			Online gaming  & \begin{tabular}{@{}l@{}}Video games played\\ over the Internet\end{tabular}             &   \begin{tabular}{@{}l@{}}Online game cheating\\(out-of-band attacks)\\DDoS (in-band attacks)\end{tabular}     & \begin{tabular}{@{}l@{}}DPI, Antispam,\\ IDS, IPS\end{tabular}    &  \begin{tabular}{@{}l@{}}Latency: 100ms\\ (first-person games~\cite{Claypool:2006:LPA:1167838.1167860})\end{tabular}   \\  \hline
			Peer-to-peer  & \begin{tabular}{@{}l@{}}File sharing over\\peer-to-peer networks\end{tabular}               &   DDoS, malware  & DPI, IDS, IPS   &   --  \\ \hline
			\begin{tabular}{@{}l@{}}Video conferencing\end{tabular}  & \begin{tabular}{@{}l@{}}Real-time audio/video\\over the Internet\end{tabular}  &   \begin{tabular}{@{}l@{}}DDoS  \end{tabular}   & \begin{tabular}{@{}l@{}}Firewall, IPS\end{tabular}   &   \begin{tabular}{@{}l@{}} Latency: 150ms~\cite{Chen:2004:QRN:1234242.1234243}  \end{tabular}   \\ \hline
			Web browsing  & \begin{tabular}{@{}l@{}}Applications for\\browsing the  WWW\end{tabular}         &   \begin{tabular}{@{}l@{}} Cross-site scripting,\\phishing, malware,\\inappropriate content \end{tabular}     & \begin{tabular}{@{}l@{}} DPI, WAF,\\ Parental control\end{tabular}    &   \begin{tabular}{@{}l@{}} Latency: 400ms~\cite{Chen:2004:QRN:1234242.1234243} \end{tabular}  \\ \hlineB{2}
		\end{tabular}
		\begin{tablenotes}
			\item $^1$Acronyms of \acp{vsnf}: \acf{dpi}, \acf{ids}, \acf{ips}, \acf{dlp}, \acf{vpn}, \acf{waf}. 
			\item $^2$PTZ: Pan, Tilt, Zoom. 
			\item $^3$Netflix Internet connection speed recommendations.
		\end{tablenotes}
	\end{threeparttable}
	\caption{Security and \ac{qos} requirements of applications.}
	\label{tab:tspapps}
	\vspace{-3mm}
\end{table*}

These are just two examples of how the security service can be tailored to the user's application requirements by appropriate selection and placement of \acp{vsnf}. A list of common classes of applications supported in the \ac{tsp} use-case is provided in Table \ref{tab:tspapps} with their corresponding security and QoS requirements and relevant \acp{vsnf}. 

One of them, the remotely accessible CCTV system, will be used in the rest of this chapter as a running example to illustrate various aspects of our work.

According to the motivations provided above, we can summarize the rationale behind the \ac{pess} approach as follows: (i) a user's application should never under-perform because of \ac{vsnf} operations and (ii) the \ac{vsnf} placement must obey the \ac{tsp}'s security best-practices in terms of application security requirements, position in the network, operational mode (stateless or stateful \ac{vsnf}), and order with respect to the direction of the traffic. In the next section, we present the \ac{pess} mathematical model for the placement of \ac{vsnf} chains based on these criteria.

\subsection{The PESS Optimal Placement Model}\label{sec:model}

\begin{figure}[!h]
	\begin{center}
		\includegraphics[width=0.75\textwidth]{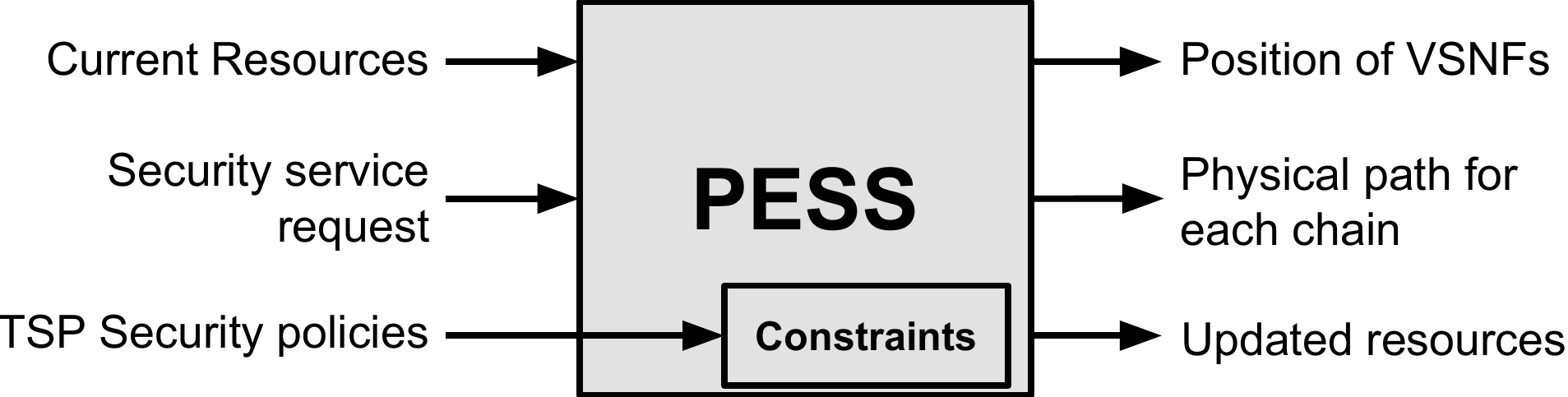}
		\caption{PESS placement model workflow.}
		\label{fig:PESS-chart}
	\end{center}
	\vspace{-5mm}
\end{figure}

The \ac{pess} model (Figure~\ref{fig:PESS-chart}) is a mathematical model to progressively embed service requests, formed by one or multiple \ac{vsnf} chains, onto a physical network substrate by considering the available resources and realistic constraints.

PESS takes as input a model of the physical network including the current status of computing and network resources of servers and links, a security service request and the \ac{tsp}'s security policies (expressed in the form of constraints for PESS). The output of PESS is the mapping of the \acp{vsnf} onto the physical network (position of the \acp{vsnf} and one or more paths between them) and an updated model of the physical network taking into account the resources used to provision the service. The updated model is used as input for the next request.

Next, we detail definitions, notations, variables, objective function and constraints that are used in the \ac{ilp} formulation of the \ac{pess} optimal placement model. Notations and variables are also summarised in Table  \ref{tab:pess-notations}.

\begin{table}[]
	\centering
	\small
	\renewcommand{\arraystretch}{1.5}
	\caption{Glossary of symbols.}
	\vspace{-2mm}
	\label{tab:pess-notations}
	\begin{tabular}{c}
		\textit{Sets}
	\end{tabular}
	\\
	\begin{tabular}{|L{2cm}|L{13cm}|}
		\hline
		\textit{N} & Set of physical nodes \\
		\hline
		\textit{E} & Set of physical links \\
		\hline
		\textit{$C$} & Set of all unidirectional chains already embedded in the network \\
		\hline
		\textit{$C_s$} & Set of all unidirectional chains in the service request $\mathcal{G}_s$\\
		\hline
		\textit{$U^c$} & Set of virtual nodes in the chain $c$ \\
		\hline
		$U^c_{pairs}$ & Set of unidirectional arcs in the chain $c$ \\
		\hline
		\textit{$A^c$} & Set of endpoints of the chain $c$. $A^c\subset U^c$ \\
		\hline
		\textit{$V^c$} & Set of \acp{vsnf} in the chain $c$. $V^c\subset U^c$ \\
		\hline
		\textit{$R_u$} & Region of $N$ where \ac{vsnf} $u$ must be placed ($region$ constraint) \\
		\hline
		\textit{$M$} & Region of $N$ where no \acp{vsnf} can be placed ($veto$ constraint) \\
		\hline
		\textit{$ep1, EP2$} & Physical endpoints of a service request. $ep1\in N, EP2\subset N$ \\
		\hline
		
	\end{tabular}
	\\
	\vspace{1mm}
	\begin{tabular}{c}
		\textit{Parameters}
	\end{tabular}
	\\
	\begin{tabular}{|L{2cm}|L{13cm}|}
		\hline
		\textit{$\gamma_i$} & Nominal computing resources of node $i$ (CPU cycles/sec)\\
		\hline
		\textit{$\gamma'_i$} & Residual computing resources of node $i$ (CPU cycles/sec) \\
		\hline
		\textit{$\gamma_u$} & CPU cycles required by $u$ to process one bit of a network packet (CPU cycles/bit)\\
		\hline
		\textit{$\gamma^c_u$} & Computing resources required by node $u$ of chain $c$ (CPU cycles/sec). $\gamma^c_u = \gamma_u\cdot \beta^c$\\
		\hline
		\textit{$\beta_{k,l}$} & Nominal capacity of link $(k,l)$ (bits/sec) \\
		\hline
		\textit{$\beta'_{k,l}$} & Residual capacity of link $(k,l)$ (bits/sec)\\
		\hline
		\textit{$\beta^c$} & Minimum bandwidth required by chain $c$  (bits/sec)\\
		\hline
		\textit{$\lambda_{k,l}$} & Propagation delay: the time spent by a packet to traverse the link $(k,l)$ (secs)\\
		\hline
		\textit{$\lambda^c_{k,l,i,j}$} & Queuing delay: time spent by a packet of chain $c$ to traverse the network devices (routers and switches) in the local networks of adjacent nodes $k$ and $l$ (secs). $\lambda^c_{k,l,i,j}>0$ iff $k=i$ or $l=j$.\\
		\hline
		\textit{$\lambda^c_{i,u}$} & Processing delay: the time spent by a packet to traverse \ac{vsnf} $u$ of chain $c$ placed on node $i$ (secs)\\
		\hline
		\textit{$\lambda^c$} & Maximum latency tolerated by chain $c$ (secs)\\
		\hline
		\textit{$\pi^c$} & Estimated latency between the \ac{tsp} network and the remote endpoint of chain $c$ (secs). $\pi^c=0$ if the endpoint belongs to the \ac{tsp} network.\\
		\hline
		\textit{$\sigma^c$} & Average packet size of chain $c$ (bits).\\
		\hline
		\textit{$b_{k,l}$} & Cost for allocating a unit of bandwidth on link \textit{(k,l)}\\
		\hline
		\textit{$c_{i}$} & Cost for allocating a unit of CPU on node \textit{i}\\
		\hline
	\end{tabular}
	\\
	\vspace{1mm}
	\begin{tabular}{c}
		\textit{Decision variables}
	\end{tabular}
	\\
	\begin{tabular}{|L{2cm}|L{13cm}|}
		\hline
		$x^c_{i,u}$ & Binary variable such that $x^c_{i,u}=1$ iff node $u\in U^c$ is mapped to $i\in N$ \\
		\hline
		$y^c_{k,l,i,j,u,v}$ & Binary variable such that $y^c_{k,l,i,j,u,v}=1$ iff physical link $(k,l)\in E$ belongs to the path between nodes $i$ and $j$ to which $u,v\in U^c$ are mapped\\
		\hline
	\end{tabular}
\end{table}

\textbf{Physical network model.} We represent the physical network as a weighted graph $\mathcal{G} = (N,E)$, i.e. a graph where weights are assigned to nodes and edges.

Without loss of generality and to simplify the model, we assume that every node $i\in N$ is a NFVI-POP (Network Function Virtualisation Infrastructure Point of Presence) \cite{nfvipop} consisting of a set of servers and a local network composed of routers and switches. Each node $i$ is characterised by the total computing resources of the servers $\gamma_i\in\mathbb{N}^+$ expressed in CPU cycles/sec.

A link $(k,l)\in E$ is a wired connection between two nodes $k$ and $l$ $\in N$. 
It is characterised by its capacity $\beta_{k,l}\in\mathbb{N}^+$ and its propagation delay $\lambda_{k,l}\in\mathbb{N}^+$. Both are expressed as positive integer numbers representing bandwidth (bits/sec) and latency (sec). 

\textit{Regions} in a physical network are defined as subsets of nodes sharing some high-level features. Examples of regions are: (i) a set of nodes in the TSP network providing the same cloud service (e.g., multimedia caching, data storage, etc.), (ii) the set of egress nodes that connect the TSP network to the Internet (called \textit{border region} in the rest of this chapter)\roberto{, or (iii) the set of nodes at the edge of the \ac{tsp} network close to a given user}.

\textbf{Security service request.} We model a security service request as a set of independent weighted directed graphs:
$$\mathcal{G}_s = \{(U^c,U_{pairs}^c): c\in C_s\}$$
where $C_s$ is the set of unidirectional chains composing the service request.  
Each graph includes nodes and arcs. Nodes $U^c=A^c\cup V^c$ comprise user and remote applications ($A^c$, the endpoints of chain $c$) as well as a subset of all \acp{vsnf} ($V^c$). Each arc in $U_{pairs}^c$ delineates the order of traversing the \acp{vsnf} $\in V^c$ between endpoints in $A^c$.

Each chain $c\in C_s$ is characterised by its requirements in terms of minimum bandwidth $\beta^c$ and maximum latency $\lambda^c$. Each endpoint in $A^c$ is characterised by an identifier, which specifies where the endpoint must be placed in the physical network. The user application is characterised by the identifier of the physical node to which the user is attached (called \textit{ep1} in the rest of the chapter). A remote application is characterised by the identifier of a region in the physical network (called \textit{EP2}). For instance, the border region if the endpoint represents a remote gaming server located outside the physical network. In this work, $ep1$ and $EP2$ are referred to as \textit{physical endpoints} of the service request $\mathcal{G}_s$.\\
A \ac{vsnf} $u\in V^c$ is characterised by its requirements in terms of CPU units $\gamma_u$ expressed in CPU cycles/bit. $u$ is also characterised by the latency $\lambda^c_{i,u}$ it introduces in the dataplane to process a packet of chain $c$ on node $i$. As formalised in Equation (\ref{eq:cpu_latency1}), this latency is a function of the residual computing capacity of the node $i$ where $u$ is placed, the computing requirements $\gamma_u$ of the \ac{vsnf}, the average packet size $\sigma^c$ of chain $c$ and the traffic load of the chain (whose upper bound is $\beta^c$). Finally, a \ac{vsnf} is characterised by its operational mode (either stateless or stateful) and by the identifier of a region in the physical network where it must be placed, if required by the \ac{tsp} security policies.

\textbf{Illustrative example.} An example of a security service request for a CCTV system (see Table~\ref{tab:tspapps}) is represented in Figure~\ref{fig:security_service_request}. The request in the example is composed of three chains (c1, c2, and c3), each one identified by the type of traffic and its direction.

\begin{figure}[h!]
	\begin{center}
		\begin{tikzpicture}[every node/.style={font=\sffamily\scriptsize}, vnf/.style={circle,thick,draw,fill={rgb,255:red,200; green,200; blue,200}},app/.style={rectangle,thick,draw,font=\sffamily\scriptsize,minimum size=1cm}]
		
		\node[app] (1) {CCTV};
		\node[vnf,text width=0.38cm] (2) at (5,0) {FW\\($\gamma_2$)};
		\node[vnf,text width=0.38cm] (3) at (2.5,-2) {IPS\\($\gamma_1$)};
		\node[app,text width=1cm] (4) at (9,0) {Remote Access};
		
		\path[->,>=triangle 45,thick]
		(1) edge [bend left=50] node[midway,fill=white] {$\beta^{c1},\lambda^{c1}$} (2)
		(2) edge [bend left=60] node[fill=white] {$\beta^{c1},\lambda^{c1}$} (4);
		\path[every node/.style={font=\sffamily\scriptsize},->,>=open triangle 45,thick]
		(1) edge [bend left] node[fill=white] {$\beta^{c2},\lambda^{c2}$} (3)
		(3) edge [bend left] node[fill=white] {$\beta^{c2},\lambda^{c2}$} (2)
		(2) edge [bend left] node[fill=white] {$\beta^{c2},\lambda^{c2}$} (4);
		\path[every node/.style={font=\sffamily\scriptsize},->,>=open triangle 45,thick]
		(4) edge [bend left] node[fill=white] {$\beta^{c3},\lambda^{c3}$} (2)
		(2) edge [bend left] node[fill=white] {$\beta^{c3},\lambda^{c3}$} (3)
		(3) edge [bend left] node[fill=white] {$\beta^{c3},\lambda^{c3}$} (1);
		
		\draw[->,>=triangle 45,thick] (5,-1.5) -- (9.5,-1.5) node[midway,fill=white] {Video Stream};
		\draw[->,>=open triangle 45,thick] (5,-1.9) -- (9.5,-1.9) node[midway,fill=white] {Camera Mgmt/Controls};
		\end{tikzpicture}
		\caption{Example of security service request for the CCTV system.}
		\label{fig:security_service_request}
	\end{center}
\end{figure}

Chain \textit{c1} is applied to the live video stream captured by the cameras and accessible over the Internet. The chain comprises a L3 firewall to ensure that the stream is only transmitted to authorised endpoints. As specified in Table~\ref{tab:tspapps}, the most relevant requirement in this case is the bandwidth ($\beta^{c1}$) which depends on the frame rate, frame size and video codec of the CCTV system. In this case, a deep inspection of the video stream packets (e.g., with an \ac{ips}) would not provide any additional protection but would possibly reduce the frame rate of the video streaming, thus compromising the detection of anomalous events. On the other hand, the bi-directional control/management traffic is inspected by the \ac{ips} and the firewall included in chains $c2$ and $c3$. Such \acp{vsnf} protect the CCTV system from attacks such as Mirai \cite{mirai} perpetrated through bots maliciously installed on Internet-connected devices, while the latency requirements $\lambda^{c2}$ and $\lambda^{c3}$ guarantee the responsiveness of the remote control of the CCTV cameras (pan, tilt, zoom, etc.).

\subsubsection{ILP Formulation}

\noindent\textbf{Definitions.} Let us first define two binary variables:
\begin{itemize}
	\item $x^c_{i,u}=1$ iff node $u\in U^c$ is mapped to $i\in N$.
	\item $y^c_{k,l,i,j,u,v}=1$ iff physical link $(k,l)\in E$ belongs to the path between nodes $i$ and $j$ to which $u,v\in U^c$ are mapped.
\end{itemize}

The residual capacity of a link, $\beta'_{k,l}$, is defined as the total amount of bandwidth available on link $(k,l)\in E$:
\begin{equation}\label{eq:residual_bandwidth}
\beta'_{k,l} = \beta_{k,l} - \sum_{\mathclap{\substack{c\in C,\ i,j\in N\\(u,v)\in U^c_{pairs}}}}\beta^c\cdot y^c_{k,l,i,j,u,v}
\end{equation}

thus, it is the nominal capacity of link $(k,l)$ minus the bandwidth required by the chains $c\in C$ already mapped on that link.

Similarly, the residual capacity of a node is defined as its nominal CPU capacity minus the computing resources used by the \acp{vsnf} $v$ instantiated on the node:
\begin{equation}\label{eq:residual_cpu}
\gamma'_i = \gamma_i - \sum_{\mathclap{c\in C, u\in V^c}} \gamma^c_u\cdot x^c_{i,u}
\end{equation}

\textbf{Problem formulation.} Given a physical network $\mathcal{G}$, for each security service request $\mathcal{G}_s$, find a suitable mapping of all its unidirectional chains on the physical network, which minimizes the physical resources of $\mathcal{G}$ expended to map $\mathcal{G}_s$, also known as the \textit{embedding cost}.\\
Hence, the solution of the problem is represented by a set of $x^c_{i,u}$ and $y^c_{k,l,i,j,u,v}$ such that the cumulative usage of physical resources for all the chains in $\mathcal{G}_s$ is minimised:

\begin{equation}\label{eq:objective_function}
\min\quad\sum_{\mathclap{\substack{c\in C_s,\ i,j\in N,\\(k,l)\in E,(u,v)\in U^c_{pairs}}}}b_{k,l}\cdot \beta^c\cdot y^c_{k,l,i,j,u,v}+\alpha\sum_{\mathclap{\substack{c\in C_s,i\in N, u\in V^c}}}c_{i}\cdot \gamma^c_u\cdot x^c_{i,u}
\end{equation}

Here, $\alpha$ is a factor that can be used to tune the relative weight of the cost components (we have used $\alpha=1$ for the experiments described in Section~\ref{sec:pess-evaluation}).\\
$b_{k,l}$ and $c_{i}$ are the costs for allocating bandwidth and CPU: 
$$b_{k,l}=\frac{1}{\beta'_{k,l}+\delta}\quad c_{i}=\frac{1}{\gamma'_i+\delta}$$
They penalize nodes and links with less residual capacity with the aim to increase the chances of accommodating more security service requests on the given physical network. $\delta\longrightarrow 0$ is a small positive constant used to avoid dividing by zero in computing the value of the function.\\

\subsubsection{Constraints}\label{sec:constraints}
\textbf{Routing Constraint} (\ref{eq:const_1}) ensures that each node $u\in U^c$ is mapped to exactly one physical node $i\in N$. With Constraint (\ref{eq:const_2}), a physical link $(k,l)$ can belong to a path between two nodes $i$ and $j$ for an arc $(u,v)\in U_{pairs}^c$ of chain $c\in C_s$ only if $u$ and $v$ are mapped to these nodes. Constraint  (\ref{eq:const_3}) ensures that the path created for arc $(u,v)$ starts at exactly one edge extending from node $i$ to where \ac{vsnf} (or start/endpoint) $u$ is mapped. Similarly, (\ref{eq:const_4}) ensures the correctness and the uniqueness of the final edges in the path. Constraints (\ref{eq:const_2}-\ref{eq:const_4}) can be easily linearised with standard techniques such as the ones presented in~\cite{Sherali2013}. Constraint (\ref{eq:const_5a}) is the classical \textit{flow conservation constraint}. That is, an outbound flow equals an inbound flow for each intermediate node $l$ (intermediate nodes cannot consume the flow). Together with Constraint (\ref{eq:const_5a}), Constraint (\ref{eq:const_5b}) prevents multiple incoming/outgoing links carrying traffic for a specific flow in the intermediate node $l$, i.e., we only consider unsplittable flows.

\begin{fleqn}[2pt]
	\begin{align}
	\begin{split}\label{eq:const_1}
	&\sum_{i\in N} x^c_{i,u}=1\qquad\forall c\in C_s,\forall u\in U^c
	\end{split}\\[2ex]
	\begin{split}\label{eq:const_2}
	& y^c_{k,l,i,j,u,v}\le x^c_{i,u}\cdot x^c_{j,v}\qquad\forall c\in C_s, \forall i,j\in N, \forall (u,v)\in U^c_{pairs}, \forall (k,l)\in E 
	\end{split}\\[2ex]
	\begin{split}\label{eq:const_3}
	&\sum_{\mathclap{\substack{(i,k)\in E\\ j\in N}}}y^c_{i,k,i,j,u,v}\cdot x^c_{i,u}\cdot x^c_{j,v}=1\qquad\forall c\in C_s, \forall (u,v)\in U^c_{pairs}
	\end{split}\\[2ex]
	\begin{split}\label{eq:const_4}
	&\sum_{\mathclap{\substack{(k,j)\in E\\ i\in N}}}y^c_{k,j,i,j,u,v}\cdot x^c_{i,u}\cdot x^c_{j,v}=1\qquad\forall c\in C_s, \forall (u,v)\in U^c_{pairs}
	\end{split}
	\end{align}
\end{fleqn}

\begin{fleqn}[2pt]
	\begin{align}
	\begin{split}\label{eq:const_5a}
	&\sum_{\mathclap{\substack{k\in N\\(k,l)\in E}}}y^c_{k,l,i,j,u,v}=\sum_{\mathclap{\substack{m\in N\\(l,m)\in E}}}y^c_{l,m,i,j,u,v}\ \forall c\in C_s, \forall i,j\in N, \forall l\in N, l\ne i, l\ne j, \forall (u,v)\in U^c_{pairs} 
	\end{split}
	\end{align}
	\begin{align}
	\begin{split}\label{eq:const_5b}
	&\sum_{\mathclap{\substack{k\in N\\(k,l)\in E}}}y^c_{k,l,i,j,u,v}\leq 1\qquad\forall c\in C_s, \forall i,j\in N, \forall l\in N, l\ne i, l\ne j, \forall (u,v)\in U^c_{pairs} 
	\end{split}
	\end{align}
\end{fleqn}


\textbf{Resource Constraints} (\ref{eq:const_8}-\ref{eq:const_9}) ensure that the resources consumed by a security service do not exceed the available bandwidth and computing capacities.
\begin{fleqn}[12pt]
	\begin{align}
	\begin{split}\label{eq:const_8}
	&\sum_{\mathclap{\substack{c\in C_s,\ i,j\in N\\(u,v)\in U^c_{pairs}}}}y^c_{k,l,i,j,u,v}\cdot \beta^c\le \beta'_{k,l}\quad \forall (k,l)\in E
	\end{split}\\[0.5ex]
	\begin{split}\label{eq:const_9}
	&\sum_{\mathclap{c\in C_s, u\in V^c}} x^c_{i,u}\cdot \gamma^c_u\le \gamma'_i\quad\forall i\in N
	\end{split}
	\end{align}
\end{fleqn}

\textbf{\ac{qos} Constraint} (\ref{eq:const_latency}) verifies that the requirements in terms of maximum end-to-end latency are met. It takes into consideration the propagation delay of physical links, the processing delay of \acp{vsnf} and the queuing delay through network devices. 
Note that the minimum bandwidth requirement is verified against the bandwidth resource Constraint (\ref{eq:const_8}).
	
\begin{equation}\label{eq:const_latency}
	\pi^c+\sum_{\mathclap{i\in N, u\in V^c}} x^c_{i,u}\cdot\lambda^c_{i,u}+\sum_{\mathclap{\substack{i,j\in N,(k,l)\in E\\(u,v)\in U^c_{pairs}}}}y^c_{k,l,i,j,u,v}\cdot (\lambda_{k,l}+\lambda^c_{k,l,i,j})\le \lambda^c\qquad\forall c\in C_s
\end{equation}

$\pi^c$ is an estimation of the propagation delay between the \ac{tsp} network and the remote endpoint of chain $c$, in case the endpoint is outside the \ac{tsp} network. We assume that this value is independent from the \ac{tsp}'s network egress node. Clearly $\pi^c$ is $0$ for those chains whose remote endpoint is part of the \ac{tsp} network (e.g., a cloud data centre managed by the \ac{tsp}). 

The processing delay $\lambda^c_{i,u}$ is the time spent by a packet to traverse \ac{vsnf} $u$ on physical node $i$. It contributes to the overall end-to-end delay of chain $c$ only if \ac{vsnf} $u$ is placed on node $i$ (i.e., $x^c_{i,u} = 1$).
$\lambda^c_{i,u}$ includes the time taken by the \ac{vsnf} to process the packet and the overhead of the virtualisation technology (VMware, KVM, QEMU virtual machines, Docker containers, etc.). For simplicity, we do not model the delays due to the CPU scheduler operations  implemented on the physical node~\cite{8704949}. 
Based on the observations in \cite{GAO2018108,oljira-qos-aware}, \cite{7842188}, $\lambda^c_{i,u}$ is modeled as a convex function of the traffic load of the chain, and its value is computed by considering the impact of other \acp{vsnf} co-located on the same physical node.
\begin{equation}\label{eq:cpu_latency1}
\lambda^c_{i,u} = \frac{\gamma_u\cdot \sigma^c}{(\gamma'_i - \gamma_u\cdot\beta^c)+\delta}  = \frac{\gamma_u\cdot \sigma^c}{(\gamma'_i - \gamma^c_u)+\delta}
\end{equation}

In Equation (\ref{eq:cpu_latency1}), $\gamma_u\cdot\sigma^c$ is the average amount of CPU cycles used by \ac{vsnf} $u$ to process a packet of chain $c$ (virtualisation overhead included). The latency overhead caused by co-located \acp{vsnf} depends on the amount of computing resources of the node they use or, equivalently, on the residual computing resources of the node $\gamma'_i$. $\gamma_u\cdot\beta^c=\gamma^c_u$ is the amount of CPU cycles/sec used by \ac{vsnf} $u$ on node $i$, which depends on the traffic load of the chain.  
$\delta$ is a small positive constant used to avoid dividing by zero in the case that $u$ consumes all the residual computing resources of node $i$.
	
The sum $\lambda_{k,l}+\lambda^c_{k,l,i,j}$ in Equation (\ref{eq:const_latency}) is the total time spent by a packet travelling between two adjacent nodes $k$ and $l$. It includes the propagation delay $\lambda_{k,l}$, proportional to the distance between $k$ and $l$, and the queuing delay $\lambda^c_{k,l,i,j}$, proportional to the number of network devices (switches and routers) the packet traverses within the local networks of $k$ and $l$. The queuing delay is influenced by the buffer size of network devices' ports and by the traffic load \cite{silo}. For the sake of simplicity, we assume that the buffers are correctly dimensioned, i.e. no dropped packets due to buffer overflow. In addition, we estimate the queuing delay $\lambda^c_{k,l,i,j}$ as a traffic-load independent value; a function of the maximum queue capacity of the ports and of the \acp{vsnf} placement (hence a function of indices $k,l,i$ and $j$). Specifically, $\lambda^c_{k,l,i,j}>0$ if at least one \ac{vsnf} is mapped either on $k$ ($k=i$), or on $l$ ($l=j$), meaning that a packet of chain $c$ must traverse the local network of either $k$, or $l$ (or both) to reach the \acp{vsnf} running on the nodes' servers. Otherwise, the local networks of $k$ and $l$ are by-passed by the traffic of $c$, resulting in $\lambda^c_{k,l,i,j}=0$.

Constraint (\ref{eq:check_latency}) ensures that the current security service $\mathcal{G}_s$ does not compromise the end-to-end latency of chains $\hat{c}\in C$ in operational security services (also called \textit{operational chains} in the rest of the chapter). 

\begin{equation}\label{eq:check_latency}
	\pi^{\hat{c}}+\sum_{\mathclap{i\in N, \hat{u}\in V^{\hat{c}}}} \bar{x}^{\hat{c}}_{i,\hat{u}}\cdot \lambda^{\hat{c}}_{i,\hat{u}}+\sum_{\mathclap{\substack{i,j\in N,(k,l)\in E\\(\hat{u},\hat{v})\in U^{\hat{c}}_{pairs}}}}\bar{y}^{\hat{c}}_{k,l,i,j,\hat{u},\hat{v}}\cdot (\lambda_{k,l}+\lambda^{\hat{c}}_{k,l,i,j})\le \lambda^{\hat{c}}\qquad\forall {\hat{c}}\in C\quad\qquad
\end{equation}

In Equation (\ref{eq:check_latency}), $\bar{x}^{\hat{c}}$ and $\bar{y}^{\hat{c}}$ are the values of decision variables $x$ and $y$ computed for the placement of chain $\hat{c}$. $\lambda^{\hat{c}}_{i,\hat{u}}$ is the updated value of the processing delay introduced to the traffic of chain $\hat{c}$ by \ac{vsnf} $\hat{u}$ when running on node $i$. 

\begin{equation}\label{eq:cpu_latency_update}
\lambda^{\hat{c}}_{i,\hat{u}} = \frac{\gamma_{\hat{u}}\cdot \sigma^{\hat{c}}}{(\gamma'_i - \sum\limits_{\makebox[0pt]{$\scriptstyle \substack{c\in C_s,u\in V^c}$}}{x^c_{i,u}\cdot\gamma^c_u)}+\delta}
\end{equation}

In Equation (\ref{eq:cpu_latency_update}), the value of $\lambda^{\hat{c}}_{i,\hat{u}}$ is updated by considering the computing resources consumed on node $i$ by \acp{vsnf} of the security service request $\mathcal{G}_s$. Approximation of Equation (\ref{eq:cpu_latency_update}) can be achieved by using piecewise linearisation techniques and Special-Ordered Set (SOS) variables and constraints available in most commercial solvers (e.g., \cite{gurobi_sos}).

\textbf{Security constraints} ensure that the \ac{tsp}'s security policies are applied. Specifically, Constraint (\ref{eq:const_stateful}) forces a subset $C'_s$ of the chains in the request to share the same \ac{vsnf} instance in case of stateful flow processing. 

\begin{equation}\label{eq:const_stateful}
x^{c_1}_{u,i}=x^{c_2}_{u,i}\quad \forall c_1,c_2\in C'_s\subset C_s, i\in N, u\in V^c 
\end{equation}

Constraint (\ref{eq:const_region}) forces the algorithm to place the \ac{vsnf} $u\in V^c$ in a specific region of the network defined as a subset of nodes $R_u\subset N$.

\begin{equation}\label{eq:const_region}
\sum_{i\in R_u} x^c_{i,u}=1\quad  R_u\subset N, R_u\neq\emptyset, u\in V^c
\end{equation}
We use Constraint (\ref{eq:const_region}) to enforce the security close to the user by placing \acp{vsnf} on \textit{ep1} ($R_u=\{ep1\}$), or to protect a portion of the \ac{tsp}'s network, such as the border region or a distributed data centre ($R_u=EP2$) from potentially malicious user traffic. Furthermore, Constraint (\ref{eq:const_region}) can be used to place a \ac{vsnf} on a physical node with special hardware characteristics (e.g., hardware acceleration for encryption). Similarly, the \textit{veto} Constraint (\ref{eq:const_veto}) can be used to prevent the placement of any \acp{vsnf} on a pre-defined subset of nodes $M\subset N$. A \ac{tsp} may choose to do this to protect specific nodes (called \textit{veto nodes}) that host sensitive data or critical functions from user traffic.
\begin{equation}\label{eq:const_veto}
\sum_{\mathclap{i\in M, u\in V^c}} x^c_{i,u}=0\quad  \forall c\in C_s, M\subset N, M\neq\emptyset
\end{equation}	
Finally, for each chain $c\in C_s$, the correct order of \acp{vsnf} in $V^c$ is ensured by Constraints (\ref{eq:const_1}-\ref{eq:const_5b}), plus Constraint (\ref{eq:const_region}) applied to user and remote applications $u\in A^c$ with $R_u=\{ep1\}$ and $R_u=EP2$ respectively.
Note that, the order can be specified per application (chain), as different applications may require the same \acp{vsnf} but in different order. 
	
These four security constraints enable fulfillment of the security policies/practices defined by the \ac{tsp} e.g., the order in which the \acp{vsnf} are executed, the position of the \acp{vsnf} in the network, and the operational mode of \acp{vsnf} (either stateful or stateless).
\subsection{The PESS Heuristic Algorithm}\label{sec:implementation}

The embedding problem presented in Section~\ref{sec:model} has been solved using a commercial solver. However, given the complexity of the \ac{ilp} model, the solver is unable to produce solutions in an acceptable time frame, as required for dynamic scenarios such as those under study. For this reason, we have also implemented a heuristic algorithm to find near optimal solutions in much shorter time. 

The logic behind the PESS heuristic is based on assuring that Constraints (\ref{eq:const_1}-\ref{eq:const_veto}) are applied in an efficient manner. In particular, the security constraint (\ref{eq:const_stateful}) ensures that a stateful \ac{vsnf} specified in two or more chains in the same service request $\mathcal{G}_s$ is placed on the same node. However, as different chains might share more than one stateful \ac{vsnf} (possibly in a different order), the correct placement of a multi-chain security service request may become a computationally expensive operation.
For this reason, given a path between $ep1$ and one of the nodes $ep2\in EP2$, the heuristic places all the \acp{vsnf} specified in $\mathcal{G}_s$ on a maximum of three nodes of the path with the following strategy: (i) place each region-specific \ac{vsnf} $u\in V^c$ ($R_u\neq\emptyset$) either on $ep1$ or on $ep2\in EP2$ depending on $R_u$ (i.e., either $R_u=\{ep1\}$ or $R_u=EP2$), (ii) place all the other \acp{vsnf} in $\mathcal{G}_s$ on the node with the highest residual capacity in the path to minimize the embedding cost (Equation \ref{eq:objective_function}). \\
The solution is obtained by selecting the candidate path between \textit{ep1} and \textit{EP2} where the embedding of all the chains in $\mathcal{G}_s$ fulfills the constraints described in Section~\ref{sec:model} at the lowest cost, as computed with the objective function (Equation \ref{eq:objective_function}).

\begin{figure*}[!h]
	\begin{center}
		\includegraphics[width=1\textwidth]{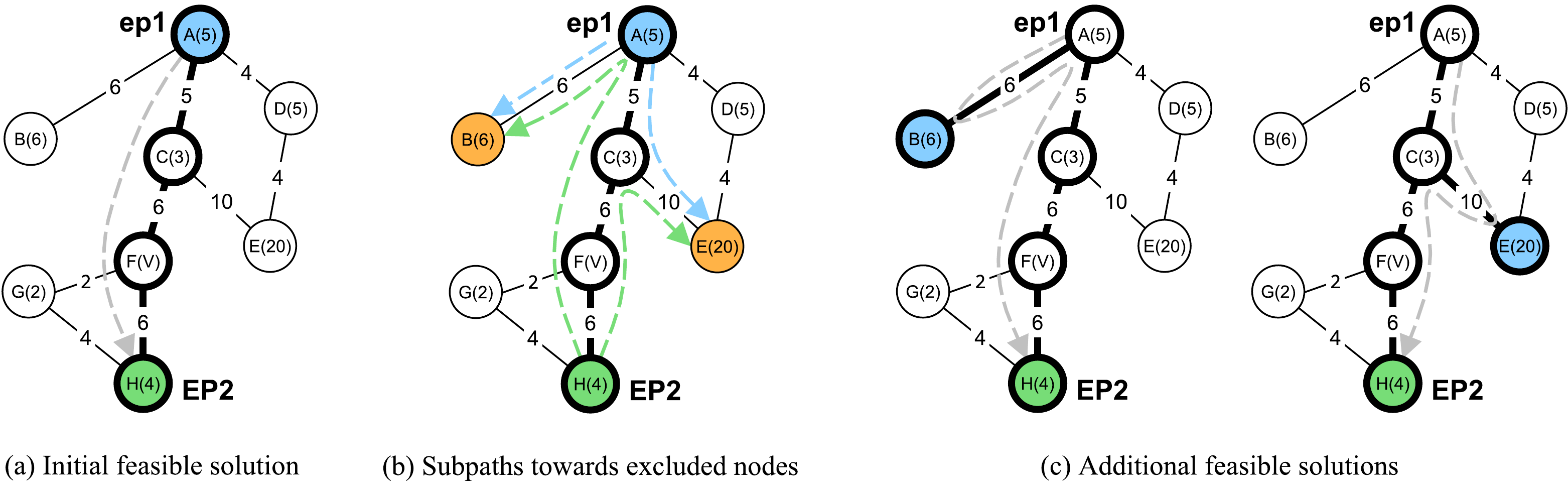}
		\caption{The main steps of the PESS heuristic. Elements in bold form a candidate solution. Blue circles represent the nodes with the highest computational capacity in the solution. The green circle is a node bound to the \textit{region} constraint, while orange circles form the set of nodes excluded from the initial solution with higher residual capacity than the best node in the initial solution. Node \textit{F} is a \textit{veto} node, while \textit{H} is a \textit{region} node.}
		\label{fig:embedding}
	\end{center}
\end{figure*}

\roberto{We first introduce the PESS heuristic with the simplified scenario illustrated in Figure~\ref{fig:embedding}. In the figure, the physical network is represented by nodes $A-H$, each annotated with the value of its computational capacity (the number inside each node), and by a set of edges with their residual bandwidth capacity. $A$ and $H$ are the two endpoints $ep1=A$ and $EP2 = \{H\}$, while $F$ is a $veto$ node where no \acp{vsnf} can be placed (Constraint~\ref{eq:const_veto}). In the example, \textit{H} also defines a region where part of the \acp{vsnf} in the request must be placed. Starting from Figure~\ref{fig:embedding}(a), PESS embeds the service on the shortest path between the two endpoints. This is the initial solution. In Figure~\ref{fig:embedding}(b), it computes the shortest path trees from \textit{ep1} and from \textit{EP2} towards the nodes not included in the initial solution but with higher residual capacity than the ones in the initial solution. The resulting trees are then used to build new paths between \textit{ep1} and \textit{EP2} (Figure~\ref{fig:embedding}(c)). The accepted solution is the path with the lowest cost that satisfies all the Constraints (\ref{eq:const_1}-\ref{eq:const_veto}).}

\begin{algorithm} [t!]
	\caption{the PESS algorithm.}\label{alg:pess_algorithm}
	\begin{algorithmic}[1]
		\renewcommand{\algorithmicrequire}{\textbf{Input:}}
		\renewcommand{\algorithmicensure}{\textbf{Output:}}
		\Require Physical network substrate ($\mathcal{G}$), security service request ($\mathcal{G}_s$), set of active chains in the network ($C$)
		\Ensure The mapping of the security service onto the physical substrate (\textit{solution}). \textit{None} if no feasible mappings are found.
		\Procedure{PESS}{$\mathcal{G}$, $\mathcal{G}_s$, $C$}
		\State $\bar{\beta}\gets\sum_{\{c\in C_s\}}\beta^c$\Comment{total required bandwidth}
		\State $\bar{\gamma}\gets\sum_{\{c\in C_s, u\in V^c\}}\gamma^c_u$\Comment{total required CPU}
		\State $P=\{p_{[ep1,ep]}\mid ep\in EP2\}\gets \Call{Dijkstra}{ep1,EP2,\bar{\beta}}$
		\If {$P=\emptyset$}
		\State \textbf{return None}
		\EndIf
		\State $S=\{s_{[ep1,ep]}\mid ep\in EP2\}\gets \Call{Embed}{P,\bar{\beta},\bar{\gamma}}$
		\State $N_S\gets \{i\in s\mid s\in S\}$\Comment{physical nodes in the initial solutions}
		\State $E\gets \{i\in N\mid i\notin N_S\cup M,\ \gamma'_i>\gamma'_j\ \forall j\in N_S\}$
		\State $\bar{s}_{[ep1,ep2]}\gets \operatorname*{argmin}\limits_{s\in S} cost(s)$\Comment{best initial solution}
		\State $P_1\gets \Call{Dijkstra}{ep1,E,\bar{\beta}}$
		\State $P_2\gets \Call{Dijkstra}{ep2,E,\bar{\beta}}$
		\State $S\gets S\cup \Call{Embed}{P_1\cup P_2,\bar{\beta},\bar{\gamma}}$\Comment{expanded solution set}
		\State $solution\gets$\textbf{None}
		\State $S\gets \Call{SortedDecreasingCost}{S}$
		\ForAll {$cs\in S$}
		\If {$\Call{LatencyOpChains}{\mathcal{G},C,cs}$ \textbf{is True}}
		\State $solution\gets cs$
		\State \textbf{break}
		\EndIf
		\EndFor
		\If {$solution$ \textbf{is None}}
		\State \textbf{return None}
		\EndIf
		\State $\Call{UpdateResources}{solution,\mathcal{G}}$
		\State $\Call{StoreSolution}{\mathcal{G},C,solution}$
		\State \textbf{return} $solution$
		\EndProcedure
	\end{algorithmic}
\end{algorithm}

\textbf{Initial solution.} The embedding process starts at line 4 in Algorithm \ref{alg:pess_algorithm} with a greedy approach based on the Dijkstra's algorithm. At this stage, we compute the shortest path tree between the two endpoints $ep1$ and $EP2$ using the residual bandwidth as link weight computed as $b_{k,l}\cdot \beta^c$ in Equation (\ref{eq:objective_function}) (Figure~\ref{fig:embedding}(a)). The Dijkstra algorithm stops when all the nodes $ep2\in EP2$ are marked as \textit{visited}, i.e. before building the whole tree of paths. For each path between \textit{ep1} and \textit{EP2}, the algorithm places the \acp{vsnf} in the chains according to the aforementioned strategy, the order of the \acp{vsnf} as specified in the service request, the latency Constraint (\ref{eq:const_latency}), and the security Constraints (\ref{eq:const_stateful}-\ref{eq:const_veto}) (line 8).
The output of this first step is a set of candidate solutions $S$ with different embedding costs. $S$ is passed as input to the next step. 

\textbf{Expanded solution set.} The algorithm now evaluates whether high-capacity nodes not included in the initial solution set $S$ can be used to build new solutions with lower embedding cost. Hence, given the initial set of solutions $S$, the algorithm identifies the physical nodes in the network with these two properties (set $E$ defined at line 10 or nodes colored in orange in Figure~\ref{fig:embedding}(b)): (i) not included in the initial set of solutions $S$ nor \textit{veto} nodes, and (ii) higher computing capacity with respect to the nodes included in $S$. 
The algorithm then computes the shortest path tree twice, once from $ep1$ to $E$ and once from $ep2\in EP2$ to $E$ (lines 12-13 and Figure~\ref{fig:embedding}(b)), where $\{ep1,ep2\}$ are the physical endpoints of the solution in $S$ with the lowest embedding cost (line 11). 

The resulting subpaths are joined to form a new set of paths between \textit{ep1} and \textit{ep2}. Afterwards, the algorithm performs the placement of the \acp{vsnf} on each of the new paths with the strategy described earlier in this section. The feasible solutions are added to the initial set $S$ (line 14 and Figure~\ref{fig:embedding}(c)).

The set of candidate solutions is sorted in descending value of embedding cost (line 16). The first one that satisfies Constraint  (\ref{eq:check_latency}) is the accepted solution (lines 18-19).
Finally, the algorithm updates the values of $\gamma'_i$ and $\beta'_i$ by removing the resources consumed with the accepted solution and stores the mapping of its chains in the set $C$ that records all the active chains in the network (lines 26-27).

\textbf{Latency of operational chains.}
Given a candidate solution $cs\in S$, function $\Call{LatencyOpChains}{}$ is invoked to verify whether embedding $cs$ compromises the end-to-end latency of operational chains (line 18 in Algorithm \ref{alg:pess_algorithm}). Instead of verifying the inequality in Equation (\ref{eq:check_latency}) for each operational chain, $\Call{LatencyOpChains}{}$ implements a heuristic approach, which reduces the time complexity of this operation from $O(n)$, with $n$ the number of operational chains, to $O(1)$.

Each time a chain $c\in C$ becomes operational, the algorithm computes $\langle\gamma\rangle^c$, a threshold value obtained from Equations (\ref{eq:const_latency}) and (\ref{eq:cpu_latency1}) as follows:

\begin{equation}\label{eq:average_residual_cpu}
	\langle\gamma\rangle^c = \frac{\sum\limits_{i\in N, u\in V^c} \bar{x}^c_{i,u}\cdot\gamma_u\cdot\sigma^c}{\lambda^c-\pi^c-\sum\limits_{\makebox[0pt]{$\scriptstyle \substack{i,j\in N,(k,l)\in E\\(u,v)\in U^c_{pairs}}$}}\bar{y}^c_{k,l,i,j,u,v}\cdot (\lambda_{k,l}+\lambda^c_{k,l,i,j})}-\delta
\end{equation}

In Equation (\ref{eq:average_residual_cpu}), $\bar{x}$ and $\bar{y}$ are the values of decision variables $x$ and $y$ used to embed $c$.
$\langle\gamma\rangle^c$ estimates the minimum average residual computing capacity necessary to satisfy the inequality in Equation (\ref{eq:const_latency}). Therefore, the algorithm records and monitors those operational chains with the highest values of $\langle\gamma\rangle^c$ to establish whether a candidate solution is feasible or not, as inequality in Equation (\ref{eq:const_latency}) is violated earlier for such chains than for the others.

The algorithm stores one operational chain per physical node in a data structure, i.e. the chain with the highest value of $\langle\gamma\rangle^c$ with at least one \ac{vsnf} mapped on that node. Hence, given the physical nodes mapped in the candidate solution $cs$, the algorithm computes Equation (\ref{eq:check_latency}) only for the operational chains in the data structure linked to such nodes by using the values of variables $x$ and $y$ of solution $cs$. If the inequality is not satisfied for one of those chains, $cs$ is rejected.

As the maximum number of physical nodes used to provision a security service is three (\textit{ep1} and \textit{EP2} to fulfill the region constraint and the node with the highest residual capacity in the path), the worst-case time complexity of this process is $O(1)$, thus constant in the number of operational chains and with respect to the size of the network.
Therefore, the overall time complexity of the \ac{pess} heuristic is $O(|E|+|N|\log(|N|))$, i.e., the worst-case time complexity of the Dijkstra's algorithm.
\subsection{Evaluation}\label{sec:pess-evaluation}
%
%
%
%
%
%
%

\pgfplotstableread[header=true]{
	LOAD	APP		BASE
	1000	0.1161	0.235
	2000	0.2304	0.4633
	4000	0.4671	0.9311
	6000	0.6904	0.994
	8000	0.9325	0.9959
	10000	0.9964	0.9967
	12000	0.9976	0.9971
	14000	0.998	0.9974
	16000	0.9983	0.9977
	18000	0.9985	0.9979
	20000	0.9986	0.998

}\AppCentricRANDOMConsResources

\pgfplotstableread[header=true]{
	LOAD	APP		BASE
	1000	0		0
	2000	0		0
	4000	0		0.0056
	6000	0		0.1112
	8000	0.0062	0.2086
	10000	0.0565	0.2812
	12000	0.1241	0.3178
	14000	0.1956	0.3668
	16000	0.2494	0.4041
	18000	0.2801	0.4314
	20000	0.3197	0.4484

}\AppCentricRANDOMBlockProb

\pgfplotstableread[header=true]{
	LOAD	APP		BASE
	1000	1007.378	1001.9667
	2000	1986.2613	1964.0259
	4000	4040.7121	3997.1643
	6000	5971.2343	5276.1338
	8000	8031.2286	6342.7492
	10000	9333.2261	7191.4012
	12000	10363.1885	8116.0247
	14000	11245.2232	8776.787
	16000	12049.8084	9475.8486
	18000	12726.1054	10228.5147
	20000	13389.2176	10722.2377
}\AppCentricRANDOMAverLoad

\pgfplotstableread[header=true]{
	LOAD	RATIO		
	1000 00001.1047
	2000 00001.1404
	4000 00001.3950
	6000 00001.5786
	8000 00001.3389
	10000 00001.1213
	12000 00001.0939
	14000 00001.0751
	16000 00001.0707
	18000 00001.0669
	20000 00001.0573
}\AppCentricRANDOMLatency

\pgfplotstableread[header=true]{
	LOAD	APP		BASE
	1000	0	0
	2000	0	0
	4000	0.0012	0.0010
	6000	0.0128	0.0199
	8000	0.0206	0.0804
	10000	0.0353	0.1411
	12000	0.0415	0.1797
	14000	0.073	0.2134
	16000	0.115	0.2469
	18000	0.1338	0.2661
	20000	0.1732	0.3209
}\AppCentricGARRBlockProb

\pgfplotstableread[header=true]{
LOAD	APP		BASE	APPreg	BASEreg
1000	0.0487	0.0987	0.0999	0.1741
2000	0.0965	0.1943	0.1739	0.3261
4000	0.1944	0.386	0.3318	0.6593
6000	0.2861	0.5747	0.4863	0.9745
8000	0.3803	0.6679	0.6409	0.9971
10000	0.4587	0.7382	0.7817	0.9983
12000	0.5557	0.8108	0.9516	0.9986
14000	0.6114	0.8655	0.9969	0.9988
16000	0.6588	0.9293	0.9987	0.999
18000	0.6871	0.9605	0.9989	0.9991
20000	0.7358	0.9627	0.9992	0.9992	
}\AppCentricGARRConsResources

\pgfplotstableread[header=true]{
	LOAD	APP		BASE
	1000	0		0
	2000	0		0
	4000	0		0.1264
	6000	0.0515	0.2213
	8000	0.1286	0.2726
	10000	0.1845	0.3086
	12000	0.2275	0.3399
	14000	0.2598	0.3804
	16000	0.2813	0.4098
	18000	0.3107	0.4386
	20000	0.3281	0.4546

}\AppCentricSTANFORDBlockProb

\pgfplotstableread[header=true]{
	LOAD	APP		BASE	APPreg	BASEreg
	1000	0.0858	0.1756	0.1971	0.4046
	2000	0.1721	0.3425	0.3914	0.7779
	4000	0.3447	0.5345	0.7794	0.9974
	6000	0.4798	0.6663	0.9971	0.9986
	8000	0.5546	0.7862	0.9987	0.9989
	10000	0.6316	0.9156	0.9991	0.9992
	12000	0.6936	0.986	0.9993	0.9992
	14000	0.7708	0.9929	0.9994	0.9993
	16000	0.8428	0.9944	0.9995	0.9994
	18000	0.9105	0.9955	0.9996	0.9994
	20000	0.9674	0.9957	0.9996	0.9994

}\AppCentricSTANFORDConsResources

\pgfplotstableread[header=true]{
	SIZE	1		3		5
	10		0.804	0.858	0.863		
	100		5.367	6.72	8.053
	250		12.782	19.876	24.922
	500		31.564	60.095	67
	750		54.811	112.452	124.999
	1000	85.024	182.723	210.269
}\ScalabilityFiniteBWRandom

\pgfplotstableread[header=true]{
	SIZE 	0
	46 	00003.0910	
}\ScalabilityFiniteBWGARR

\pgfplotstableread[header=true]{
SIZE	0		10		25		50
10		0.863	0.86	0.917	0.996
100		8.053	7.027	7.929	10.421
250		24.922	23.439	25.767	32.04
500		67		69.668	74.541	83.771
750		124.999	135.388	148.941	172.147
1000	210.269	218.456	233.288	263.879
}\ScalabilityRegionSizeRandom

\pgfplotstableread[header=true]{
SIZE	0
46 		00002.3820
}\ScalabilityRegionSizeRandomGARR

We first assess the PESS heuristic by comparing its solutions against the optimal embeddings as computed by a commercial solver (Gurobi~\cite{gurobi}). 
We then prove the benefits of the proposed application-aware approach against the baseline (the application-agnostic approach adopted, for instance, in \cite{7592416}), in which security services are provided without taking into account the specific requirements of applications. We finally analyze the scalability of PESS by measuring the average embedding time on different network sizes.

\subsubsection{Test Configuration}\label{sec:test_config}
The \ac{pess} heuristic has been implemented as a single-threaded Python program, while the ILP model formalised in Section \ref{sec:model} has been implemented with the Gurobi Python API version 7.5~\cite{gurobi_api}. All experiments are performed on a server-class computer equipped with 2 Intel Xeon Silver 4110 CPUs (16 cores each running at 2.1\,GHz) and 64\,GB of RAM. 
\subsubsection{Topology} 
The simulations are performed on synthetic topologies randomly generated based on the Barab\'asi-Albert model~\cite{barabasi}. We generate topologies of different sizes and densities to evaluate the performance of the \ac{pess} heuristic in a variety of generic network scenarios.

\begin{figure*}[!h]
	\vspace{-1mm}
	\begin{minipage}{.485\textwidth}
		\centering
		\includegraphics[width=1\textwidth]{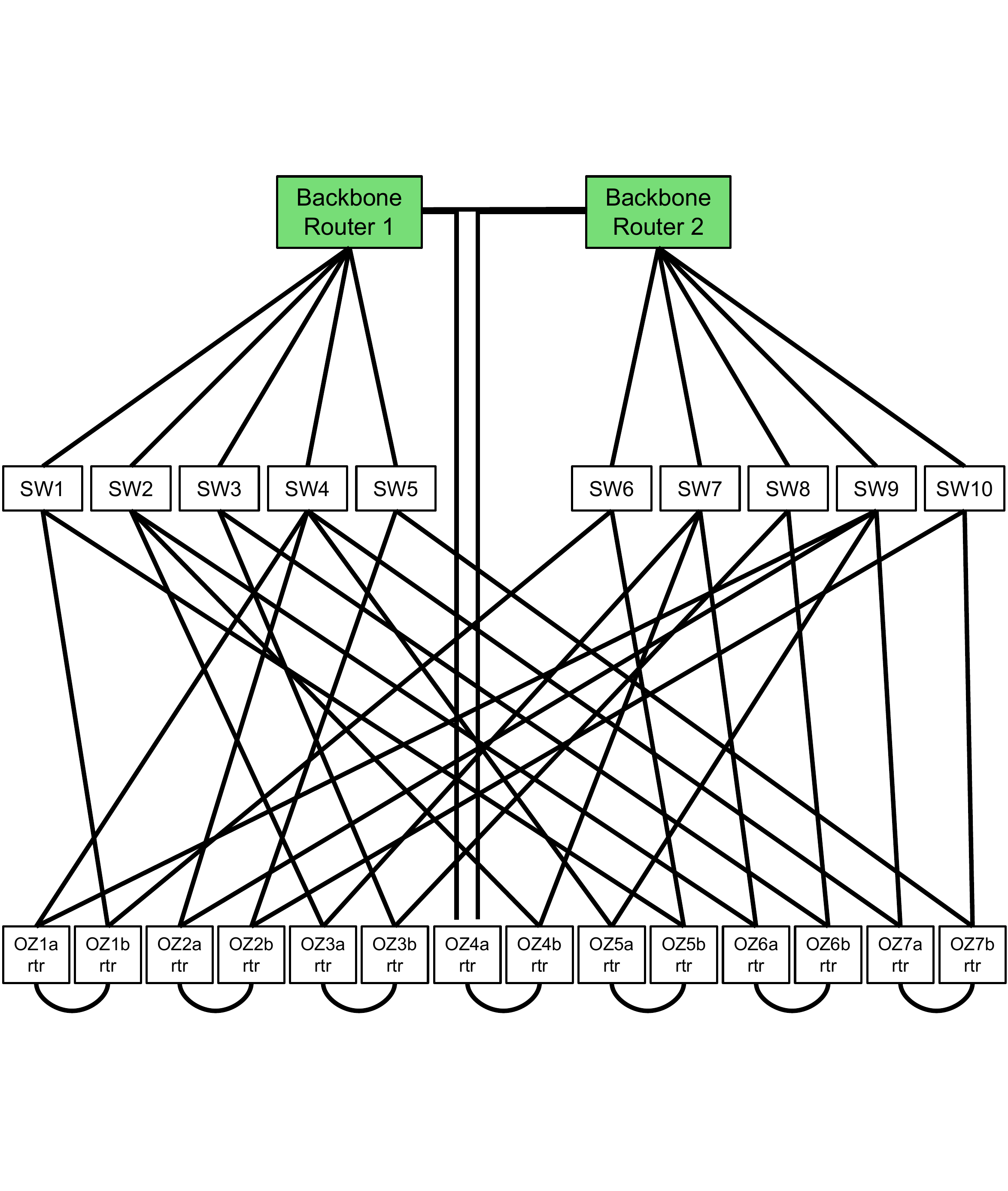}
		\vspace{-7mm}
		\caption{Representation of the Stanford network (26 nodes and 46 links). The nodes in green form the \textit{border} region.}
		\label{fig:stanford_topology}
	\end{minipage}
	\hfill
	\begin{minipage}{.49\textwidth}
		\centering
		\includegraphics[width=1\textwidth]{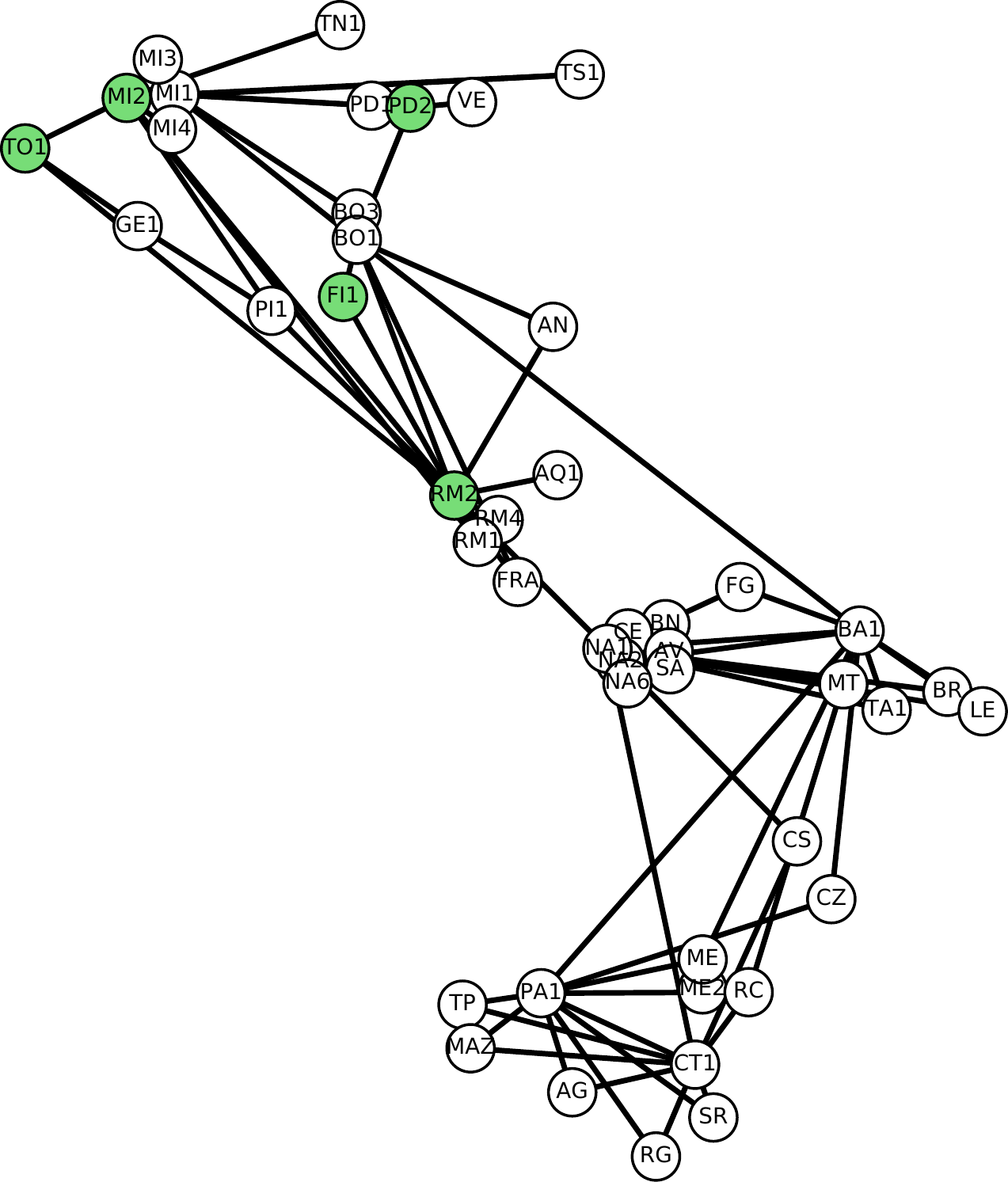}
		\vspace{-7mm}
		\caption{Representation of the GARR topology (46 nodes and 83 links). The nodes in green form the \textit{border} region.}
		\label{fig:garr_topology}
	\end{minipage}
\end{figure*}

We also validate \ac{pess} with two realistic network models. One is the Stanford University backbone \cite{headerspace} (represented in Figure \ref{fig:stanford_topology}), a medium-scale campus network consisting of 46 links, 14 operational zone routers, 10 Ethernet switches, and 2 border routers connecting the University to the Internet. We assume one NFVI-POP for each network device and 10\,Gbps links. The second model is the Italian education and research network (consortium GARR~\cite{garr-whitepaper}). The GARR network, represented in Figure \ref{fig:garr_topology}, covers the entire Italian national territory (\cite{garr-topology}), comprising 83 links and 46 nodes. Along with the actual view of the physical topology, \cite{garr-topology} provides the specification of the egress nodes, i.e. the nodes that connect the GARR network to the Internet and that compose the \textit{border} region in our evaluation (nodes \textit{FI1, MI2, PD2, RM2} and \textit{TO1}, as indicated in \cite{garr-topology}). As we have no information related to data centre distribution in the GARR network, we have assumed one NFVI-POP for each node. In addition, we set the nominal capacity of the links $\beta_{k,l}$ using the values specified in \cite{garr-backbone}.
	
Given the relatively small size of the Stanford network, we assume no propagation delay between its nodes, i.e. $\lambda_{k,l}=0\ \forall k,l$. For the other networks, random and GARR, we compute the propagation delay of each link with the following formula:

$$\lambda_{k,l} = \frac{d_{k,l}\cdot r\_index}{C}$$

where $r\_index=1.5$ is an approximation of the refractive index of optical fibers, $C\simeq 3\cdot 10^8\ m/s$ is the speed of light in the vacuum and $d_{k,l}$ is the distance between two nodes $k$ and $l$. In the case of random networks, $d_{k,l}$ is a random positive value ranging from 10 to 100\,Km, while for the GARR network $d_{k,l}$ is computed by approximating the coordinates of the nodes based on the information available on the web site.

As introduced in Section \ref{sec:constraints}, we estimate the worst-case queuing delay $\lambda^c_{k,l,i,j}$ as a traffic-load independent value using the queue capacity of switch ports reported in \cite{silo} ($80\,\mu s$ for 10\,Gbps ports with a 100\,KB buffer). Specifically, we assume a three-tier local network at each node of GARR and random topologies, resulting in a maximum of $12\times80\,\mu s$ queuing delay introduced at each node. This reflects the maximum queuing delay experienced by each packet crossing a node to be processed by one or more \acp{vsnf} mapped on the node, which involves traversing three network devices (hence, six 10\,Gbps ports) to reach the servers where the \acp{vsnf} are running, and traversing three network devices before leaving the node (six further 10\,Gbps ports).   

For the campus scenario, implemented using the Stanford University topology, we instead assume only one network device per node; the device specified in the network topology. Hence, the maximum queuing latency for a packet crossing a Stanford node is  $4\times80\,\mu s$. 

For each node of the three evaluation scenarios we assume one server with computing capacity of 32x2.1\,GHz (a 32-core CPU running at 2.1\,GHz).

\subsubsection{Security Service Requests}\label{sec:service_requests}
As introduced in Section \ref{sec:model}, a security service request is configured by the \ac{tsp} to provision security for user applications (see the CCTV example in Section  \ref{sec:model}). For evaluation purposes, we automatically generate requests composed of a random number of chains, ranging between 1 and 5. Each chain comprises a random subset of \acp{vsnf} from the list presented in Table \ref{tab:vsnf_cpu_requirements}, with a maximum of 3 \acp{vsnf} per chain (i.e., up to 15 \acp{vsnf} per user application). Based on the use case scenarios illustrated in Sections \ref{sec:motivation} and \ref{sec:model} (web browsing, online gaming, CCTV system), we believe these are reasonable values.

\begin{table}[h!]
	\vspace{5mm}
	\centering 
	\begin{threeparttable}
		\begin{tabular}{>{\bfseries}lcc} \toprule[\heavyrulewidth]
			\textbf{\ac{vsnf}} & \textbf{Virtualisation} & \begin{tabular}{@{}c@{}}{\boldmath $\gamma_u$}\\ \textbf{(cycles/bit)}\tnote{1}\end{tabular} \\ \midrule[\heavyrulewidth]
			Snort IDS/IPS & VirtualBox &9.5~\cite{SHAH2018157} \\ \midrule
			Suricata IDS/IPS & VirtualBox &8.2~\cite{SHAH2018157} \\ \midrule
			\begin{tabular}{@{}c@{}}OpenVPN with AES-NI\end{tabular}& KVM/QEMU &31~\cite{7973470} \\ \midrule
			\begin{tabular}{@{}c@{}}strongSwan with AES-NI\end{tabular}& KVM/QEMU &16~\cite{7973470}\\ \midrule
			\begin{tabular}{@{}c@{}}Fortigate-VM NGFW\end{tabular} & \begin{tabular}{@{}c@{}}FortiOS \end{tabular}& 9~\cite{fortigate}\\  \midrule
			\begin{tabular}{@{}c@{}}Fortigate-VM SSL VPN\end{tabular} & \begin{tabular}{@{}c@{}}FortiOS \end{tabular}& 13.6~\cite{fortigate}\\ \midrule
			\begin{tabular}{@{}c@{}}Fortigate-VM IPSec VPN\end{tabular} & \begin{tabular}{@{}c@{}}FortiOS \end{tabular}& 14.5~\cite{fortigate}\\ \midrule
			\begin{tabular}{@{}c@{}}Fortigate-VM Threat protection\end{tabular} & \begin{tabular}{@{}c@{}}FortiOS \end{tabular}& 11.3~\cite{fortigate}\\ \midrule
			\begin{tabular}{@{}c@{}}Cisco ASAv Stateful IDS\end{tabular} & \begin{tabular}{@{}c@{}}VMware ESX/ESXi \end{tabular}& 4.2~\cite{cisco_asav}\\ \midrule
			\begin{tabular}{@{}c@{}}Cisco ASAv AES VPN\end{tabular} & \begin{tabular}{@{}c@{}}VMware ESX/ESXi \end{tabular}& 6.9~\cite{cisco_asav}\\ \midrule
			Juniper vSRX FW & \begin{tabular}{@{}c@{}}VMware VMXNET3 \end{tabular}& 2.3~\cite{juniper_vsrx}\\ \midrule
			Juniper vSRX IPS & \begin{tabular}{@{}c@{}}VMware VMXNET3 \end{tabular}& 2.4~\cite{juniper_vsrx}\\ \midrule
			\begin{tabular}{@{}c@{}}Juniper vSRX AppMonitor\end{tabular}& \begin{tabular}{@{}c@{}}VMware VMXNET3 \end{tabular}& 1.5~\cite{juniper_vsrx}\\ \bottomrule[\heavyrulewidth]
		\end{tabular}
		\begin{tablenotes}
			\item[1] $\gamma_u$=(CPU clock)*(CPU usage)/Throughput. CPU usage is set to 1 (i.e. 100\%) when the value is not specified.
		\end{tablenotes}
	\end{threeparttable}
	\caption{CPU requirements for some \ac{vsnf} implementations.}
	\label{tab:vsnf_cpu_requirements}
\end{table}

The CPU requirements for the VSNFs are presented in Table \ref{tab:vsnf_cpu_requirements}. It should be noted that the values of $\gamma_u$ (cycle/bit) reported in Table \ref{tab:vsnf_cpu_requirements} are estimated based on the results of experiments reported in scientific papers or product datasheets and obtained under optimal conditions, with only one \ac{vsnf} running at a time. The impact on the network traffic caused by concurrent \acp{vsnf} running on the same node are estimated with Equations (\ref{eq:cpu_latency1}) and (\ref{eq:cpu_latency_update}). 
These values of $\gamma_u$ have been used to perform the evaluation tests described in the remainder of this section, with the aim of enabling interested readers to replicate the experiments in similar conditions. However, we also obtained comparable results using random values.

\subsubsection{Comparison Between Solver and Heuristic}\label{sec:performance_comparison}
\textbf{Methodology.} In this experiment, we compare the PESS ILP-based algorithm implemented with the solver and the PESS heuristic on the Stanford and GARR network models, and on Barab{\'a}si-Albert random topologies with 20 nodes and 36 links.

The security service requests are generated using a Poisson process with exponential distribution of inter-arrival and holding times. Once a service expires, the resources allocated to it are released.

We start by simulating the processing of $10^5$ service requests using the PESS heuristic. Once a stable network utilisation (load) is reached, we save the subsequent service requests along with the network state and the heuristic solution. In a second stage, we run the solver to compute the optimal solution for each of the requests saved in the previous stage and we compare the results with the recorded heuristic solutions. This process is repeated with values of network load ranging between 1000 and 20000 Erlang.

\textbf{Metrics.} (i) Heuristic embedding cost overhead over optimal solutions and (ii) embedding time.

\textbf{Discussion.}  As explained in Section~\ref{sec:implementation}, the PESS heuristic places all the chains of a service request on a single path to efficiently guarantee that the \ac{qos} Constraint (\ref{eq:const_latency}) and the region Constraint (\ref{eq:const_region}) are respected. Once the path is found, the heuristic places the \acp{vsnf} of all the chains on a maximum of three nodes in the chosen path: the one with the highest residual computing capacity and the ones specified with the region constraint (if any). Such implementation choices reduce the solution space in case of requests with multiple chains and \acp{vsnf}. On the other hand, Constraints (\ref{eq:const_latency}) and (\ref{eq:const_region}) also narrow down the solution space for the solver, often resulting in single-path optimal solutions. As a result, we measure a marginal embedding cost overhead of the heuristic solutions with respect to the optimal solutions on all three evaluation scenarios (see Table \ref{tab:heuris_solver}).

\begin{table}[h!]
	\vspace{5mm}
	\centering 
	\begin{threeparttable}
		\begin{tabular}{lccc} \toprule[\heavyrulewidth]
			\textbf{Network model} & \begin{tabular}{@{}c@{}}\textbf{Heuristic embedding}\\ \textbf{cost overhead}\tnote{1}\end{tabular} & \multicolumn{2}{c}{\begin{tabular}{@{}c@{}}\textbf{Average Time (sec)}\\ \textbf{Heuristic}\ \qquad \textbf{Solver}\ \ \ \end{tabular}} \\ \midrule[\heavyrulewidth]
			\textbf{Random}& 0.06\% & \quad 0.002 & \quad 150\\ \midrule
			\textbf{Stanford}& 0.07\% & \quad 0.003 & \quad 700 \\ \midrule
			\textbf{GARR}& 0.5\% & \quad 0.003 & \quad 1500\\ \bottomrule[\heavyrulewidth]
		\end{tabular}
		\begin{tablenotes}
			\item[1]Average overhead with respect to the solver embedding cost.
		\end{tablenotes}
	\end{threeparttable}
	\caption{Comparison between PESS heuristic and PESS ILP on three network scenarios.}
	\label{tab:heuris_solver}
\end{table}

It is worth analysing the reason behind nearly one order of magnitude difference between the GARR topology and the other two network scenarios. When the initial solution is computed, the heuristic algorithm selects the endpoint $ep2\in EP2$ to further explore the solution space, thus excluding the other endpoints in $EP2$ (line 11 in Algorithm \ref{alg:pess_algorithm}). This strategy improves the scalability of the heuristic in case of large endpoint sets $EP2$, at the cost of slightly reducing the quality of the solutions.

In this regard, on the GARR network the border region is used as endpoint $EP2$ for 80\% of the requests, to simulate a real-world \ac{tsp} network where most of the traffic is directed towards the Internet. Hence, good solutions involving four of the five nodes in the border are not considered during the second stage of the heuristic, possibly leading to less accurate solutions. Conversely, a border region of only two nodes is defined in the Stanford topology (the two border routers), while no special regions at all are configured for the random networks (thus, always $|EP2|=1$), resulting in more precise embeddings.

As reported in Table \ref{tab:heuris_solver}, the embedding time measured for the heuristic is 3\,ms, on average, with the Stanford and GARR topologies, and below 3\,ms, on average, with the random topologies. In contrast, the solver takes between 150 and 1500\,s, on average, to find the optimal solutions on the three network scenarios. Please note that, the results related to the GARR network are limited to service requests with less than 10 \acp{vsnf}. Due to the size of the GARR topology (46 nodes and 83 links), above this threshold the solver runs out of memory and it is terminated by the operating system.

\subsubsection{PESS vs Application-agnostic Provisioning}\label{sec:app-base-comparison}
\textbf{Methodology.}  We start two experiments in parallel using two identical copies of the same physical network graph. At each iteration, we generate a service request with application-specific \ac{qos} and security requirements. In \textit{Experiment 1}, the security service is provisioned on one copy of the network with the \ac{pess} heuristic. In \textit{Experiment 2}, the service is provisioned on the second copy of the network by simulating the standard approach (adopted, for instance, in \cite{7592416} and used in this test as baseline), where two application-agnostic chains of \acp{vsnf} (one for each direction of the traffic) are applied to the user traffic to fulfill all the security requirements regardless of the specific needs of the applications. At the end of each iteration, the two copies of the network are updated according to the resources consumed by the respective provisioning approach.

As in the previous experiment, security service requests are generated using a Poisson process with exponential distribution of inter-arrival and holding times. We run $10^5$ iterations, starting to collect statistics after the first $8\cdot 10^4$ requests (once a stable network load is reached). The two parallel experiments are repeated with different network load values.

\textbf{Metrics.} Blocking probability, consumption of computing resources, end-to-end latency of the chains and number of active services in the network.

\begin{figure*}[h!]
	\subfigure[Consumed CPU resources.\label{fig:consumed_resources_random}]{  
		\begin{tikzpicture}
		\begin{axis}[  
		legend columns=1,
		legend pos=south east,
		height=5 cm,
		width=0.48\textwidth,
		grid = both,
		xlabel={Load (Erlang)},
		ylabel={Consumed resources (\%)},
		scaled y ticks=false,
		scaled x ticks=false,
		xtick={1000,4000,8000,12000,16000,20000},
		xticklabels={1K,4K,8K,12K,16K,20K},
		ymin=0,
		ytick={0,0.2,0.4,0.6,0.8,1},
		yticklabels={0,20,40,60,80,100},
		]
		\addplot [color=skyblue,mark=square*,mark size=1.5] table[x index=0,y index=1] {\AppCentricRANDOMConsResources};
		\addplot [color=orange,mark=triangle*,mark size=2.2] table[x index=0,y index=2] {\AppCentricRANDOMConsResources};
		\legend{PESS, Base}
		\end{axis}
		\end{tikzpicture}
	}
	\hfill
	\subfigure[Blocking probability.\label{fig:block_prob_random}]{  
		\begin{tikzpicture}
		\begin{semilogyaxis}[  
		legend columns=1,
		legend pos=south east,
		height=5 cm,
		width=0.48\textwidth,
		grid = both,
		xlabel={Load (Erlang)},
		ylabel={Blocking Probability},
		scaled y ticks=false,
		scaled x ticks=false,
		xmin=0,
		xtick={1000,4000,8000,12000,16000,20000},
		xticklabels={1K,4K,8K,12K,16K,20K},
		ymax=1.3,
		ymin=0.0008,
		ytick={0.001,0.01,0.1,1},
		yticklabels={$\mathsf{10^{-3}}$,$\mathsf{10^{-2}}$,$\mathsf{10^{-1}}$,1},
		]
		\addplot [color=skyblue,mark=square*,mark size=1.5] table[x index=0,y index=1] {\AppCentricRANDOMBlockProb};
		\addplot [color=orange,mark=triangle*,mark size=2.2] table[x index=0,y index=2] {\AppCentricRANDOMBlockProb};
		\legend{PESS, Base}
		\end{semilogyaxis}
		\end{tikzpicture}
	}\\

	\subfigure[Average load.\label{fig:average_load_random}]{  
		\begin{tikzpicture}
		\begin{axis}[  
		legend columns=1,
		legend pos=north west,
		height=5 cm,
		width=0.48\textwidth,
		grid = both,
		xlabel={Load (Erlang)},
		ylabel={Active services},
		scaled y ticks=false,
		scaled x ticks=false,
		xmin = 0,
		xtick={1000,4000,8000,12000,16000,20000},
		xticklabels={1K,4K,8K,12K,16K,20K},
		ymin = 0,
		ymax=15000,
		ytick={1000,4000,8000,12000,16000,20000},
		yticklabels={1K,4K,8K,12K,16K,20K},
		]
		\addplot [color=skyblue,mark=square*,mark size=1.5] table[x index=0,y index=1] {\AppCentricRANDOMAverLoad};
		\addplot [color=orange,mark=triangle*,mark size=2.2] table[x index=0,y index=2] {\AppCentricRANDOMAverLoad};
		\legend{PESS, Base}
		\end{axis}
		\end{tikzpicture}
	}
	\hfill
	\subfigure[Processing delay ratio (Base/PESS).\label{fig:latency_random}]{  
		\begin{tikzpicture}
		\begin{axis}[  
		legend pos=north east,
		height=5 cm,
		width=0.48\textwidth,
		grid = both,
		xlabel={Load (Erlang)},
		ylabel={Delay ratio},
		scaled y ticks=false,
		scaled x ticks=false,
		xtick={1000,4000,8000,12000,16000,20000},
		xticklabels={1K,4K,8K,12K,16K,20K},
		ymin=0.9, 
		ymax=2.1,
		ytick={1,1.2,1.4,1.6,1.8,2},
		yticklabels={1,1.2,1.4,1.6,1.8,2},
		legend columns=1,
		]
		\addplot [color=red,mark=*,mark size=1.8] table[x index=0,y index=1] {\AppCentricRANDOMLatency};
		\legend{Base/PESS}
		\end{axis}
		\end{tikzpicture}
	}
	\caption{Comparison between the baseline (Base) and the PESS approaches on random networks (20 nodes and 36 links).}
	\label{fig:app_base_random}
\end{figure*}
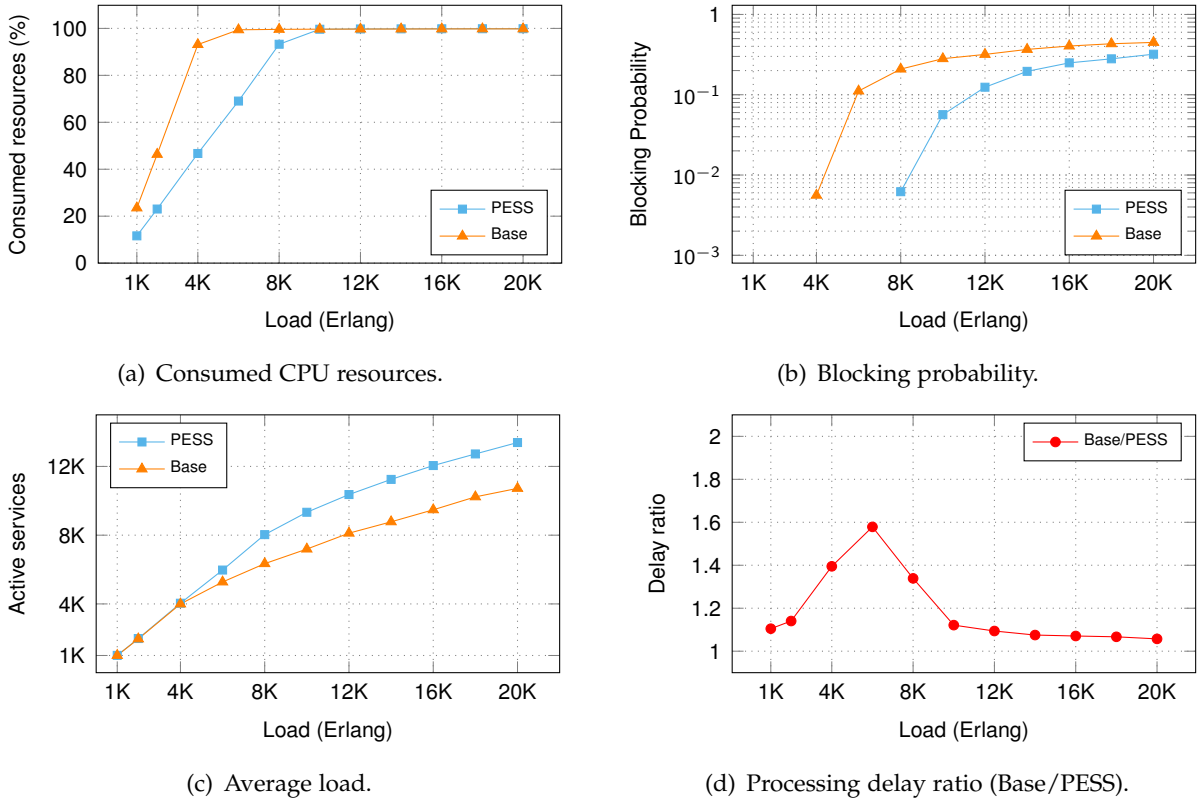

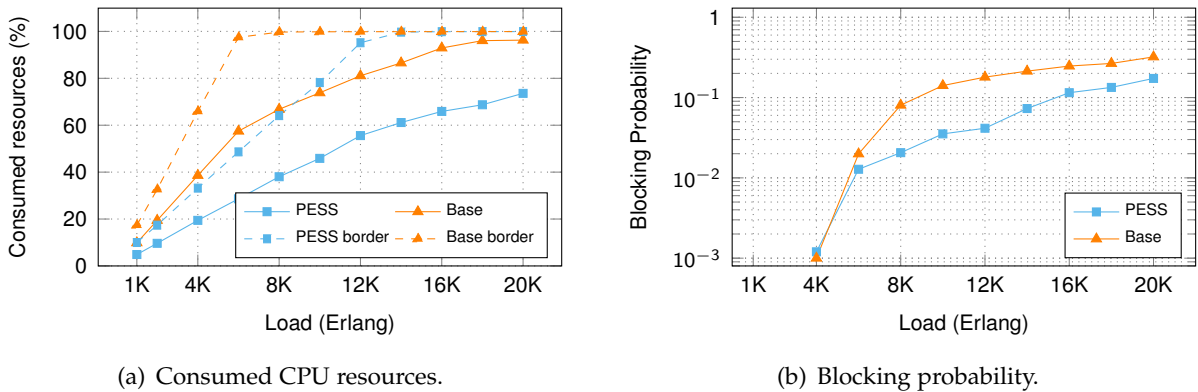
\begin{figure*}[h!]
	\subfigure[Consumed CPU resources.\label{fig:consumed_resources_garr}]{  
		\begin{tikzpicture}
		\begin{axis}[  	
		legend pos=south east,
		legend columns=2,
		height=5 cm,
		width=0.48\textwidth,
		grid = both,
		xlabel={Load (Erlang)},
		ylabel={Consumed resources (\%)},
		scaled y ticks=false,
		scaled x ticks=false,
		xtick={1000,4000,8000,12000,16000,20000},
		xticklabels={1K,4K,8K,12K,16K,20K},
		ymin=0, 
		ytick={0,0.2,0.4,0.6,0.8,1},
		yticklabels={0,20,40,60,80,100},
		]
		\addplot [color=skyblue,mark=square*,mark size=1.5] table[x index=0,y index=1] {\AppCentricGARRConsResources};
		\addplot [color=orange,mark=triangle*,mark size=2.2] table[x index=0,y index=2] {\AppCentricGARRConsResources};
		\addplot [color=skyblue,dashed,mark=square*,mark size=1.5] table[x index=0,y index=3] {\AppCentricGARRConsResources};
		\addplot [color=orange,dashed,mark=triangle*,mark size=2.2] table[x index=0,y index=4] {\AppCentricGARRConsResources};
		\legend{PESS,Base,PESS border, Base border}
		\end{axis}
		\end{tikzpicture}
	}
	\hfill
	\subfigure[Blocking probability.\label{fig:block_prob_garr}]{  
		\begin{tikzpicture}
		\begin{semilogyaxis}[  
		legend pos=south east,
		height=5 cm,
		width=0.48\textwidth,
		grid = both,
		xlabel={Load (Erlang)},
		ylabel={Blocking Probability},
		scaled y ticks=false,
		scaled x ticks=false,
		xmin=0,
		xtick={1000,4000,8000,12000,16000,20000},
		xticklabels={1K,4K,8K,12K,16K,20K},
		ymax=1.3,
		ymin=0.0008,
		ytick={0.0001,0.001,0.01,0.1,1},
		yticklabels={$\mathsf{10^{-4}}$,$\mathsf{10^{-3}}$,$\mathsf{10^{-2}}$,$\mathsf{10^{-1}}$,1},
		legend columns=1,
		]
		\addplot [color=skyblue,mark=square*,mark size=1.5] table[x index=0,y index=1] {\AppCentricGARRBlockProb};
		\addplot [color=orange,mark=triangle*,mark size=2.2] table[x index=0,y index=2] {\AppCentricGARRBlockProb};
		\legend{PESS, Base}
		\end{semilogyaxis}
		\end{tikzpicture}
	}
	\caption{Comparison between the baseline (Base) and the PESS approaches on the GARR network.}
	\label{fig:app_base_garr}
\end{figure*}

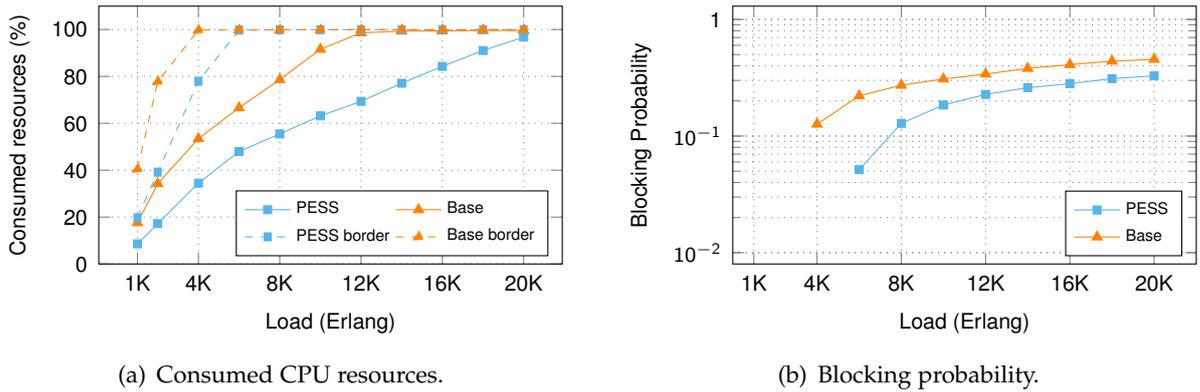
\begin{figure*}[h!]
	\subfigure[Consumed CPU resources.\label{fig:consumed_resources_stanford}]{  
		\begin{tikzpicture}
		\begin{axis}[  	
		legend pos=south east,
		legend columns=2,
		height=5 cm,
		width=0.48\textwidth,
		grid = both,
		xlabel={Load (Erlang)},
		ylabel={Consumed resources (\%)},
		scaled y ticks=false,
		scaled x ticks=false,
		xtick={1000,4000,8000,12000,16000,20000},
		xticklabels={1K,4K,8K,12K,16K,20K},
		ymin=0, 
		ytick={0,0.2,0.4,0.6,0.8,1},
		yticklabels={0,20,40,60,80,100},
		]
		\addplot [color=skyblue,mark=square*,mark size=1.5] table[x index=0,y index=1] {\AppCentricSTANFORDConsResources};
		\addplot [color=orange,mark=triangle*,mark size=2.2] table[x index=0,y index=2] {\AppCentricSTANFORDConsResources};
		\addplot [color=skyblue,dashed,mark=square*,mark size=1.5] table[x index=0,y index=3] {\AppCentricSTANFORDConsResources};
		\addplot [color=orange,dashed,mark=triangle*,mark size=2.2] table[x index=0,y index=4] {\AppCentricSTANFORDConsResources};
		\legend{PESS,Base,PESS border, Base border}
		\end{axis}
		\end{tikzpicture}
	}
	\hfill
	\subfigure[Blocking probability.\label{fig:block_prob_stanford}]{  
		\begin{tikzpicture}
		\begin{semilogyaxis}[  
		legend pos=south east,
		height=5 cm,
		width=0.48\textwidth,
		grid = both,
		xlabel={Load (Erlang)},
		ylabel={Blocking Probability},
		scaled y ticks=false,
		scaled x ticks=false,
		xmin=0,
		xtick={1000,4000,8000,12000,16000,20000},
		xticklabels={1K,4K,8K,12K,16K,20K},
		ymax=1.3,
		ymin=0.008,
		ytick={0.01,0.1,1},
		yticklabels={$\mathsf{10^{-2}}$,$\mathsf{10^{-1}}$,1},
		legend columns=1,
		]
		\addplot [color=skyblue,mark=square*,mark size=1.5] table[x index=0,y index=1] {\AppCentricSTANFORDBlockProb};
		\addplot [color=orange,mark=triangle*,mark size=2.2] table[x index=0,y index=2] {\AppCentricSTANFORDBlockProb};
		\legend{PESS, Base}
		\end{semilogyaxis}
		\end{tikzpicture}
	}
	\caption{Comparison between the baseline (Base) and the PESS approaches on the Stanford backbone network.}
	\label{fig:app_base_stanford}
\end{figure*}

\textbf{Discussion.} Figure~\ref{fig:app_base_random} compares the performance of the \ac{pess} application-aware service provisioning algorithm (\textit{PESS} in the figure) and the baseline approach (\textit{Base}) on random networks. The experimental results are plotted as functions of the network load, which is expressed in terms of the average number of security service requests in the network (Erlang).

The efficient usage of the computing resources reported in Figure~\ref{fig:consumed_resources_random} is a major benefit of the application-aware provisioning  mechanism proposed in this chapter. In particular, \ac{pess} avoids inefficiencies, such as a high bandwidth video stream being processed by a high resource demanding \ac{ips} (see the CCTV example in Section \ref{sec:model}), ultimately leading to a lower blocking probability and to a higher number of active services in the network, as shown in Figure~\ref{fig:block_prob_random} and \ref{fig:average_load_random} respectively.\\

The benefits of \ac{pess} in terms of reduced end-to-end latency are reported in Figure~\ref{fig:latency_random}. The plot illustrates the ratio between the average end-to-end latency of the chains in \textit{Experiment 2} (Baseline), and the average end-to-end latency of the chains in \textit{Experiment 1} (\ac{pess}). At low loads, when the nodes in the networks of both experiments are only partially busy, the value of this ratio is between 1.1 and 1.4. In other words, under typical operational conditions, the average end-to-end latency of chains provisioned with our approach is 10-40\% lower than the baseline. Moreover, when the nodes in \textit{Experiment 2} are heavily loaded, the processing delay introduced by busy nodes becomes very high, as modelled with Equation (\ref{eq:const_latency}). This phenomenon produces high ratios, represented by the spike in the plot, which gradually decrease at high loads when the nodes in the network of \textit{Experiment 1} also become fully loaded.

Figure~\ref{fig:app_base_garr} reports the results of the simulations performed with the GARR network. In this case, we are particularly interested in observing the behaviour of our approach in the presence of a critical region (from the security viewpoint) such as the border of the network. In order to analyse this, we empirically configure the random generator of service requests to generate 80\% of requests directed towards the Internet (i.e., crossing the border of the network). In Figure~\ref{fig:block_prob_garr}, it can be noted that both PESS and the baseline have similar blocking probability at low loads (below 6000). This is a consequence of the bandwidth usage on links towards the border region, which is almost always identical for \textit{Experiment 1} and \textit{Experiment 2}. The two curves start diverging at load 6000, i.e. when the border region runs out of computing resources with the baseline approach (as shown with dashed curves in Figure~\ref{fig:consumed_resources_garr}). The probability curves in Figure~\ref{fig:block_prob_garr} begin to re-converge at load 12000, when the border region with PESS also becomes full. Solid curves in Figure~\ref{fig:consumed_resources_garr} indicate that, between loads 1000 and 6000, when the blocking probability of the two experiments is comparable, \ac{pess} requires around 50\% less computing resources than the baseline to provision the security services.

The results obtained with the Stanford network model are presented in Figure~\ref{fig:app_base_stanford}. In contrast to the GARR network, where busy links in sparsely connected areas cause rejected requests at low loads, in these experiments we see a non-zero blocking probability only when the border region of the Stanford network runs out of computing resources, i.e. at loads 4000 and 6000 for the baseline and \ac{pess}, respectively. \\

Similar to the random networks scenario, we can observe a higher number of active services and a lower end-to-end latency with \ac{pess} in both GARR and Stanford networks. The plots are omitted due to space constraints.

\subsubsection{Scalability Evaluation}\label{sec:scalability}
\textbf{Methodology.} We evaluate the scalability of the \ac{pess} heuristic on Barab{\'a}si-Albert random topologies of between 10 and 1000 nodes. For each of these topologies, we simulate the processing of 1000 service requests and report the average execution time.

\textbf{Metrics.} Average execution time.

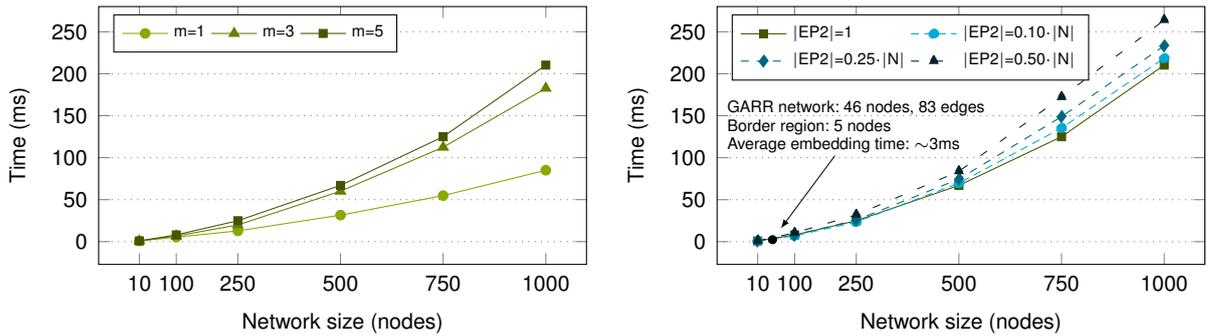
\begin{figure*}[!t]
	\parbox{0.4\linewidth}{
		\begin{tikzpicture}[define rgb/.code={\definecolor{mycolor}{RGB}{#1}}, rgb color/.style={define rgb={#1},mycolor}]
		\begin{axis}[
		legend pos=north west,
		legend columns=3,
		height=5 cm,
		width=0.5\textwidth,
		xlabel=Network size (nodes),
		ylabel=Time (ms),
		ymax=280,
		ymajorgrids,
		ytick={0,50,100,150,200,250},
		yticklabels={0,50,100,150,200,250},
		xtick={10,100,250,500,750,1000},
		xticklabels={10,100,250,500,750,1000}
		]
		
		\addplot [rgb color={136,170,0},mark=*,mark size=1.8] table[x index=0,y index=1] {\ScalabilityFiniteBWRandom};
		\addplot [rgb color={102,128,0},mark=triangle*,mark size=2.2] table[x index=0,y index=2] {\ScalabilityFiniteBWRandom};
		\addplot [rgb color={68,85,0},mark=square*,mark size=1.5] table[x index=0,y index=3] {\ScalabilityFiniteBWRandom};
		\legend{m=1,m=3,m=5};
		\end{axis}
		\end{tikzpicture}
	} 
	\hspace{1.5cm}
	\parbox{0.4\linewidth}{
		\begin{tikzpicture}[define rgb/.code={\definecolor{mycolor}{RGB}{#1}}, rgb color/.style={define rgb={#1},mycolor}]
		\begin{axis}[
		legend pos=north west,
		legend columns=2,
		height=5 cm,
		width=0.5\textwidth,
		xlabel=Network size (nodes),
		ylabel=Time (ms),
		ymax=280,
		ymajorgrids,
		ytick={0,50,100,150,200,250},
		yticklabels={0,50,100,150,200,250},
		xtick={10,100,250,500,750,1000},
		xticklabels={10,100,250,500,750,1000}
		]
		
		\addplot [rgb color={68,85,0},mark=square*,mark size=1.5] table[x index=0,y index=1] {\ScalabilityRegionSizeRandom};
		\addplot [rgb color={0,170,212},densely dashed,mark=*,mark size=1.8] table[x index=0,y index=2] {\ScalabilityRegionSizeRandom};
		\addplot [rgb color={0,102,128},dashed,mark=diamond*,mark size=2.4] table[x index=0,y index=3] {\ScalabilityRegionSizeRandom};
		\addplot [rgb color={0,34,43},loosely dashed, mark=triangle*,mark size=2.2] table[x index=0,y index=4] {\ScalabilityRegionSizeRandom};
		\addplot [color=black,mark=*,mark size=1.5] table[x index=0,y index=1] {\ScalabilityRegionSizeRandomGARR} node [pos=1, pin={[inner sep=0pt, pin edge={black, latex-},xshift=1.1cm,yshift=0.37cm] 90:{\sffamily\tiny 
				\begin{tabular}{l}
				GARR network: 46 nodes, 83 edges \\
				Border region: 5 nodes \\
				Average embedding time: $\sim$3ms \\
				\end{tabular}}
		}]{};
		\legend{$\mid$EP2$\mid$=1,$\mid$EP2$\mid$=0.10$\cdot|$N$|$,$\mid$EP2$\mid$=0.25$\cdot|$N$|$,$\mid$EP2$\mid$=0.50$\cdot|$N$|$};
		\end{axis}
		\end{tikzpicture}
	} 
	\caption{Heuristic execution time as a function of the number of physical nodes. The plot on the left reports the results with $m=1,3,5$ in the Barab{\'a}si-Albert model when generating random graphs, while the one on the right side shows the measurements with fixed $m=5$ at different sizes of endpoint $EP2$.}
	\label{fig:scalability-plots}
\end{figure*}

\textbf{Discussion.} In the first experiment (reported in the leftmost plot of Figure~\ref{fig:scalability-plots}) we used $|EP2|=1$ for all the service requests and we varied the attachment parameter $m$, which determines the number of edges to attach from a new node to existing nodes when generating the random network. This influences the execution time of the shortest path algorithm. For instance, $m=1$ produces tree-like topologies with $|E|=|N|-1$. The general rule for computing the number of edges in Barab{\'a}si-Albert networks is $|E|=m\cdot|N|-m^2$.
As illustrated in Figure \ref{fig:scalability-plots}, even for very large networks with 1000 nodes and 4975 edges ($m=5$ in the figure), on average, the PESS heuristic can provision a security service in around 200\,ms.

In the second experiment, we used a fixed value of $m=5$ (the worst case in the first experiment) and we varied the size of endpoint $EP2$, as the number of nodes in $EP2$ determines how long PESS takes to compute the initial solution. In the rightmost plot in Figure~\ref{fig:scalability-plots}, the black solid curve is the reference measurement from the first experiment. As shown by the dashed curves in the plot, the average execution time increases linearly with the size of endpoint $|EP2|$, up to around 250\,ms in the worst case with $|N|=1000$, $|E|=4975$ and $|EP2|=500$.

The \acp{vsnf} placement model and heuristic presented in this chapter target NFV-enabled systems where security services are dynamically provisioned and updated based on users' applications and their security and \ac{qos} requirements. Such systems require efficient provisioning strategies to minimize the exposure of such applications to cyber attacks. With respect to these objectives, the experimental results from the PESS scalability evaluation are encouraging and clearly indicate the potential for practical implementation of the proposed application-aware approach in real-world scenarios. 

\subsection{Related Work}\label{sec:sota}
With the recent ``softwarisation'' of network resources, a plethora of research initiatives has emerged in the last few years to address the problem of the optimal placement of chained \acp{vnf}. Most of these tackle the problem by using linear programming techniques and by proposing heuristic algorithms to cope with large scale problems. In this section, we classify and review the most relevant works for our studies.

\subsubsection{QoS-driven VNF Placement}
\ac{qos}-driven approaches primarily focus on the \ac{qos} requirements of specific services without considering network security aspects. In this regard, the proposed mathematical models include bandwidth and latency constraints (similar to Constraints (\ref{eq:const_8}) and (\ref{eq:const_latency}) presented in Section \ref{sec:model}) or define objective functions that require minimisation of the total bandwidth and latency of created chains.

The \ac{ilp} model in \cite{7469866} considers computing and bandwidth constraints to minimize the costs related to (i) \ac{vnf} deployment, (ii) energy consumption of the servers, and (iii) forwarding traffic. The end-to-end delay requirement is formulated as a penalty in the objective function. However, the computation of the end-to-end delay only considers link propagation delays without including the processing delay introduced at each \ac{vnf}. In~\cite{6968961}, the placement problem is formulated as a Mixed Integer Quadratically Constrained Problem with respect to bandwidth, number of used nodes and latency. The processing delay at each \ac{vnf} is also not considered in this work. 
The study in \cite{qos-driven} proposes an \ac{ilp} formulation and a heuristic algorithm for the \ac{vnf} placement problem focusing on \ac{qos} parameters such as end-to-end delay and \ac{nsc} availability. The \ac{ilp} model formulation presented in the paper does not discuss how the processing delay introduced by the \acp{vnf} is computed. This limitation is reflected in the assumptions made for the evaluation, where the processing delay is considered independent from the \ac{vnf} type/implementation and from the computing capacity of the physical node where \acp{vnf} are placed. 

In \cite{8480442}, Tajiki et al. present a resource allocation architecture for softwarised networks. The proposed architecture includes two resource allocation modules whose goal is configuring the network while satisfying \ac{qos} constraints and optimising the energy consumption and the number of flow entries in the network. Although the authors tackle the problem of progressively allocating resources for newly arrived flows, neither the \ac{ilp} formulation nor the heuristic algorithm consider the effects of the resource allocation on servers whose computing capacity is close to the limit. As discussed in Sections \ref{sec:constraints} and \ref{sec:implementation}, this may lead to a degradation of the \ac{qos} of existing services in terms of higher end-to-end latency.


\subsubsection{Placement of VNFs/VSNFs}\label{sec:relsecb}
In addition to the research work on QoS-driven \acp{vnf} placement, there are a number of works that specifically consider the placement of \acp{vsnf}. 


The method proposed in~\cite{Park:2017:DDP:3040992.3041005} is based on light-weight, protocol-specific intrusion detection \acp{vnf}. The system dynamically invokes a chain of these \acp{ids} according to the traffic characteristics. The placement of the chains is based on a user-defined or common shortest-path algorithm such as Dijkstra, without consideration of the application \ac{qos} requirements or available network/computing resources. 

In~\cite{7899497}, the authors argue that reactive mechanisms used by cloud providers to deploy \acp{vsnf} do not ensure an optimal resource allocation. To address this, the authors propose a novel resource allocation scheme, which estimates the behaviour of the traffic load by monitoring the history of the current \acp{vsnf}, and pro-actively provisions new instances of those \acp{vsnf} as a countermeasure to any incoming resource pressure. The proposed algorithm does not tackle the problem of \ac{vsnf} chaining. Instead, it focuses on the optimal placement of new instances of \acp{vsnf}, which are part of existing chains. It also assumes infinite network and computing resources.

In~\cite{8530989}, Dermici et al. tackle the \acp{vsnf} placement problem by proposing an \ac{ilp} formulation whose objective is the minimisation of the energy consumption of servers. This solution does not consider any security nor \ac{qos} constraints. The aim of the \ac{nsc} embedding model presented in \cite{8254344} is to minimize the end-to-end latency of cross-domain chains of \acp{vsnf}. The main limitation of the proposed \ac{ilp} formulation is that it only considers link propagation delays, while ignoring the processing delay introduced at each \ac{vsnf}.

\subsubsection{Security-driven VSNF Placement}
Although the literature reviewed in Section \ref{sec:relsecb} addresses the placement of \acp{vsnf}, few solutions have been proposed with a focus on the network security requirements of the \ac{vsnf} placement. In \cite{7592416}, the authors propose a model for the placement of \acp{vsnf} that takes into account security deployment constraints. Such constraints are necessary to avoid incorrect deployment of security functions such as placing an \ac{ids} on an encrypted channel. The authors propose an \ac{ilp} formulation of the problem and validate their model by measuring the execution time in four different scenarios and by comparing the model with other heuristics in terms of placement cost. However, the proposed optimisation algorithm is always computed for all flows in the network. Therefore, it does not scale well. The authors mitigate the problem by partitioning the network into independent blocks. Nevertheless, the partitioning scheme is limited to fat-tree topologies. Furthermore, the end-to-end latency is not considered among the constraints of the proposed model, which limits its application space.
The authors of \cite{8466784} propose an \ac{ilp} formulation and a heuristic algorithm for efficiently composing chains of virtual security functions. The \ac{ilp} formulation includes a single security-related constraint to ensure that the security level of each deployed \ac{vsnf} instance is higher than the security level required by the service request. However, this work does not take into account basic security aspects, such as order and operational mode (stateful/stateless) of the chained \acp{vsnf}. Moreover, the proposed formulation does not consider the mutual interference between security services caused by the concurrent access to the (finite) computing resources available in the infrastructure. The latter aspect is particularly relevant in a \ac{tsp} scenario (see also part II of \cite{huang}), where the security services are provisioned in a dynamic manner based on the incoming customers' requests.

	
\newcommand{\features}{11}

\graphicspath{{ddos-detection-cnn/artworks/}}
\DeclareGraphicsExtensions{.pdf,.jpeg,.png}

\section{DDoS Detection with Deep Learning}\label{sec:detection-cnn}

\lettrine[findent=2pt]{\textbf{T}}{ }he challenge of \ac{ddos} detection is the combination of attack approaches coupled with the volume of live traffic to be analysed. In this chapter, we present a practical, lightweight deep learning \ac{ddos} detection system called \acs{lucid} (\aclu{lucid}), which exploits the properties of \acfp{cnn} to classify traffic flows as either malicious or benign.

\ac{lucid} is a lightweight \ac{dl}-based \ac{ddos} detection architecture suitable for online resource-constrained environments, which leverages \acp{cnn} to learn the behaviour of \ac{ddos} and benign traffic flows with both low processing overhead and attack detection time. \ac{lucid} has been trained with the latest datasets consisting of several days of network activity, including \ac{ddos} attack traffic generated with well-known tools widely used by hackers groups such as Anonymous~\cite{hoic}. These datasets have been pre-processed to produce traffic observations consistent with those collected in online systems, where the detection algorithms must cope with segments of traffic flows collected over pre-defined time windows. 

We demonstrate that \ac{lucid} matches state-of-the-art detection accuracy whilst presenting a 40x reduction in processing time. With our evaluation results, we prove that the proposed approach is suitable for effective \ac{ddos} detection in resource-constrained operational environments, such as edge computing facilities, where devices possess limited computing capabilities.   

The research work presented in this chapter has been carried out in collaboration with the Queen's University Belfast's Centre for Secure Information Technologies. Moreover, the results have been accepted for publication in the IEEE Transactions on Network and Service Management~\cite{lucid-tnsm-paper}.

The remainder of this chapter is structured as follows: Section \ref{sec:lucid-motivation} provides the motivation behind this work. Section \ref{sec:methodology} details the methodology with respect to the network traffic processing and the \ac{lucid} \ac{cnn} model architecture. Section \ref{sec:setup} describes the experimental setup detailing the datasets and the development of \ac{lucid} with the hyper-parameter tuning process. In Section \ref{sec:lucid-evaluation}, \ac{lucid} is evaluated and compared with the state-of-the-art approaches. Section \ref{sec:analysis} introduces our kernel activation analysis for explainability of \ac{lucid}'s classification process. Section \ref{sec:usecase} presents the experiment and results for the \ac{ddos} detection at the edge. Section \ref{sec:lucid-related} reviews and discusses the related work.


\subsection{Motivation}\label{sec:lucid-motivation}
\ac{ddos} attacks are one of the most harmful threats in today's Internet, disrupting the availability of essential services in production systems and everyday life. Although \ac{ddos} attacks have been known to the network research community since the early 1980s, our network defences against these attacks still prove inadequate. 

In late 2016, the attack on the Domain Name Server (DNS) provider, Dyn, provided a worrying demonstration of the potential disruption from targeted \ac{ddos} attacks \cite{ddos-spotify}. This particular attack leveraged a botnet (Mirai) of unsecured IoT (Internet of Things) devices affecting more than 60 services. At the time, this was the largest \ac{ddos} attack recorded, at 600\,Gbps. This was exceeded in February 2018 with a major \ac{ddos} attack towards Github \cite{memcached}. At its peak, the victim saw incoming traffic at a rate of 1.3\,Tbps. The attackers leveraged a vulnerability present in memcached, a popular database caching tool. In this case, an amplification attack was executed using a spoofed source IP address (the victim IP address). If globally implemented, BCP38 ``Network Ingress Filtering'' \cite{bcp38} could mitigate such an attack by blocking packets with spoofed IP addresses from progressing through the network. However, these two examples illustrate that scale rather than sophistication enables the \ac{ddos} to succeed. 

In recent years, \ac{ddos} attacks have become more difficult to detect due to the many combinations of attack approaches. For example, multi-vector attacks where an attacker uses a combination of multiple protocols for the \ac{ddos} are common. In order to combat the diversity of attack techniques, more nuanced and more robust defence techniques are required. Traditional signature-based intrusion detection systems cannot react to new attacks. Existing statistical anomaly-based detection systems are constrained by the requirement to define thresholds for detection. \acp{nids} using machine learning techniques are being explored to address the limitations of existing solutions. In this category, \ac{dl} systems have been shown to be very effective in discriminating \ac{ddos} traffic from benign traffic by deriving high-level feature representations of the traffic from low-level, granular features of packets \cite{7946998,8482019}. However, many existing \ac{dl}-based approaches described in the scientific literature are too resource-intensive from the training perspective, and lack the pragmatism for real-world deployment. Specifically, current solutions are not designed for online attack detection within the constraints of a live network where detection algorithms must process traffic flows that can be split across multiple capture time windows.

\revised{\acfp{cnn}, a specific \ac{dl} technique, have grown in popularity in recent times leading to major innovations in computer vision \cite{resnet, imagenet,sabokrou} and Natural Language Processing \cite{KimConvolutionalNN}, as well as various niche areas such as protein binding prediction \cite{deepbind, DanQ}, machine vibration analysis \cite{janssens} and medical signal processing \cite{sleepCNN}.  Whilst their use is still under-researched in cybersecurity generally, the application of \acp{cnn} has advanced the state-of-the-art in certain specific scenarios such as malware detection \cite{AndroidCNNCODA, MultiModelDLAndroid, wang, yeo}, code analysis \cite{sourcecodecnn}, network traffic analysis \cite{7946998, wu2018novel, potluri2018convolutional, vinayakumar,liu} and intrusion detection in industrial control systems~\cite{aads}.}  These successes, combined with the benefits of \ac{cnn} with respect to reduced feature engineering and high detection accuracy, motivate us to employ \acp{cnn} in our work.  

While large \ac{cnn} architectures have been proven to provide state-of-the-art detection rates, less attention has been given to minimise their size while maintaining competent performance in limited resource environments. As observed with the Dyn attack and the Mirai botnet, the opportunity for launching \ac{ddos} attacks from unsecured IoT devices is increasing as we deploy more IoT devices on our networks. This leads to consideration of the placement of the defence mechanism. Mitigation of attacks such as the Mirai and Memcached examples include the use of high-powered appliances with the capacity to absorb volumetric \ac{ddos} attacks. These appliances are located locally at the enterprise or in the Cloud. With the drive towards edge computing to improve service provision, it becomes relevant to consider the ability to both protect against attacks closer to the edge and on resource-constrained devices. Indeed, even without resource restrictions, it is valuable to minimize resource usage for maximum system output.
\subsection{Methodology}\label{sec:methodology}

In this chapter we present \ac{lucid}, a \revised{\ac{cnn}-based solution} for \ac{ddos} detection that can be deployed in online resource-constrained environments.  Our \ac{cnn} encapsulates the learning of malicious activity from traffic to enable the identification of \ac{ddos} patterns regardless of their temporal positioning. This is a fundamental benefit of \acp{cnn}; to produce the same output regardless of where a pattern appears in the input. This encapsulation and learning of features whilst training the model removes the need for excessive feature engineering, ranking and selection.  To support an online attack detection system, we use a novel preprocessing method for \revised{the network traffic} that generates a spatial data representation used as input to the \ac{cnn}. In this section, we introduce the \revised{network traffic} preprocessing method, the \revised{\ac{cnn}} model architecture, and the learning procedure.

\begin{table}[!h]
	\centering
	\scriptsize
	\renewcommand{\arraystretch}{1.3}
	\begin{adjustbox}{width=0.75\linewidth}
		\begin{tabular}{|l|l|l|l|}
			\hline
			\textit{$\alpha$} & Learning rate & \textit{n} & Number of packets per sample \\
			\hline
			\textit{f} & Number of features per packet & \textit{s} & Batch size \\
			\hline
			\textit{h} & Height of convolutional filters & \textit{t} & Time window duration \\
			\hline
			\textit{id} & 5-tuple flow identifier & $\tau$ & Time window start time\\
			\hline
			\textit{k} & Number of convolutional filters & $\mathcal{E}$ & Array of labelled samples \\
			\hline
			\textit{m} & Max pooling size & $\mathcal{L}$ & Set of labels \\
			\hline
		\end{tabular}
	\end{adjustbox}
	\caption{Glossary of symbols.}
	\label{tab:lucid-notations}
\end{table}

\subsubsection{Network Traffic Preprocessing}\label{sec:pcap-processing}
Network traffic is comprised of data flows between endpoints. Due to the shared nature of the communication link, packets from different data flows are multiplexed resulting in packets from the same flow being separated for transmission. This means that the processing for live presentation of traffic to a \ac{nids} is quite different to the processing of a static dataset comprising complete flows. For the same reason, the ability to generate flow-level statistics, as relied upon by many of the existing works described in Section \ref{sec:lucid-related}, is not feasible in an online system.

In order to develop our online \ac{nids}, we created a tool that converts the traffic flows extracted from network traffic traces \revised{of a dataset} into array-like data structures and splits them into sub-flows based on time windows. Shaping the input as packet flows in this manner creates a spatial data representation, which allows the \ac{cnn} to learn the characteristics of \ac{ddos} attacks and benign traffic through the convolutional filters sliding over such input to identify salient patterns. This form of input is compatible with traffic captured in online deployments.  The process is illustrated in Algorithm \ref{lst:pcap-algorithm} and described next. The symbols are defined in Table \ref{tab:lucid-notations}. 

\begin{algorithm}[t!]
	\caption{Network traffic preprocessing algorithm}
	\label{lst:pcap-algorithm}
	\begin{algorithmic}[1]
		
		\renewcommand{\algorithmicrequire}{\textbf{Input:}}
		\renewcommand{\algorithmicensure}{\textbf{Output:}}
		\Require Network traffic trace $(NTT)$, flow-level labels $(\mathcal{L})$, time window $(t)$, max packets/sample $(n)$
		\Ensure List of labelled samples $(\mathcal{E})$
		\Procedure{PreProcessing}{$NTT$, $\mathcal{L}$, $t$, $n$}
		\State $\mathcal{E}\gets\emptyset$ \Comment Initialise the set of samples
		\State $\tau \gets -1$ \Comment Initialise the time window start-time 
		\ForAll {$pkt \in NTT $} \Comment Loop over the packets
		\State $id\gets pkt.tuple$ \Comment 5-tuple flow identifier
		\If {$\tau == -1$ \textbf{or} $pkt.time > \tau+t$} 
		\State $\tau\gets pkt.time$ \Comment Time window start time
		\EndIf
		\If {$\big|\mathcal{E}[\tau,id]\big|<n$} \Comment Max $n$ pkts/sample
		\State $\mathcal{E}[\tau,id].pkts.append(pkt.features)$
		\EndIf
		\EndFor
		\State $\mathcal{E}\gets normalisation\_padding(\mathcal{E})$
		\ForAll {$e \in \mathcal{E} $} \Comment Labelling
		\State $e.label\gets \mathcal{L}[e.id]$ \Comment Apply the label to the sample
		\EndFor
		\State \textbf{return} $\mathcal{E}$
		\EndProcedure
		
	\end{algorithmic}
\end{algorithm}

\textbf{Feature extraction.}
Given a traffic trace file from the dataset and a pre-defined time window of length $t$ seconds, the algorithm collects all the packets from the file with capture time between $t_0$, the capture time of the first packet, and time $t_0+t$. From each packet, the algorithm extracts \features\ attributes (see Table \ref{tab:features}). We intuitively exclude those attributes that would be detrimental to the generalisation of the model, such as \revised{IP addresses and TCP/UDP ports (specific to the end-hosts and user applications), link layer encapsulation type (linked to the network interfaces) and application-layer attributes (e.g., IRC or HTTP protocol attributes).}

\textbf{Data processing algorithm.}
This procedure, described in Algorithm \ref{lst:pcap-algorithm} at lines 4-12, simulates the traffic capturing process of online \acp{ids}, where the traffic is collected for a certain amount of time $t$ before being sent to the anomaly detection algorithms. Hence, such algorithms must base their decisions on portions of traffic flows, without the knowledge of their whole life. To simulate this process, the attributes of the packets belonging to the same bi-directional traffic flow are grouped in chronological order to form an example of shape $[n,f]$ (as shown in Table \ref{tab:features}), where $f$ is the number of features (\features) and $n$ is the maximum number of packets the parsing process collects for each flow within the time window. $t$ and $n$ are hyper-parameters for our \ac{cnn}. Flows longer than $n$ are truncated, while shorter flows are zero-padded at the end during the next stage after normalisation. The same operations are repeated for the packets within time window $[t_0+t,t_0+2t]$ and so on, until the end of the file. 

Logically, we hypothesize that short time windows enable the online systems to detect \ac{ddos} attacks within a very short time frame. Conversely, higher values of $t$ and $n$ offer more information on flows to the detection algorithms, which we expect to result in higher detection accuracy. The sensitivity of our \ac{cnn} to the values of $t$ and $n$ is evaluated in Section \ref{sec:setup}.

\begin{figure}[!h]
	\begin{center}
		\includegraphics[width=0.85\linewidth]{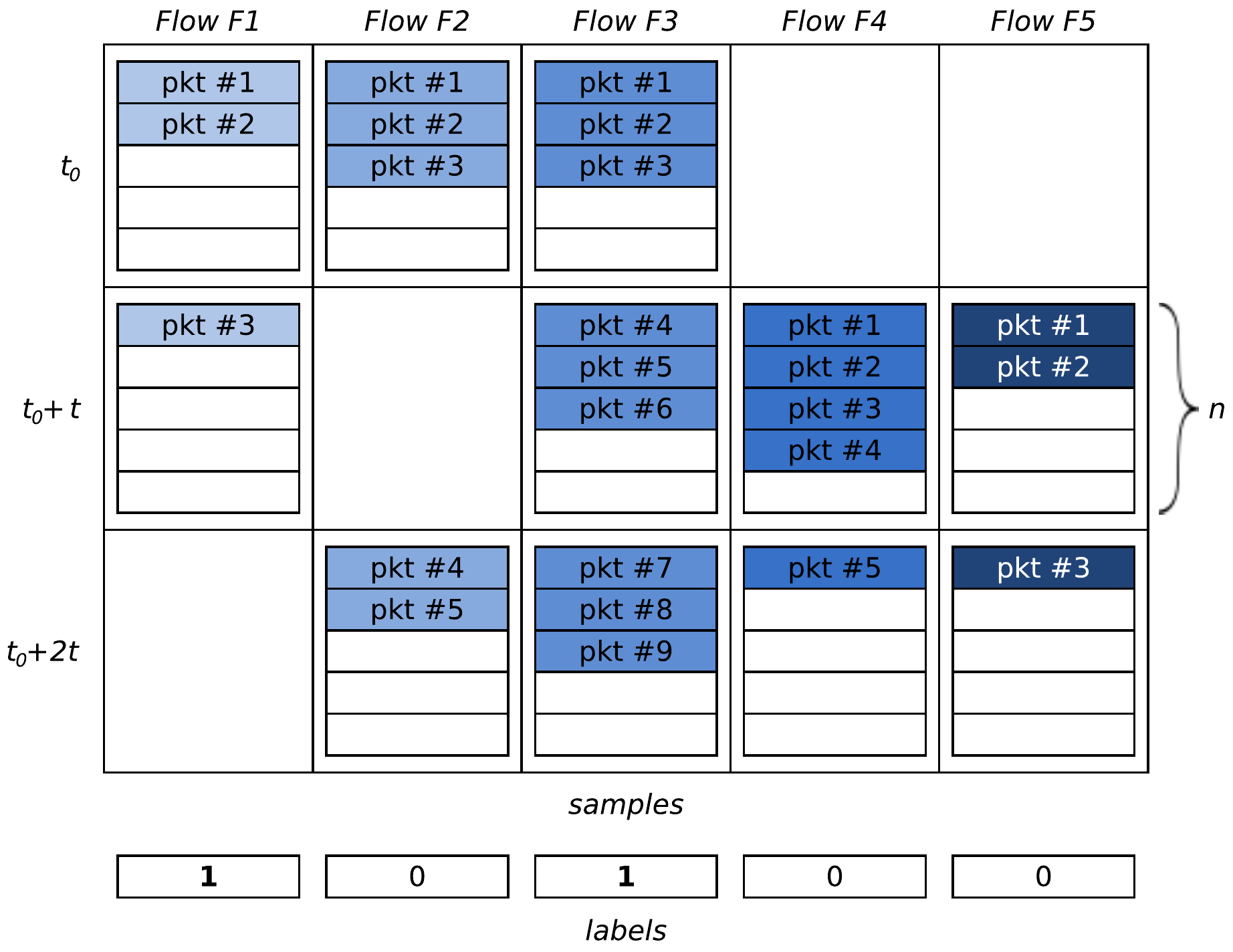}
		\caption{Graphical representation of $\mathcal{E}$.}
		\label{fig:dataset-format}
	\end{center}
	\vspace{-3mm}
\end{figure}

The output of this process can be seen as a bi-dimensional array of samples ($\mathcal{E}[\tau,id]$ in Algorithm \ref{lst:pcap-algorithm}). A row of the array represents the samples whose packets have been captured in the same time window, whilst a column represents the samples whose packets belong to the same bi-directional flow. A graphical representation of array $\mathcal{E}$ is provided in Figure~\ref{fig:dataset-format}.

\begin{table*}[!t]
	\centering
	\scriptsize
	\renewcommand{\arraystretch}{1.5}
	\begin{threeparttable}
		\begin{tabular}{rc|c|c|c|c|c|c|c|c|c|c|c}
			\Cline{1.2pt}{2-13}
			&
			\textbf{Pkt \#} & \begin{tabular}{@{}c@{}}\textbf{Time}\\\textbf{(sec)}\tnote{1}\end{tabular} & \begin{tabular}{@{}c@{}}\textbf{Packet}\\\textbf{Len}\end{tabular} & \begin{tabular}{@{}c@{}}\textbf{Highest}\\\textbf{Layer}\tnote{2}\end{tabular} &
			\begin{tabular}{@{}c@{}}\textbf{IP}\\\textbf{Flags}\end{tabular} &  \textbf{Protocols}\tnote{3} & \begin{tabular}{@{}c@{}}\textbf{TCP}\\\textbf{Len}\end{tabular} & \begin{tabular}{@{}c@{}}\textbf{TCP}\\\textbf{Ack}\end{tabular} & \begin{tabular}{@{}c@{}}\textbf{TCP}\\\textbf{Flags}\end{tabular} & \begin{tabular}{@{}c@{}}\textbf{TCP}\\\textbf{WSize}\end{tabular} & \begin{tabular}{@{}c@{}}\textbf{UDP}\\\textbf{Len}\end{tabular} & \begin{tabular}{@{}c@{}}\textbf{ICMP}\\\textbf{Type}\end{tabular}  \\ \Cline{0.8pt}{2-13}
			
			\multirow{4}*{\hspace{-1em}\rotatebox[origin=c]{90}{\textit{Packets}}\vspace{1em}$\left\{\begin{matrix}\vspace{5.8em}\end{matrix}\right.$\hspace{-1em}}
			& 0 & 0 & 151 & 99602525 & 0x4000 & 0011010001000b & 85 & 336 & 0x018 & 1444 &  0 & 0 \\ \cline{2-13}
			& 1 & 0.092 & 135 & 99602525  & 0x4000 & 0011010001000b & 69 & 453 & 0x018 & 510 & 0 & 0 \\ \cline{2-13}
			& \vdots & \vdots & \vdots & \vdots & \vdots & \vdots & \vdots & \vdots & \vdots & \vdots & \vdots & \vdots \\  \cline{2-13}
			& $j$ & 0.513 &  66 & 78354535  & 0x4000 & 0010010001000b & 0 & 405 & 0x010 & 1444 & 0 & 0 \\ \cline{2-13}
			\multirow{3}*{\hspace{-1em}\rotatebox[origin=c]{90}{\textit{Padding}}\hspace{-0.2em}\vspace{0.3em}$\left\{\begin{matrix}\vspace{4em}\end{matrix}\right.$\hspace{-1em}} 
			& $j+1$ & 0 & 0 & 0 & 0 & 0000000000000b & 0 & 0 & 0 & 0 & 0 & 0 \\ \cline{2-13}
			& \vdots & \vdots & \vdots & \vdots & \vdots & \vdots & \vdots & \vdots & \vdots & \vdots & \vdots & \vdots \\  \cline{2-13}
			& $n$ & 0 & 0 & 0 & 0 & 0000000000000b & 0 & 0 & 0 & 0 & 0 & 0 \\ 
			\Cline{0.8pt}{2-13}
		\end{tabular}
		\vspace{0.2em}
		\begin{tablenotes}
			\scriptsize
			\setlength{\itemindent}{0.5em}
			\item[1] Relative time from the first packet of the flow.
			\item[2] Numerical representation of the highest layer recognised in the packet.
			\item[3] Binary representation of the list of protocols recognised in the packet using the well-known \ac{bow} model.
			\item[] It includes protocols from Layer 2 (\textit{arp}) to common clear text application layer protocols such as \textit{http}, \textit{telnet}, \textit{ftp} and \textit{dns}. 
		\end{tablenotes}
	\end{threeparttable}	
 	\caption{A TCP flow sample before normalisation.}
 	\label{tab:features}
\end{table*}

\textbf{Normalisation and padding.}
Each attribute value is normalised to a $[0,1]$ scale and the samples are zero-padded so that each sample is of fixed length $n$, since having samples of fixed length is a requirement for a CNN to be able to learn over a full sample set.  In Figure~\ref{fig:dataset-format}, each non-empty element of the array $\mathcal{E}$ is a compact graphical representation of a sample. 
In each $\mathcal{E}$ element, coloured rows are the packets in the form of \features\ normalised attributes (i.e., the upper part of Table \ref{tab:features}), while the white rows represent the zero-padding (i.e., the lower part of Table \ref{tab:features}). Please note that, empty elements in Figure~\ref{fig:dataset-format} are for visualisation only and are not included in the dataset. An empty $\mathcal{E}[\tau,id]$ means that no packets of flow $id$ have been captured in time window $[\tau,\tau+t]$ (e.g., $\mathcal{E}[t_0,F4]$).

\textbf{Labelling.} Each example $\mathcal{E}[\tau,id]$ is labelled by matching its flow identifier $id$ with the labels provided with the original dataset (lines 14-16 in Algorithm \ref{lst:pcap-algorithm}). This also means that the value of the label is constant along each column of array $\mathcal{E}$, as represented in Figure~\ref{fig:dataset-format}.

\subsubsection{\acs{lucid} Model Architecture}\label{sec:architecture}
\begin{figure}[t!]
	\begin{center}
		\includegraphics[width=0.9\linewidth]{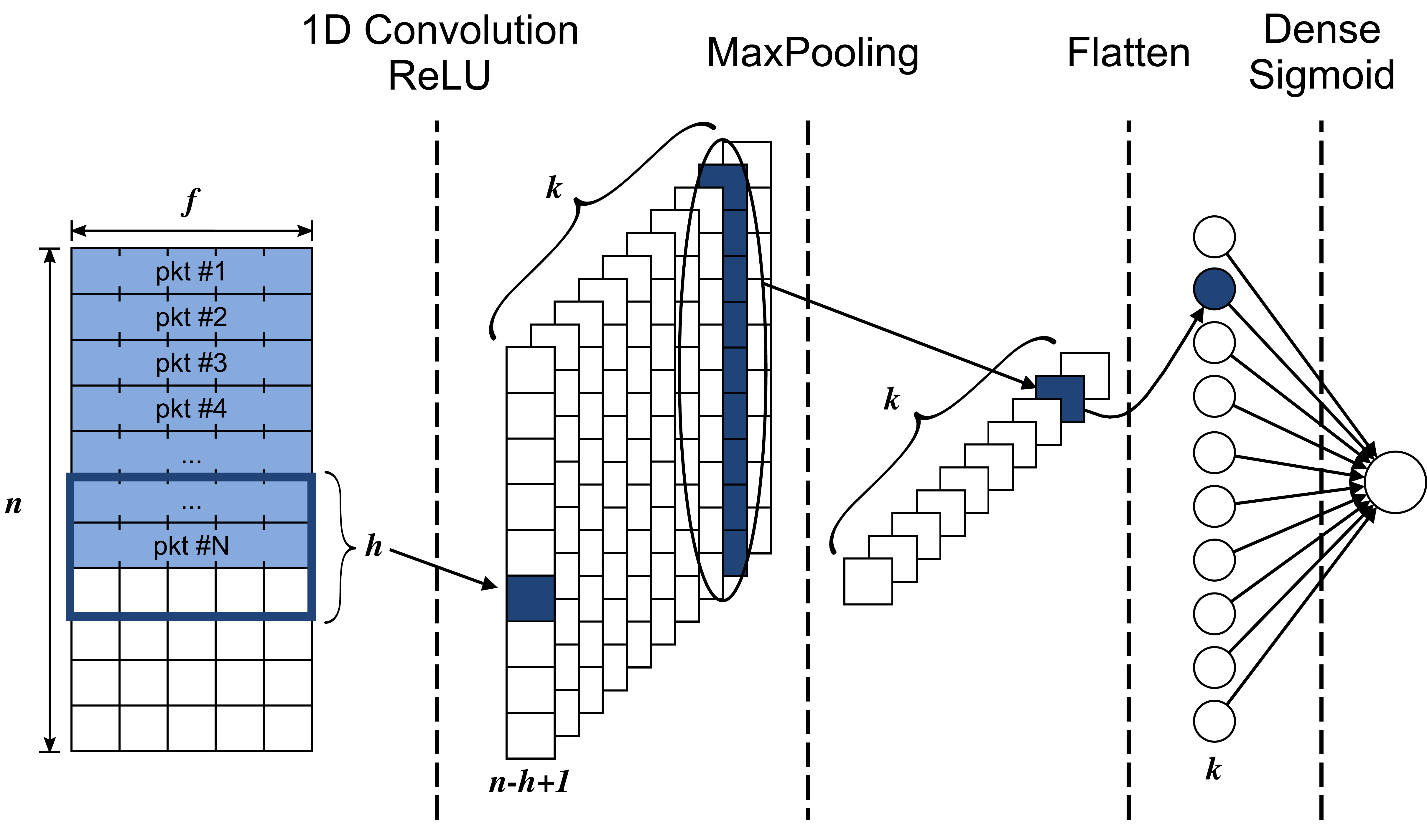}
		\caption{\ac{lucid} architecture.}
		\label{fig:CNN-architecture}
	\end{center}
\end{figure}

We take the output from Algorithm \ref{lst:pcap-algorithm} as input to our \ac{cnn} model for the purposes of online attack detection.  \ac{lucid} classifies traffic flows into one of two classes, either malicious (\ac{ddos}) or benign.  Our objective is to minimise the complexity and performance time of this \ac{cnn} model for feasible deployment on resource-constrained devices. To achieve this, the proposed approach is a lightweight, supervised \revised{detection system} that incorporates a \ac{cnn}\revised{, similar to that of \cite{KimConvolutionalNN} from the field of Natural Language Processing.} CNNs have shared and reused parameters with regard to the weights of the kernels, whereas in a traditional neural network every weight is used only once. This reduces the storage and memory requirements of our model.  The complete architecture is depicted in Figure~\ref{fig:CNN-architecture} and described in the next sections, with the hyper-parameter tuning and ablation studies being discussed in Section \ref{sec:setup}.

\textbf{Input layer.}
Recall that each traffic flow has been reshaped into a 2-D matrix of packet features as per Section \ref{sec:pcap-processing}, creating a novel spatial representation that enables the \ac{cnn} to learn the correlation between packets of the same flow.  Thus, this first layer takes as input a traffic flow represented by a matrix $F$ of size $n\times f$.  $F$ contains $n$ individual packet vectors, such that $F$ = \{$pkt_1$, ... , $pkt_n$\} where $pkt_n$ is the $n$th packet in a flow, and each packet vector has length $f$ = \features\ features.

\textbf{CNN layer.} As per Figure~\ref{fig:CNN-architecture}, each input matrix $F$ is operated on by a single convolutional layer with $k$ filters of size $h\times f$, with $h$ being the length of each filter, and again $f$ = \features.  Each filter, also known as a kernel or sliding window, convolves over $F$ with a step of 1 to extract and learn local features that contain useful information for detection of \ac{ddos} and benign flows.  Each of the $k$ filters generates an activation map $a$ of size ($n-h+1$), such that $a_k = ReLU(Conv(F)W_k,b_k)$, where $W_k$ and $b_k$ are the weight and bias parameters of the $k$th filter that are learned during the training stage.  To introduce non-linearity among the learned filters, we use the rectified linear activation function $ReLU(x) = max\{0,x\}$, as per convention for CNNs.  All activation maps are stacked, creating an activation matrix $A$ of size $(n-h+1)\times k$, such that $A=[a_1 | ... | a_k]$.  

There are two main benefits of including a \ac{cnn} in our architecture.  Firstly, it allows the model to benefit from efficiency gains compared to standard neural networks, since the weights in each filter are reused across the whole input. Sharing weights, instead of the full end-to-end connectivity with a standard neural net, makes the model more lightweight and reduces its memory footprint as the number of learnable parameters is greatly reduced.  Secondly, during the training phase, the \ac{cnn} automatically learns the weights and biases of each filter such that the learning of salient characteristics and features is encapsulated inside the resulting model during training. This reduces the time-consuming feature engineering and ranking involved in statistical and traditional machine learning methods, which relies on expert human knowledge. As a result, this model is more adaptable to new subtleties of \ac{ddos} attack, since the training stage can be simply repeated anytime with fresh training data without having to craft and rank new features.

\textbf{Max pooling layer.}
For max pooling, we down-sample along the first dimension of $A$, which represents the temporal nature of the input.  A pool size of $m$ produces an output matrix $m_o$ of size $((n-h+1)/m)\times k$, which contains the largest $m$ activations of each learned filter, such that $m_o = [max(a_1) | ... | max(a_k)]$.  In this way, the model disregards the less useful information that produced smaller activations, instead paying attention to the larger activations.  This also means that we dispose of the positional information of the activation, i.e. where it occurred in the original flow, giving a more compressed feature encoding, and, in turn, reducing the complexity of the network.  $m_o$ is then flattened to produce the final one-dimensional feature vector $v$ to be input to the classification layer.

\textbf{Classification layer.}
$v$ is input to a fully-connected layer of the same size, and the output layer has a sole node.  This output $x$ is passed to the sigmoid activation function such that $\sigma(x) = 1/(1+e^{-x})$.  This constrains the activation to a value of between 0 and 1, hence returning the probability $p\in[0,1]$ of a given flow being a malicious \ac{ddos} attack. The flow is classified as \ac{ddos} when $p>0.5$, and benign otherwise.

\subsubsection{The Learning Procedure}\label{sec:learning}

When training \ac{lucid}, the objective is to minimise its cost function through iteratively updating all the weights and biases contained within the model.  These weights and biases are also known as trainable, or learnable, parameters.  The cost function calculates the cost, also called the error or the loss, between the model's prediction, and the ground truth of the input.  Hence by minimising this cost function, we reduce the prediction error.  At each iteration in training, the input data is fed forward through the network, the error calculated, and then this error is back-propagated through the network. This continues until convergence is reached, when further updates don't reduce the error any further, or the training process reaches the set maximum number of epochs.  With two classes in our problem the binary cross-entropy cost function is used.  Formally this cost function $c$ that calculates the error over a batch of $s$ samples can be written as: 
\begin{equation}
     c = -\frac{1}{s}\sum_{j=1}^{s}(y_j\log p_j + (1-y_j)\log(1-p_j)) 
\end{equation}

where $y_j$ is the ground truth target label for each flow $j$ in the batch of $s$ samples, and $p_j$ is the predicted probability flow $j$ is malicious \ac{ddos}.  This is supervised learning because each flow in our datasets is labelled with the ground truth, either \ac{ddos} or benign.  To reduce bias in our learning procedure, we ensure that these datasets are balanced with equal numbers of malicious and benign flows, which gives a greater degree of confidence that the model is learning the correct feature representations from the patterns in the traffic flows.  As previously highlighted, the learning is encapsulated inside the model by all the weights and biases, meaning that our approach does not require significant expert input to craft bespoke features and statistically assess their importance during preprocessing, unlike many existing methods, as outlined in Section \ref{sec:lucid-related}. \\


\subsection{Experimental Setup}\label{sec:setup}
\pgfplotstableread[header=true]{
P	T=1		T=10	T=100
1	0.94244	0.91464	0.88712
2	0.97988	0.98657	0.98975
3	0.98695	0.99104	0.9942
4	0.99045	0.99051	0.99379
5	0.9912	0.99202	0.99367
10	0.99022	0.99172	0.99406
20	0.98998	0.9923	0.99393
50	0.99023	0.99451	0.99472
100	0.98844	0.99153	0.99501

}\FScoreP

\pgfplotstableread[header=true]{
T	P=1		P=2		P=10	P=100
1	0.94244	0.98975	0.99022	0.98844
2	0.92951	0.98569	0.99179	0.98957
3	0.93132	0.98486	0.99202	0.98887
4	0.93276	0.98657	0.98756	0.98911
5	0.9272	0.97988	0.98678	0.98967
10	0.91464	0.98448	0.99172	0.99153
20	0.90821	0.98295	0.99325	0.9937
50	0.89324	0.98202	0.9946	0.99383
100	0.88712	0.98531	0.99406	0.99501

}\FScoreT

\pgfplotstableread[header=true]{
	K	100s100p
	1	0.98132
	2	0.98707
	4	0.99128
	8	0.99267
	16	0.9943
	32	0.99495
	64	0.995
		
}\FScoreK

\pgfplotstableread[header=true]{
	H	100s100p
	1	0.99344
	2	0.99456
	3	0.99501
	4	0.99510
	5	0.99520
	
}\FScoreH

\pgfplotstableread[header=true]{
	f	100s100p
	1	0.87896
	2	0.97490
	3	0.99192
	4	0.99245
	5	0.99265
	6	0.99235
	7	0.99210
	8	0.99407
	9	0.99475
	10	0.99495
	11  0.99501
	
}\FeatureAnalysis

\pgfplotstableread[header=true]{
	BATCHSIZE	PKTSECGPU	SAMPLESECGPU	PKTSECCPU	SAMPLESECCPU
	64	504809	18780	346298	12883
	128	458845	16710	396203	14740
	256	562915	20942	445924	16589
	512	591070	21989	479257	17829
	1024	614937	22877	521698	19408
	2048	587934	21872	496299	18463
	4096	507707	18888	448861	16699
	8192	501544	18658	416435	15492

}\PredictionTX

\pgfplotstableread[header=true]{
	BATCHSIZE	TIMEGPU		TIMECPU 	TIMESERVER
	64			43.259		78.27075	26.5802
	128			36.80025	48.01475	20.1674
	256			28.56475	43.246		11.0942
	512			27.05625	41.37		10.10325
	1024		26.389		40.47675	9.1022
	2048		26.05425	41.33		7.9872
	4096		25.811		42.20275	9.484
	8192		26.6555		43.2305		12.407

}\TrainingBatchSizeTX

\pgfplotstableread[header=true]{
	SETSIZE	TIMESAMPLECPU	TOTALTIME	SCORE
	25000	6.625			1689		0.99361
	50000	12.203			1977		0.99430
	75000	17.526			3014		0.99470
	100000	23.2			3666		0.99485
	125000	28.437			5630		0.99485
	150000	34.371			6771		0.99490
	181551	41.303			7600		0.99501
				
}\TrainingSetSizeTX

\subsubsection{Datasets}\label{sec:datasets}
Our \revised{\ac{cnn}} model is validated with recent datasets ISCX2012 \cite{ISCXIDS2012}, CIC2017 \cite{CICIDS2017} and CSECIC2018 \cite{CSECICIDS2018} provided by the Canadian Institute for Cybersecurity of the \ac{unb}, Canada. They consist of several days of network activity, normal and malicious, including \ac{ddos} attacks. The three datasets are publicly available in the form of traffic traces in \textit{pcap} format including full packet payloads, plus supplementary text files containing the labels and statistical details for each traffic flow. 

The \ac{unb} researchers have generated these datasets by using profiles to accurately represent the abstract properties of human and attack behaviours. One profile characterises the normal network activities and provides distribution models for applications and  protocols (HTTP, SMTP, SSH, IMAP, POP3 and FTP) produced with the analysis of real traffic traces. Other profiles describe a variety of attack scenarios based on recent security reports. They are used to mimic the behaviour of the malicious attackers by means of custom botnets and well-known \ac{ddos} attacking tools such as High Orbit Ion Cannon (HOIC)~\cite{hoic} and its predecessor, the Low Orbit Ion Cannon (LOIC)~\cite{loic}. HOIC and LOIC have been widely used by Anonymous and other hacker groups in some highly-publicised attacks against PayPal, Mastercard, Visa, Amazon, Megaupload, among others \cite{paypal}.

Table \ref{tab:unb-datasets} shows the parts of the three datasets used in this work. In the table, the column \textit{Traffic trace} specifies the name of the trace, according to \cite{ISCXIDS2012}, \cite{CICIDS2017} and \cite{CSECICIDS2018}. \revised{Specifically, the \textit{ISCX2012-Tue15} trace contains a DDoS attack based on an IRC botnet. The \textit{CIC2017-Fri7PM} trace contains a HTTP \ac{ddos} generated with LOIC, while the \textit{CSECIC2018-Wed21} trace contains a HTTP \ac{ddos} generated with HOIC.} With respect to the original file, the trace \textit{CIC2017-Fri7PM} is reduced to timeslot 3.30PM-5.00PM to exclude malicious packets related to other cyber attacks (port scans and backdoors). 

\begin{table}[h!]
	\vspace{5mm}
	\centering 
	\begin{tabular}{llccc} \toprule[\heavyrulewidth]
		\textbf{Dataset} & \textbf{Traffic trace} & \textbf{\#Flows} & \textbf{\#Benign} & \textbf{\#\ac{ddos}} \\ \midrule[\heavyrulewidth]
		ISCX2012	& Tue15 & 571698 & 534320 & 37378\\ \midrule
		CIC2017	& Fri7PM & 225745 & 97718 & 128027 \\ \midrule
		CSECIC2018& Wed21 & 1048575 & 360832 & 687743\\ \bottomrule[\heavyrulewidth]
	\end{tabular}
	\caption{The datasets from \ac{unb} \cite{unb-datasets}.} 
	\label{tab:unb-datasets}
\end{table}

In an initial design, the model was trained and validated on the ISCX2012 dataset producing high accuracy results. However, testing the model on the CIC2017 dataset confirmed the generally held observation that a model trained on one dataset will not necessarily perform well on a completely new dataset. \revised{In particular, we obtained a false negative rate of about 17\%. This can be attributed to the different attacks represented in the two datasets, as previously described.} What we attempt in this work is to develop a model that when trained and validated across a mixed dataset can reproduce the high performance results on completely unseen test data. To achieve this, a combined training dataset is generated as described in Sec. \ref{sec:data-preparation}. 


\subsubsection{Data Preparation}\label{sec:data-preparation}
We extract the 37378 \ac{ddos} flows from ISCX2012, plus randomly select 37378 benign flows from the same year to balance. We repeat this process with 97718/97718 benign/\ac{ddos} flows for CIC2017 and again with 360832/360832 benign/\ac{ddos} flows for CSECIC2018. 

After the pre-preprocessing stage, where flows are translated into array-like data structures (Section \ref{sec:pcap-processing}), each of the three datasets is split into training (90\%) and test (10\%) sets, with 10\% of the training set used for validation. Please note that, the split operation is performed on a per-flow basis to ensure that samples obtained from the same traffic flow end up in the same split, hence avoiding the ``contamination'' of the validation and test splits with data used for the training. We finally combine the training splits from each year by balancing them with equal proportions from each year to produce a single training set. We do the same with the validation and test splits, to obtain a final dataset referred to as \textit{UNB201X} in the rest of the chapter. UNB201X training and validation sets are only used for training the model and tuning the hyper-parameters (Section \ref{sec:tuning}), while the test set is used for the evaluation presented in Sections \ref{sec:lucid-evaluation} and \ref{sec:usecase}, either as a whole combined test set, or as individual per-year test sets for state-of-the-art comparison.  



A summary of the final UNB201X splits is presented in Table \ref{tab:datasets-splits}, which reports the number of samples as a function of time window duration $t$. As illustrated in Table \ref{tab:datasets-splits}, low values of this hyper-parameter yield larger numbers of samples. Intuitively, using short time windows leads to splitting traffic flows into many small fragments (ultimately converted into samples), while long time windows produce the opposite result. In contrast, the value of $n$ has a negligible impact on the final number of samples in the dataset.

\begin{table}[h!]
	\vspace{5mm}
	\centering 
	\begin{tabular}{lcccc} \toprule[\heavyrulewidth]
		\begin{tabular}{@{}l@{}}\textbf{Time}\\\textbf{Window}\end{tabular} & \begin{tabular}{@{}c@{}}\textbf{Total}\\\textbf{Samples}\end{tabular} & \begin{tabular}{@{}c@{}}\textbf{Training}\end{tabular} & \begin{tabular}{@{}c@{}}\textbf{Validation}\end{tabular}& \begin{tabular}{@{}c@{}}\textbf{Test}\end{tabular} \\ \midrule[\heavyrulewidth]
	
		$t$=1s & 480519 & 389190 & 43272 & 48057\\   
		$t$=2s & 353058 & 285963 & 31782 & 35313\\ 
		$t$=3s & 310590 & 251574 & 27957 & 31059\\ 
		
		$t$=4s & 289437 & 234438 & 26055 & 28944 \\   
		$t$=5s & 276024 & 223569 & 24852 & 27603\\ 
		$t$=10s & 265902 & 215379 & 23931 & 26592\\ 
	
		$t$=20s & 235593 & 190827 & 21204 & 23562\\   
		$t$=50s & 227214 & 184041 & 20451 & 22722\\ 
		$t$=100s & 224154 & 181551 & 20187 & 22416\\ \bottomrule[\heavyrulewidth]
	\end{tabular}
	\caption{UNB201X dataset splits.} 
	\label{tab:datasets-splits}
\end{table}

\subsubsection{Evaluation Methodology}
\revised{As per convention in the literature, we report the metrics \textit{Accuracy (ACC)}, \textit{\ac{fpr}}, \textit{Precision} (or \textit{\ac{ppv}}), \textit{Recall} (or \textit{\ac{tpr}}) and \textit{F1 Score (F1)}, with a focus on the latter. \textit{Accuracy} is the percentage of correctly classified samples (both benign and \ac{ddos}). \textit{\ac{fpr}} represents the percentage of samples that are falsely classified as \ac{ddos}. \textit{\ac{ppv}} is the ratio between the correctly detected \ac{ddos} samples and all the detected \ac{ddos} samples (true and false). \textit{\ac{tpr}} represents the percentage of \ac{ddos} samples that are correctly classified as such. The \textit{F1 Score} is an overall measure of a model's performance; that is the harmonic mean of the \textit{\ac{ppv}} and \textit{\ac{tpr}}. These metrics are formally defined as follows:\\
	
	\begin{center}
		$\displaystyle ACC=\frac{TP+TN}{TP+TN+FP+FN}\qquad FPR=\frac{FP}{FP+TN}$\\ \vspace{1.5em}
		$\displaystyle PPV=\frac{TP}{TP+FP}\qquad TPR=\frac{TP}{TP+FN}\qquad F1=2\cdot\frac{PPV\cdot TPR}{PPV + TPR}$
		\vspace{3mm}
	\end{center}
	
\noindent where \textit{TP=True Positives}, \textit{TN=True Negatives}, \textit{FP=False Positives}, \textit{FN=False Negatives}.}

The output of the training process is a combination of trainable and hyper parameters that maximizes the \textit{F1 Score} on the validation set or, in other words, that minimizes the total number of False Positives and False Negatives. 

Model training and validation have been performed on a server-class computer equipped with two 16-core Intel Xeon Silver 4110 @2.1\,GHz CPUs and 64\,GB of RAM. The models have been implemented in Python v3.6 using the Keras API v2.2.4 \cite{keras} on top of Tensorflow 1.13.1 \cite{tensorflow}.

\subsubsection{Hyper-parameter Tuning}\label{sec:tuning}
Tuning the hyper-parameters is an important step to optimise the model's accuracy, as their values influence the model complexity and the learning process.  Prior to our experiments, we empirically chose the hyper-parameter values based on the results of preliminary tuning and on the motivations described per parameter. We then adopted a grid search strategy to explore the set of hyper-parameters using \textit{F1 score} as the performance metric.  At each point in the grid, the training continues indefinitely and stops when the loss does not decrease for a consecutive 25 times. Then, the search process saves the \textit{F1 score} and moves to the next point.

As per Section \ref{sec:data-preparation}, UNB201X is split into training, validation and testing sets.  For hyper-parameter tuning, we use only the validation set. It is important to highlight that we do not tune to the test set, as that may artificially improve performance. 
The test set is kept completely unseen, solely for use in generating our experimental results, which are reported in Section \ref{sec:lucid-evaluation}.  

\textbf{Learning Rate.}
The learning rate $\alpha\in(0,1]$ controls the speed at which the model learns. Common practice is to start with $\alpha=0.1$ and then progressively reduce the order of magnitude (0.01, 0.001, etc.). We trained our model using the Adam optimizer \cite{adam} starting with $\alpha=0.1,0.01,0.001$ and no learning rate decay. As the optimizer could not converge with $\alpha=0.1$ in the case of 32 or more convolutional filters, and it converged too slowly with $\alpha=0.001$, we set $\alpha=0.01$.

\textbf{Batch Size.}
The batch size $s$ is the number of training samples used in one training iteration.  The value of the batch size is usually increased by a power of two (e.g., 1, 2, 4, etc.). Lower values of $s$ mean higher number of forward and backward propagations for each epoch, hence possibly higher accuracy but also longer learning time. As we did not experience any substantial variation in the \textit{F1 Score} while varying $s$ in the preliminary tests, we empirically limited the tuning of the batch size to $s=1024,2048$. We experimented with both values in this tuning phase.

\textbf{Maximum number of packets/sample.}
$n$ is important for the characterisation of the traffic and for capturing the temporal patterns of traffic flows. The value of $n$ indicates the maximum number of packets of a flow recorded in chronological order in a sample. 

\begin{figure}[h!]
	\centering
	\captionsetup{justification=centering}
	\begin{tikzpicture}
	\begin{semilogxaxis}[  
	legend pos=south east,
	legend columns=1,
	height=6 cm,
	width=0.75\linewidth,
	grid = both,
	xlabel={Value of hyper-parameter $n$ (packets/example) in logarithmic scale},
	ylabel={F1 Score},
	scaled y ticks=false,
	scaled x ticks=false,
	xmin=1,
	xmax=100,
	xtick={1,2,3,4,5,10,20,50,100},
	xticklabels={1,2,3,4,5,10,20,50,100},
	xtick pos=left,
	ymin=0.88, ymax=1,
	ytick={0.88,0.90,0.95,0.98,0.99,1},
	yticklabels={0.88,0.90,0.95,0.98,0.99,1},
	ytick pos=left,
	enlargelimits=0.02,
	]
	\addplot [color=red,style=thick,mark=*,mark size=2.5] table[x index=0,y index=1] {\FScoreP};
	\addplot [color=blue,style=thick,mark=x,mark size=3.5] table[x index=0,y index=2] {\FScoreP};
	\addplot [color=orange,style=thick,mark=triangle*,mark size=3.3] table[x index=0,y index=3] {\FScoreP};
	\node[fill=white,anchor=west] at (axis cs: 4,.96) {\sffamily\scriptsize $\alpha=0.01$, $s=2048$, $k=64$, $h=3$, $m=n-h+1$};
	\legend{$t$=1, $t$=10,$t$=100}
	\end{semilogxaxis}
	\end{tikzpicture}
	\caption{Sensitivity of our model to hyper-parameter $n$.}
	\label{fig:sensitivity_n}
\end{figure}
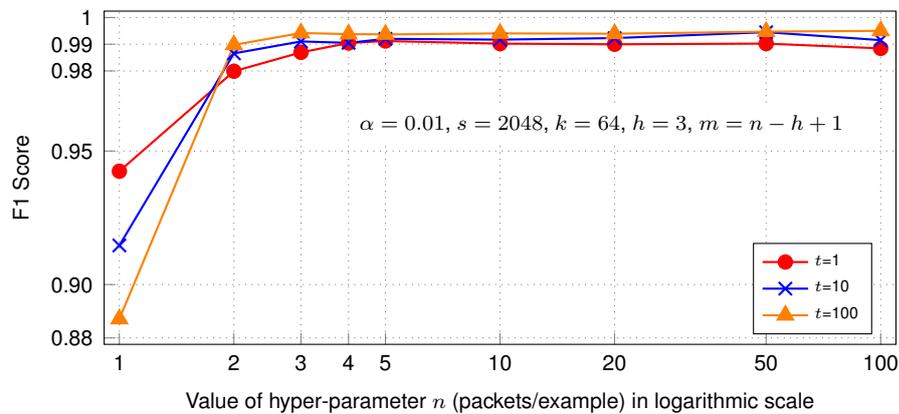

The resulting set of packets describes a portion of the life of the flow in a given time window, including the (relative) time information of packets. Repetition-based \ac{ddos} attacks use a small set of messages at approximately constant rates, therefore a small value of $n$ is sufficient to spot the temporal patterns among the packet features, hence requiring a limited number of trainable parameters. On the other hand, more complex attacks, such as the ones performed with the HOIC tool, which uses multiple HTTP headers to make the requests appear legitimate, might require a larger number of packets to achieve the desired degree of accuracy. Given the variety of \ac{ddos} tools used to simulate the attack traffic in the dataset (IRC-based bot, LOIC and HOIC), we experimented with $n$ ranging between 1 and 100, and we compared the performance in terms of \textit{F1 score}. The results are provided in Figure~\ref{fig:sensitivity_n} for different durations of time window $t$, but at fixed values of the other hyper-parameters for the sake of visualisation. 

The \textit{F1 score }steadily increases with the value of $n$ when $n<5$, and then stabilises when $n\ge 5$. However, an increase in \textit{F1 score} is still observed up to $n=100$. 
Although, a low value of $n$  can be used to speed up the detection time (less convolutions) and to reduce the requirements in terms of storage and RAM (smaller sample size), which links to our objective of a lightweight implementation, we wish to balance high accuracy with low resource consumption. This will be demonstrated in Section \ref{sec:usecase}.

\begin{figure}[h!]
	\centering
	\captionsetup{justification=centering}
	\begin{tikzpicture}
	\begin{semilogxaxis}[  
	legend pos=south west,
	legend columns=2,
	height=6 cm,
	width=0.75\linewidth,
	grid = both,
	xlabel={Value of hyper-parameter $t$ (seconds) in logarithmic scale},
	ylabel={F1 Score},
	scaled y ticks=false,
	scaled x ticks=false,
	xmin=1,
	xmax=100,
	xtick={1,2,3,4,5,10,20,50,100},
	xticklabels={1,2,3,4,5,10,20,50,100},
	xtick pos=left,
	ymin=0.88, ymax=1,
	ytick={0.88,0.90,0.95,0.98,0.99,1},
	yticklabels={0.88,0.90,0.95,0.98,0.99,1},
	ytick pos=left,
	enlargelimits=0.02,
	]
	\addplot [color=red,style=thick,mark=*,mark size=2.5] table[x index=0,y index=1] {\FScoreT};
	\addplot [color=reddish-purple,style=thick,mark=square*,mark size=2.3] table[x index=0,y index=2] {\FScoreT};
	\addplot [color=blue,style=thick,mark=x,mark size=3.5] table[x index=0,y index=3] {\FScoreT};
	\addplot [color=orange,style=thick,mark=triangle*,mark size=3.3] table[x index=0,y index=4] {\FScoreT};
	\node[fill=white,anchor=west] at (axis cs: 4,.96) {\sffamily\scriptsize $\alpha=0.01$, $s=2048$, $k=64$, $h=3$, $m=n-h+1$};
	\legend{$n$=1,$n$=2,$n$=10,$n$=100}
	\end{semilogxaxis}
	\end{tikzpicture}
	\caption{Sensitivity of our model to hyper-parameter $t$.}
	\label{fig:sensitivity_t}
	\vspace{-3mm}
\end{figure}
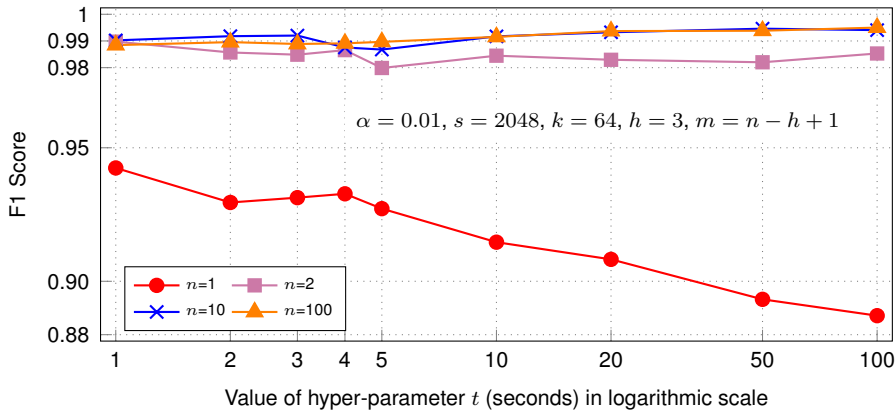

\textbf{Time Window.}
The time window $t$ is used to simulate the capturing process of online systems (see Section \ref{sec:pcap-processing}). We evaluated the \textit{F1 score} for time windows ranging between 1 and 100 seconds (as in the related work e.g., \cite{7946998}) at different values of $n$. The results are shown in Figure~\ref{fig:sensitivity_t}. 

Although the number of samples in the training set decreases when $t$ increases (see Table \ref{tab:datasets-splits}), the \ac{cnn} is relatively insensitive to this hyper-parameter for $n>1$. With $n=1$, the traffic flows are represented by samples of shape $[1,f]$, i.e. only one packet/sample, irrespective of the duration of the time window. In such a corner case, since the \ac{cnn} cannot correlate the attributes of different packets within the same sample, the \textit{F1 score} is more influenced by the number of samples in the training set (the more samples, the better).

\textbf{Number of convolutional filters.}
The higher the number of convolutional filters $k$, the more features are learned by the \ac{cnn}, but also the larger the number of trainable parameters in the model (hence longer training time). Common practice is to experiment by increasing the value of this hyper-parameter by powers of 2 (e.g., $k=1,2,4,8,16,32,64$).    

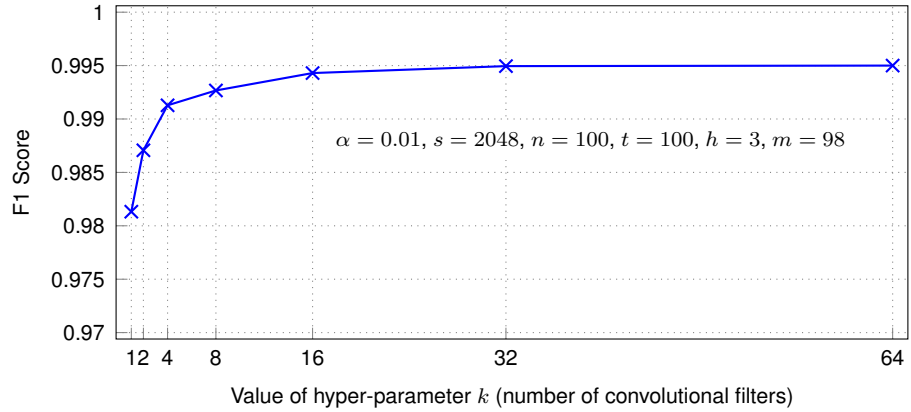
\begin{figure}[h!]
	\centering
	\captionsetup{justification=centering}
	\begin{tikzpicture}
	\begin{axis}[  
	height=6 cm,
	width=0.75\linewidth,
	grid = both,
	xlabel={Value of hyper-parameter $k$ (number of convolutional filters)},
	ylabel={F1 Score},
	scaled y ticks=false,
	scaled x ticks=false,
	xmin=1,
	xmax=64,
	xtick={1,2,4,8,16,32,64},
	xticklabels={1,2,4,8,16,32,64},
	xtick pos=left,
	ymin=0.97, ymax=1,
	ytick={0.97,0.975,0.98,0.985,0.99,0.995,1},
	yticklabels={0.97,0.975,0.98,0.985,0.99,0.995,1},
	ytick pos=left,
	enlargelimits=0.02,
	]
	\addplot [color=blue,style=thick,mark=x,mark size=3.5] table[x index=0,y index=1] {\FScoreK};
	\node[fill=white,anchor=west] at (axis cs: 17,.988) {\sffamily\scriptsize $\alpha=0.01$, $s=2048$, $n=100$, $t=100$, $h=3$, $m=98$};
	\end{axis}
	\end{tikzpicture}
	\caption{Sensitivity of our model to hyper-parameter $k$.}
	\label{fig:sensitivity_k}
\end{figure}

Figure~\ref{fig:sensitivity_k} shows the \textit{F1 score} as a function of $k$. For the sake of readability, the plot in the figure only reports the results obtained with one single set of the other hyper-parameters. However, similar trends have been observed with other combinations. It can be concluded that the performance improved with the number of filters up to a point (32/64 filters) where no performance gain is obtained, and increasing the number of filters will only increase the computational time.

\begin{figure}[h!]
	\centering
	\begin{tikzpicture}
	\begin{axis}[  
	height=6 cm,
	width=0.75\linewidth,
	grid = both,
	xlabel={Value of hyper-parameter $h$ (height of convolutional filters)},
	ylabel={F1 Score},
	scaled y ticks=false,
	scaled x ticks=false,
	xmin=1,
	xmax=5,
	xtick={1,2,3,4,5},
	xticklabels={1,2,3,4,5},
	xtick pos=left,
	ymin=0.98, ymax=1,
	ytick={0.98,0.985,0.99,0.995,1},
	yticklabels={0.98,0.985,0.99,0.995,1},
	ytick pos=left,
	enlargelimits=0.02,
	]
	\addplot [color=red,style=thick,mark=*,mark size=2.5] table[x index=0,y index=1] {\FScoreH};
	\node[fill=white,anchor=west] at (axis cs: 2,.992) {\sffamily\scriptsize $\alpha=0.01$, $s=2048$, $n=100$, $t=100$, $k=64$, $m=98$};
	\end{axis}
	\end{tikzpicture}
	\caption{Sensitivity of our model to hyper-parameter $h$.}
	\label{fig:sensitivity_h}
\end{figure}
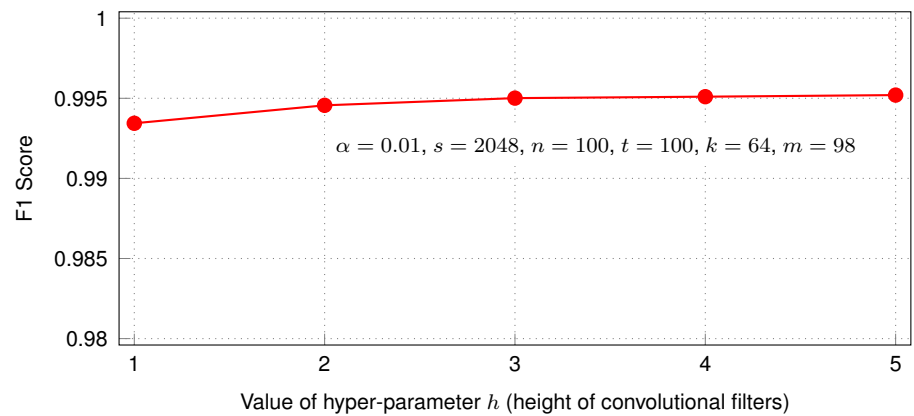

\revised{
\textbf{Height of convolutional filters.}
$h$ determines the height of the filters (the width is fixed to \features, the number of features), i.e. the number of packets to involve in each matrix operation.  
Testing with $h=1,2,3,4,5$, we observed a small, but noticeable, difference in the \textit{F1 score} between $h=1$ (0.9934) and $h=3$ (0.9950), with no major improvement beyond $h=3$ (Figure~\ref{fig:sensitivity_h}).
}

\textbf{Pooling Size.}
$m$ determines the pooling size of the max-pooling operation applied to the output of the convolution. By using $m=n-h+1$ we max pooled over the whole length of each activation map generated by each filter, known as global max pooling.  The global max-pooling reduces the shape of the next hidden layer to $[1,1,k]$, i.e. one single output unit per convolutional filter.

\textbf{Resulting hyper-parameter set.}
After conducting a comprehensive grid search on \revised{more than 5000} combinations of hyper-parameters, we have selected the \ac{cnn} model configuration that maximises the \textit{F1 score} on the UNB201X validation set (Table \ref{tab:training_results}). That is: 
$$\textbf{n}=100,\ \textbf{t}=100,\ \textbf{k}=64,\ \textbf{h}=3,\ \textbf{m}=98$$
The resulting model, trained with learning rate $\alpha=0.01$ and batch size $s=2048$, consists of 2241 trainable parameters, 2176 for the convolutional layer ($h\cdot f$ units for each filter plus bias, multiplied by the number of filters K) and 65 for the fully connected layer (64 units plus bias).  

As previously noted, other configurations may present lower resource requirements at the cost of a minimal decrease in \textit{F1 score}. For example, using $k=32$ would reduce the number of convolutions by half, while $n=10,20,50$ would also require fewer convolutions and a smaller memory footprint. However, setting $n=100$ not only maximises the \textit{F1 score}, but also enables a fair comparison with state-of-the-art approaches such as DeepDefense \cite{7946998} (Section \ref{sec:lucid-evaluation}), where the authors trained their neural networks using $n=100$ (in \cite{7946998}, the hyper-parameter is denoted as $T$). Furthermore, the chosen configuration enables a worst-case analysis for resource-constrained scenarios such as that presented in Section \ref{sec:usecase}.

These hyper-parameters are kept constant throughout our experiments presented in Sections \ref{sec:lucid-evaluation} and \ref{sec:usecase}.\\

\begin{table}[h!]
	\centering 
	\begin{threeparttable}
		\begin{tabular}{lccccc} \toprule[\heavyrulewidth]
			\textbf{Validation set}  & \textbf{ACC} & \revised{\textbf{FPR}} & \textbf{PPV} & \textbf{TPR} & \textbf{F1} \\ \midrule[\heavyrulewidth]
			UNB201X & 0.9950 & \revised{0.0083} & 0.9917 & 0.9983 & 0.9950  \\ \bottomrule[\heavyrulewidth]
		\end{tabular}
	\end{threeparttable}
	\caption{Scores obtained on the UNB201X validation set.} 
	\label{tab:training_results}
\end{table}

\subsection{Results}\label{sec:lucid-evaluation}

In this section, we present a detailed evaluation of the proposed approach with the datasets presented in Sec. \ref{sec:datasets}.
Evaluation metrics of  \textit{Accuracy (ACC)}, \textit{\acf{fpr}}, \textit{Precision} (\textit{\ac{ppv}}), \textit{Recall} (\textit{\ac{tpr}}) and \textit{F1 Score (F1)} have been used for performance measurement and for comparison with state-of-the-art models. 

\subsubsection{Detection Accuracy}\label{sec:evaluation-datasets}
In order to validate our approach and the results obtained on the validation dataset, we measure the performance of \ac{lucid} in classifying unseen traffic flows as benign or malicious (\ac{ddos}). Table \ref{tab:detection_performance} summarizes the results obtained on the various test sets produced through the procedure described in Section \ref{sec:data-preparation}. As illustrated, the very high performance is maintained across the range of test datasets indicating the robustness of the \ac{lucid} design. These results are further discussed in Section \ref{sec:evaluation-sota}, where we compare our solution with state-of-the-art works reported in the scientific literature.

\begin{table}[h!]
	\vspace{5mm}
	\centering 
	\begin{threeparttable}
		\begin{tabular}{lccccc} \toprule[\heavyrulewidth]
			\textbf{Test set}  & \textbf{ACC} & \revised{\textbf{FPR}} & \textbf{PPV} & \textbf{TPR} & \textbf{F1} \\ \midrule[\heavyrulewidth]
			ISCX2012 & 0.9888 & \revised{0.0179} & 0.9827 & 0.9952 & 0.9889 \\ \midrule
			CIC2017  & 0.9967 & \revised{0.0059} & 0.9939 & 0.9994 & 0.9966 \\ \midrule
			CSECIC2018 & 0.9987 & \revised{0.0016} & 0.9984 & 0.9989 & 0.9987 \\ \midrule
			UNB201X & 0.9946 & \revised{0.0087} & 0.9914 & 0.9979 & 0.9946  \\ \bottomrule[\heavyrulewidth]
		\end{tabular}
	\end{threeparttable}
	\caption{\ac{lucid} detection performance on the test sets.} 
	\label{tab:detection_performance}
\end{table}

The results show that thanks to the properties of its \ac{cnn}, \ac{lucid} learns to distinguish between patterns of malicious \ac{ddos} behaviour and benign flows.  Given the properties of convolutional methods, these patterns are recognised regardless of the position they occupy in a flow, demonstrating that our spatial representation of a flow is robust.  Irrespective of whether the \ac{ddos} event appears at the start or the end of the input, \ac{lucid} will produce the same representation in its output.  Although the temporal dynamics in \ac{ddos} attacks might suggest that alternative \ac{dl} architectures may seem more suitable (e.g., \ac{lstm}), our novel preprocessing method combined with the \ac{cnn} removes the requirement for the model to maintain temporal context of each whole flow as the data is pushed through the network.   \revised{In comparison, \acp{lstm} are known to be very difficult to train, and their performance is inherently slower for long sequences compared to \acp{cnn}.}

\subsubsection{State-Of-The-Art Comparison}\label{sec:evaluation-sota}
For a fair comparison between \ac{lucid} and the state-of-the-art, we focus our analysis on solutions that have validated the \ac{unb} datasets for \ac{ddos} attack detection. 

We have paid particular attention to DeepDefense \cite{7946998} as, similar to our approach, the model is trained with packet attributes rather than flow-level statistics used in other works. DeepDefense  translates the \textit{pcap} files of ISCX2012 into arrays that contain packet attributes collected within sliding time windows. The label assigned to a sample is the label of the last packet in the time window, according to the labels provided with the original dataset. The proposed data preprocessing technique is similar to \ac{lucid}'s. However, in \ac{lucid}, a sample corresponds to a single traffic flow, whereas in DeepDefense a sample represents the traffic collected in a time window. 

Of the four \ac{dl} models presented in the DeepDefense paper, the one called 3LSTM produces the highest scores in the classification of \ac{ddos} traffic. Therefore, we have implemented 3LSTM for comparison purposes. The architecture of this model includes 6 LSTM layers of 64 neurons each, 2 fully connected layers of 128 neurons each, and 4 batch normalisation layers. To directly compare the \ac{dl} models, we have trained 3LSTM on the UNB201X training set with $n=100$ and $t=100$ as done with \ac{lucid}.  We have compared our implementation of 3LSTM with \ac{lucid} on each of the four test sets, and present the F1 score results in Table \ref{tab:deepdefense_f1_comparison}. 

\begin{table}[h!]
	\vspace{5mm}
	\centering 
	\begin{threeparttable}
			\begin{tabular}{lccccc} \toprule[\heavyrulewidth]
				\textbf{Model} & \begin{tabular}{@{}c@{}}\textbf{Trainable}\\\textbf{Parameters}\end{tabular} & \begin{tabular}{@{}c@{}}\textbf{ISCX}\\\textbf{2012}\end{tabular} & \begin{tabular}{@{}c@{}}\textbf{CIC}\\\textbf{2017}\end{tabular} & \begin{tabular}{@{}c@{}}\textbf{CSECIC}\\\textbf{2018}\end{tabular} & \begin{tabular}{@{}c@{}}\textbf{UNB}\\\textbf{201X}\end{tabular} \\ \midrule[\heavyrulewidth]
				\textbf{\ac{lucid}} & 2241 & 0.9889 & 0.9966 & 0.9987 & 0.9946  \\ \midrule
				3LSTM & 1004889 & 0.9880 & 0.9968 & 0.9987 & 0.9943 \\\bottomrule[\heavyrulewidth]
			\end{tabular}
	\end{threeparttable}
	\caption{\ac{lucid}-DeepDefense comparison (F1 score).} 
	\label{tab:deepdefense_f1_comparison}
\end{table}

The results presented in Table \ref{tab:deepdefense_f1_comparison} show that \ac{lucid} and 3LSTM are comparable in terms of F1 score across the range of test datasets. However, in terms of computation time, \ac{lucid} 
outperforms 3LSTM in detection time. Specifically, as measured on the Intel Xeon server in these experiments, \ac{lucid} can classify more than 55000 samples/sec on average, while 3LSTM barely reaches 1300 samples/sec on average (i.e., more than 40 times slower). Indeed, \ac{lucid}'s limited number of hidden units and trainable parameters contribute to a much lower computational complexity compared to 3LSTM.

As previously noted, there are a number of solutions in the literature that present performance results for the ISCX2012 and CIC2017 datasets. Notably, these works do not all specify whether the results presented are based on a validation dataset or a test dataset. For \ac{lucid}, we reiterate that the results presented in this section are based on a test set of completely unseen data. 

\begin{table}[h!]
	\vspace{5mm}
	\centering 
	\captionsetup{width=0.65\textwidth}
	\begin{threeparttable}
		\begin{tabular}{lccccc} \toprule[\heavyrulewidth]
			\textbf{Model}  & \textbf{ACC} & \revised{\textbf{FPR}} & \textbf{PPV} & \textbf{TPR} & \textbf{F1} \\ \midrule[\heavyrulewidth]
			\textbf{\ac{lucid}} & 0.9888 & \revised{0.0179} & 0.9827 & 0.9952 & 0.9889  \\ \midrule
			\begin{tabular}{@{}l@{}}DeepDefense \\3LSTM \cite{7946998} \end{tabular}  & 0.9841 & \revised{N/A} & 0.9834 & 0.9847 & 0.9840 \\ \midrule
			TR-IDS \cite{tr-ids}  & 0.9809 & \revised{0.0040} & N/A & 0.9593 & N/A \\ \midrule
			E3ML \cite{8343104}  & N/A & \revised{N/A} & N/A & 0.9474 & N/A \\ \bottomrule[\heavyrulewidth]
		\end{tabular}
	\end{threeparttable}
	\caption{Performance comparison with State-Of-The-Art approaches using the ISCX2012 dataset for \ac{ddos} detection.} 
	\label{tab:ISCXIDS2012_comparison}
	\vspace{2mm}
\end{table}

In Table \ref{tab:ISCXIDS2012_comparison}, we compare the performance of \ac{lucid} against state-of-the-art works validated on ISCX2012. Table \ref{tab:ISCXIDS2012_comparison} also includes the performance of 3LSTM as reported in the DeepDefense paper \cite{7946998}. With respect to our version of 3LSTM, the scores are slightly lower, which we propose is due to the different \textit{pcap} preprocessing mechanisms used in the two implementations. This indicates a performance benefit when using the \ac{lucid} preprocessing mechanism.

TR-IDS \cite{tr-ids} is an \ac{ids} which adopts a text-\ac{cnn} \cite{KimConvolutionalNN} to extract features from the payload of the network traffic. These features, along with a combination of 25 packet and flow-level attributes, are used for traffic classification by means of a Random Forest algorithm. Accuracy and \ac{tpr} scores of TR-IDS are above 0.99 for all the attack profiles available in ISCX2012 except the \ac{ddos} attack, for which the performance results are noticeably lower than \ac{lucid}. 

E3ML \cite{8343104} uses 20 entropy-based traffic features and three \ac{ml} classifiers (a \ac{rnn}, a Multilayer Perceptron and an Alternating Decision Tree) to classify the traffic as normal or \ac{ddos}. Despite the complex architecture, the \ac{tpr} measured on ISCX2012 shows that E3ML is inclined to false negatives.

For the CIC2017 dataset, we present the performance comparison with state-of-the-art solutions in Table \ref{tab:CICIDS2017_comparison}.
 
\begin{table}[h!]
	\vspace{5mm}
	\centering 
	\captionsetup{width=0.65\textwidth}
	\begin{threeparttable}
			\begin{tabular}{lccccc} \toprule[\heavyrulewidth]
				\textbf{Model}  & \textbf{ACC} & \revised{\textbf{FPR}} & \textbf{PPV} & \textbf{TPR} & \textbf{F1} \\ \midrule[\heavyrulewidth]
				\textbf{\ac{lucid}} & 0.9967 & \revised{0.0059} & 0.9939 & 0.9994 & 0.9966  \\ \midrule
				DeepGFL \cite{8599821}  & N/A & \revised{N/A} & 0.7567 & 0.3024 & 0.4321 \\ \midrule
				MLP \cite{8666588}  & 0.8634 & \revised{N/A} & 0.8847 & 0.8625 & 0.8735 \\ \midrule
				1D-CNN \cite{8666588}  & 0.9514 & \revised{N/A} & 0.9814 & 0.9017 & 0.9399 \\ \midrule
				LSTM \cite{8666588}  & 0.9624 & \revised{N/A} & 0.9844 & 0.8989 & 0.8959 \\ \midrule
				\begin{tabular}{@{}l@{}}1D-CNN +\\LSTM \cite{8666588}\end{tabular} & 0.9716 & \revised{N/A} & 0.9741 & 0.9910 & 0.9825  \\ \bottomrule[\heavyrulewidth]
			\end{tabular}
	\end{threeparttable}
	\caption{Performance comparison with State-Of-The-Art approaches using the CIC2017 dataset for \ac{ddos} detection.} 
	\label{tab:CICIDS2017_comparison}
\end{table}

DeepGFL \cite{8599821} is a framework designed to extract high-order traffic features from low-order features forming a hierarchical graph representation. To validate the proposed framework, the authors used the graph representation of the features to train two traffic classifiers, namely Decision Tree and Random Forest, and tested them on CIC2017. Although the \ac{ppv} scores on the several attack types are reasonably good (between 0.88 and 1 on any type of traffic profile except \ac{ddos}), the results presented in the paper reveal that the proposed approach is prone to false negatives, leading to very low F1 scores.

The authors of \cite{8666588} propose four different \ac{dl} models for \ac{ddos} attack detection in \ac{iot} networks. The models are built with combinations of LSTM, \ac{cnn} and fully connected layers. The input layer of all the models consists of 82 units, one for each flow-level feature available in CIC2017, while the output layer returns the probability of a given flow being part of a \ac{ddos} attack. The model 1D-CNN+LSTM produces good classification scores, while the others seem to suffer from high false negatives rates. 

To the best of our knowledge, no \ac{ddos} attack detection solutions validated on the CSECIC2018 dataset are available yet in the scientific literature.

\subsubsection{Discussion}
From the results presented and analysed in the previous sections, we can conclude that using packet-level attributes of network traffic is more effective, and results in higher classification accuracy, than using flow-level features or statistic information such as the entropy measure. This is not only proved by the evaluation results obtained with \ac{lucid} and our implementation of DeepDefense (both based on packet-level attributes), but also by the high classification accuracy of TR-IDS, which combines flow-level features with packet attributes, including part of the payload. 

In contrast, E3ML, DeepGFL and most of the solutions proposed in \cite{8666588}, which all rely on flow-level features, seem to be more prone to false negatives, and hence to classify \ac{ddos} attacks as normal activity. The only exception is the model 1D-CNN+LSTM of \cite{8666588}, which produces a high \ac{tpr} by combining \ac{cnn} and \ac{rnn} layers. 

Furthermore, we highlight that \ac{lucid} has not been tuned to the individual datasets but rather to the validation portion of a combined dataset, and still outperforms the state-of-the-art on totally unseen test data.


\subsection{Analysis}\label{sec:analysis}
We now present interpretation and explanation of the internal operations of \ac{lucid} by way of proving that the model is learning the correct domain information. We do this by analysing the features used in the dataset and their activations in the model. 

This approach is inspired by a similar study \cite{jacovi-etal-2018-understanding} to interpret \acp{cnn} in the rather different domain of natural language processing.  However, the kernel activation analysis technique is transferable to our work.  As each kernel has the same width as the input matrix, it is possible to remove the classifier, push the \ac{ddos} flows through the convolutional layer and capture the resulting activations per kernel.  For each flow, we calculate the total activations per feature, which in the spatial input representation means per column, resulting in 11 values that map to the 11 features.  This is then repeated for all kernels, across all \ac{ddos} flows, with the final output being the total column-wise activation of each feature.  \revised{The intuition is that the higher a feature's activation when a positive sample i.e. a \ac{ddos} flow is seen, the more importance the \ac{cnn} attaches to that particular feature.  Conversely, the lower the activation, the lower the importance of the feature, and since our model uses the conventional rectified linear activation function, $ReLU(x) = max\{0,x\}$, this means that any negative activations become zero and hence have no impact on the Sigmoid classifier for detecting a DDoS attack.} 

Summing these activations over all kernels is possible since they are of the same size and operate over the same spatial representations.  \revised{We analyse \ac{ddos} flows from the same UNB201X test set used in Sec. V-A.}  
Table \ref{tab:col-activ} presents the ranking of the 11 features based on \revised{the post-$ReLU$ average column-wise feature activation sums}, and highlights two features that activate our \ac{cnn} the most, across all of its kernels.  

\begin{table}[h!]
	\vspace{5mm}
	\centering 
	\captionsetup{width=0.73\textwidth}
	\renewcommand{\arraystretch}{1.1}
	\begin{tabular}{lc|lc} \toprule[\heavyrulewidth]
		\textbf{Feature} & \begin{tabular}{@{}c@{}}\textbf{Total Kernel}\\\textbf{Activation}\end{tabular} & \textbf{Feature} & \begin{tabular}{@{}c@{}}\textbf{Total Kernel}\\\textbf{Activation}\end{tabular} \\ 
		\midrule[\heavyrulewidth]
		Highest Layer  & \revised{0.69540} & Time & \revised{0.11108} \\\hline
		IP Flags  & \revised{0.30337} & TCP Win Size & \revised{0.09596} \\\hline
		TCP Flags  & \revised{0.19693} & TCP Ack  & \revised{0.00061} \\\hline
		TCP Len & \revised{0.16874} & UDP Len  & \revised{0.00000} \\\hline
		Protocols  & \revised{0.14897} & ICMP Type  & \revised{0.00000} \\ \hline
		Pkt Len & \revised{0.14392} & & \\
		
		\bottomrule[\heavyrulewidth]
		
	\end{tabular}
	\caption{\revised{Ranking of the total column-wise feature kernel activations for the UNB201X dataset}} 
	\label{tab:col-activ}
\end{table}

\textbf{Highest Layer.}
\revised{We assert that the \ac{cnn} may be learning from the highest layer at which each \ac{ddos} flow operates.  Recall that highest layer links to the type of \ac{ddos} attack e.g. network, transport, or application layer attack. We propose that this information could be used to extend \ac{lucid} to predict the specific type of \ac{ddos} attack taking place, and therefore, to contribute to selection of the appropriate protection mechanism. We would achieve the prediction by extending the dataset labeling, which we consider for future work.} 

\textbf{\revised{IP Flags.}}
In our design, this attribute is a 16-bit integer value which includes three bits representing the flags \textit{Reserved Bit}, \textit{Don't Fragment} and \textit{More Fragments}, plus 13 bits for the \textit{Fragment offset} value, which is non-zero only if bit \textit{``Don't Fragment''} is unset. Unlike the \textit{IP fragmented flood} \ac{ddos} attacks, in which the \textit{IP flags} are manipulated to exploit the datagram fragmentation mechanisms, $99.99\%$ of \ac{ddos} packets in the \ac{unb} datasets present an \textit{IP flags} value of 0x4000, with only the \textit{``Don't Fragment''} bit set to $1$. A different distribution of \textit{IP flags} is observed in the \ac{unb} benign traffic, with the \textit{``Don't Fragment''} bit set to $1$ in about $92\%$ of the packets. Thus, the pattern of \textit{IP flags} is slightly different between attack and benign traffic, and we are confident that \ac{lucid} is indeed learning their significance in \ac{ddos} classification, as evidenced by its 2nd place in our ranking.	\\

Even given this activation analysis, there is no definitive list of features that exist for detecting \ac{ddos} attacks with which we can directly compare our results. Analysing the related work, we identify a wide range of both stateless and stateful features highlighted for their influence in a given detection model, which is not unexpected as the features of use vary depending on the attack traffic. This is highlighted by the 2014 study \cite{FIMU}, which concludes that different classes of attack have different properties, leading to the wide variance in features identified as salient for the attack detection. The authors also observe that the learning of patterns specific to the attack scenario would be more valuable than an effort to produce an attack-agnostic finite list of features. We, therefore, conclude from our analysis that \ac{lucid} appears to be learning the importance of relevant features for \ac{ddos} detection, which gives us confidence in the prediction performance.

\subsection{Use-case: DDoS Detection at the Edge}\label{sec:usecase}
Edge computing is an emerging paradigm adopted in a variety of contexts (e.g., fog computing \cite{Bonomi2014}, edge clouds \cite{6849256}), with the aim of improving the performance of applications with low-latency and high-bandwidth requirements. Edge computing complements centralised data centres with a large number of distributed nodes that provide computation services close to the sources of the data.

The proliferation of attacks leveraging unsecured \ac{iot} devices (e.g., the Mirai botnet \cite{mirai} and its variants) demonstrate the potential value in edge-based \ac{ddos} attack detection. Indeed, with edge nodes close to the \ac{iot} infrastructure, they can detect and block the \ac{ddos} traffic as soon as it leaves the compromised devices. However, in contrast to cloud high-performance servers, edge nodes cannot exploit sophisticated solutions against \ac{ddos} attacks, due to their limited computing and memory resources. Although recent research efforts have demonstrated that the mitigation of \ac{ddos} attacks is feasible even by means of commodity computers \cite{smartnic-ddos,hoiland2018express}, edge computing-based \ac{ddos} detection is still at an early stage.

\begin{figure}[!h]
	\begin{center}
		\includegraphics[width=0.75\textwidth]{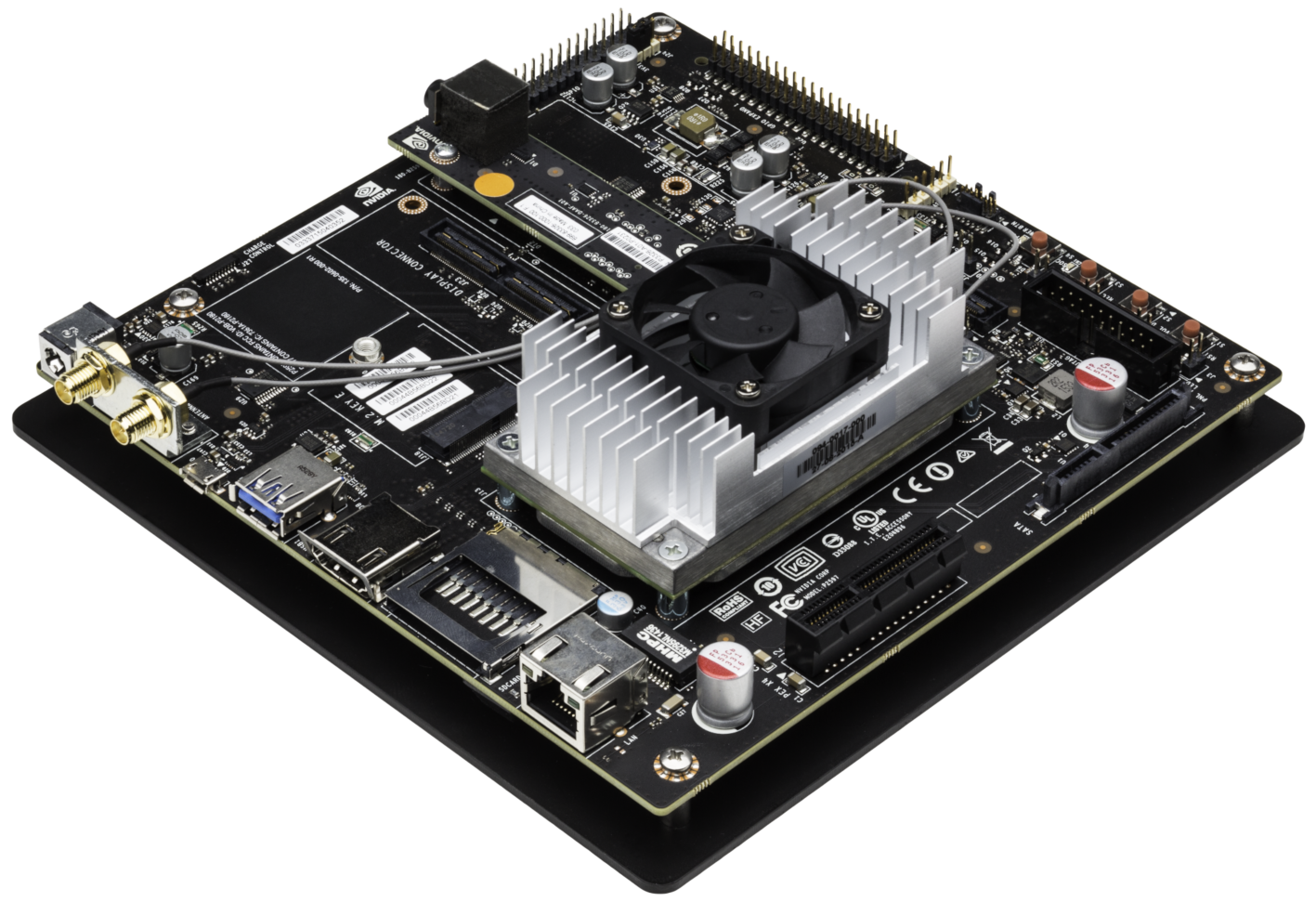}
		\caption{The NVIDIA Jetson TX2 development board.}
		\label{fig:TX2}
	\end{center}
\end{figure}

In this section, we demonstrate that our \revised{\ac{ddos} detection solution} can be deployed and effectively executed on resource-constrained devices, such as edge nodes or \ac{iot} gateways, by running \ac{lucid} on an NVIDIA Jetson TX2 development board \cite{jetson-tx2} (Figure \ref{fig:TX2}), equipped with a quad-core ARM Cortex-A57@2\,GHz CPU, 8\,GB of RAM and a 256-core Pascal@1300\,MHz \ac{gpu}. For the experiments, we used Tensorflow 1.9.0 with GPU support enabled by cuDNN, a GPU-accelerated library for deep neural networks \cite{cuDNN}.
	
\subsubsection{Detection}
In the first experiment, we analyse the applicability of our approach to online edge computing environments by estimating the prediction performance in terms of samples processed per second. As we are aware that edge nodes do not necessarily mount a \ac{gpu} device, we conduct the experiments with and without the \ac{gpu} support on the UNB201X test set and discuss the results. 

\revised{We note that in an online system, our preprocessing tool presented in Section \ref{sec:pcap-processing} can be integrated into the server/edge device. The tool would process the live traffic collected from the NICs of the server/edge device, collecting the packet attributes, organising them into flows and, after a predefined time interval, $T$, pass the data structure to the CNN for inference. We acknowledge that the speed of this process will influence the overall system performance. However, as we have not focused on optimising our preprocessing tool, rather on optimising detection, its evaluation is left as future work. Instead, in these experiments, we load the \ac{unb} datasets from the hard disk rather than processing live traffic.}

With respect to this, one relevant parameter is the batch size, which configures how many samples are processed by the \ac{cnn} in parallel at each iteration. Such a parameter influences the speed of the detection, as it determines the number of iterations  and, as a consequence, the number of memory reads required by the \ac{cnn} to process all the samples in the test set (or the samples collected in a time window, in the case of online detection).

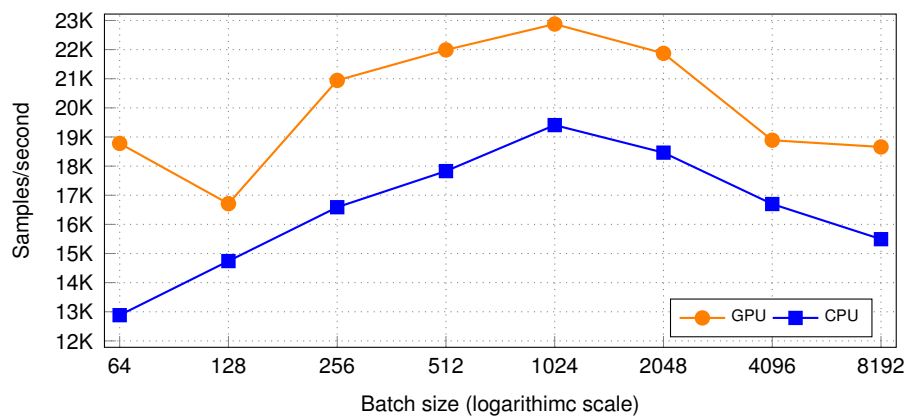
\begin{figure}[h!]
	\centering
	\captionsetup{justification=centering}
	\begin{tikzpicture}
	\begin{semilogxaxis}[  
	legend pos=south east,
	legend columns=2,
	height=6 cm,
	width=0.75\linewidth,
	grid = both,
	xlabel={Batch size (logarithimc scale)},
	ylabel={Samples/second},
	scaled y ticks=false,
	scaled x ticks=false,
	xmin=64,
	xmax=8192,
	xtick={64,128,256,512,1024,2048,4096,8192},
	xticklabels={64,128,256,512,1024,2048,4096,8192},
	xtick pos=left,
	ymin=12000, ymax=23000,
	ytick={12000,13000,14000,15000,16000,17000,18000,19000,20000,21000,22000,23000},
	yticklabels={12K,13K,14K,15K,16K,17K,18K,19K,20K,21K,22K,23K},
	ytick pos=left,
	enlargelimits=0.02,
	]
	\addplot [color=orange,style=thick,mark=*,mark size=2.5] table[x index=0,y index=2] {\PredictionTX};
	\addplot [color=blue,style=thick,mark=square*,mark size=2.3] table[x index=0,y index=4] {\PredictionTX};
	\legend{GPU,CPU}
	\end{semilogxaxis}
	\end{tikzpicture}
	\caption{Inference performance on the NVIDIA Jetson TX2 board.}
	\label{fig:prediction_time}
\end{figure}

Figure~\ref{fig:prediction_time} shows the performance of \ac{lucid} on the development board in terms of processed samples/second. As the shape of each sample is $[n,f]=[100,\features]$, i.e. each sample can contain the features of up to 100 packets, we can estimate that the maximum number of packets per second (pps) that the device can process without the \ac{gpu} and using a batch size of 1024 samples is approximately 1.9\,Mpps. As an example, the content of the UNB201X test set is 602,547 packets distributed over 22,416 samples, which represents a processing requirement of 500\,Kpps without the \ac{gpu}, and 600\,Kpps when the \ac{gpu} is enabled. This illustrates the ability to deploy \ac{lucid} on a resource-constrained platform.

The second measurement regarding resource-constrained systems is the memory requirement to store all the samples collected over a time window. The memory occupancy per sample is 8,800 bytes, i.e. $100\cdot\features=1100$ floating point values of 8 bytes each. As per Figure~ \ref{fig:prediction_time}, the \ac{cnn} can process around 23K samples/second with the help of the \ac{gpu} and using a batch size of 1024. To cope with such a processing speed, the device would require approximately 20\,GB RAM for a $t=100$ time window. However, this value greatly exceeds the typical amount of memory available on edge nodes, in general (e.g., 1\,GB on Raspberry Pi 3 \cite{raspberry}, 2\,GB on the ODROID-XU board \cite{8039523}), and on our device, in particular. Indeed, the memory resources of nodes can represent the real bottleneck in an edge computing scenario.

Therefore, assuming that our edge node is equipped with 1\,GB RAM, the maximum number of samples that can be stored in RAM is approximately 100K (without taking into account RAM used by the operating system and applications). We have calculated that this memory size would be sufficient for an attack such as the HTTP-based \ac{ddos} attack in the CSECIC2018 dataset, for which we measured approximately 30K samples on average over a 100\,s time window. For more aggressive attacks, however, a strategy to overcome the memory limitation would be to configure the \ac{cnn} model with lower values of $t$ and $n$. For instance, setting the value of both parameters to 10 can reduce the memory requirement by a factor of 100, with a low cost in detection accuracy (F1 score 0.9928 on the UNB201X test set, compared to the highest score obtained with $t=n=100$, i.e. 0.9946). 

The measurements based on our test datasets demonstrate that \ac{lucid} is usable on a resource-constrained platform both with respect to processing and memory requirements. \revised{These results are promising for effective deployment of \ac{lucid} in a variety of edge computing scenarios, including those where the nodes execute latency-sensitive services. A major challenge in this regard is balancing between resource usage of \ac{lucid} (including traffic collection and preprocessing) and detection accuracy, i.e. ensuring the required level of protection against \ac{ddos} attacks without causing delays to the services. A deep study of this trade-off is out of scope of this thesis and is reserved for future work.
}

\subsubsection{Training Time}
In a real-world scenario, the \ac{cnn} model will require re-training with new samples of benign and malicious traffic to update all the weights and biases. In edge computing environments, the traditional approach is to send large amounts of data from edge nodes to remote facilities such as private or commercial data centres. However, this can result in high end-to-end latency and bandwidth usage. In addition, it may raise security concerns, as it requires trust in a third party entity (in the case of commercial cloud services) regarding the preservation of data confidentiality and integrity. 

A solution to this issue is to execute the re-training task locally on the edge nodes. In this case, the main challenge is to control the total training time, as this time determines how long the node remains exposed to new \ac{ddos} attacks before the detection model can leverage the updated parameters. 

To demonstrate the suitability of our model for this situation, we have measured the convergence training time of \ac{lucid} on the development board using the UNB201X training and validation sets with and without the \ac{gpu} support. We have experimented by following the learning procedure described in Section \ref{sec:learning}, thus with a training termination criterion based on the loss value measured on the validation set. The results are presented in Table \ref{tab:training_tx} along with the performance obtained on the server used for the study in Section \ref{sec:tuning}. 

\begin{table}[h!]
	\vspace{5mm}
	\centering 
	\begin{threeparttable}
		\begin{tabular}{lcc} \toprule[\heavyrulewidth]
			\textbf{Setup}  & \begin{tabular}{@{}c@{}}\textbf{Time/epoch} \\\textbf{(sec)} \end{tabular} & \begin{tabular}{@{}c@{}}\textbf{Convergence} \\\textbf{time (sec)} \end{tabular} \\ \midrule[\heavyrulewidth]
			\textbf{\ac{lucid}}\ Server  & 10.2 & 1880 \\ \midrule
			\textbf{\ac{lucid}}\ Dev. board (\ac{gpu}) & 25.8 & 4500  \\ \midrule
			\textbf{\ac{lucid}}\ Dev. board (CPU) & 40.5 & 7450   \\ \midrule
			3LSTM Dev. board (\ac{gpu}) & 1070 & $>$90000   \\ \bottomrule[\heavyrulewidth]
		\end{tabular}
	\end{threeparttable}
	\caption{Training convergence time.} 
	\label{tab:training_tx}
\end{table}

As shown in Table \ref{tab:training_tx}, the \ac{cnn} training time on the development board without using the \ac{gpu} is around 2 hours (184 epochs). This is approximately 4 times slower than training on the server, but clearly outperforms the training time of our implementation of DeepDefense 3LSTM, which we measured at more than 1000 sec/epoch \textit{with} the \ac{gpu} (i.e., 40 times slower than \ac{lucid} under the same testing conditions).  

In application scenarios where a faster convergence is required, the time can be further reduced by either terminating the training process early after a pre-defined number of epochs, or limiting the size of the training/validation sets. As adopting one or both of such strategies can result in a lower detection accuracy, the challenge in such scenarios is finding the trade-off between convergence time and detection accuracy that meets the application requirements.


\subsection{Related Work}\label{sec:lucid-related}

\ac{ddos} detection and mitigation techniques have been explored by the network research community since the first reported \ac{ddos} attack incident in 1999 \cite{criscuolo2000distributed}. In this section, we review and discuss anomaly-based \ac{ddos} detection techniques categorised by statistical approaches and machine learning approaches, with a specific focus on deep learning techniques.

\subsubsection{Statistical Approaches to DDoS Detection}\label{sec:related-stats}
Measuring statistical properties of network traffic attributes is a common approach to \ac{ddos} detection, and generally involves monitoring the entropy variations of specific packet header fields. By definition, the entropy is a measure of the diversity or the randomness in a data set. Entropy-based \ac{ddos} detection approaches have been proposed in the scientific literature since the early 2000s, based on the assumption that during a volumetric \ac{ddos} attack, the randomness of traffic features is subject to sudden variations. The rationale is that volumetric \ac{ddos} attacks are typically characterised by a huge number of attackers (in the order of hundreds of thousands \cite{8367750}), often utilising compromised devices that send a high volume of traffic to one or more end hosts (the victims). As a result, these attacks usually cause a drop in the distribution of some of the traffic attributes, such as the destination IP address, or an increase in the distribution of other attributes, such as the source IP address. The identification of a \ac{ddos} attack is usually determined by means of thresholds on these distribution indicators.

In one of the first published works using this approach, Feinstein et al. \cite{1194894} proposed a \ac{ddos} detection technique based on the computation of source IP address entropy and Chi-square distribution. The authors observed that the variation in source IP address entropy and chi-square statistics due to fluctuations in legitimate traffic was small, compared to the deviations caused by \ac{ddos} attacks. Similarly, \cite{BOJOVIC201984} combined entropy and volume traffic characteristics to detect volumetric \ac{ddos} attacks, while the authors of \cite{8466805} proposed an entropy-based scoring system based on the destination IP address entropy and dynamic combinations of IP and TCP layer attributes to detect and mitigate DDoS attacks.

A common drawback to these entropy-based techniques is the requirement to select an appropriate detection threshold. Given the variation in traffic type and volume across different networks, it is a challenge to identify the appropriate detection threshold that minimizes false positive and false negative rates in different attack scenarios. One solution is to dynamically adjust the thresholds to auto-adapt to the normal fluctuations of the network traffic, as proposed in \cite{10.1007/978-981-13-2622-6_31,8423699}. 

Importantly, monitoring the distribution of traffic attributes does not provide sufficient information to distinguish between benign and malicious traffic. To address this, some approaches apply a rudimentary threshold on the packet rate \cite{Jun:2014:DAD:2554850.2555109} or traceback techniques \cite{5467062,7345297}. 

An alternative statistical approach is adopted in \cite{8522048}, where Ahmed et al. use packet attributes and traffic flow-level statistics to distinguish between benign and \ac{ddos} traffic. However, this solution may not be suitable for online systems, since some of the flow-level statistics used for the detection e.g. total bytes, number of packets from source to destination and from destination to source, and flow duration, cannot be computed when the traffic features are collected within observation time windows. Approaches based on flow-level statistics have also been proposed in \revised{ \cite{7997375,8599821,tr-ids,10.1007/978-3-030-00018-9_15,8666588,homayoun}, among many others. In particular, \cite{tr-ids,10.1007/978-3-030-00018-9_15,8666588,homayoun} use flow-level statistics} to feed \acp{cnn} and other \ac{dl} models, as discussed in Sec. \ref{sec:related-dl}. To overcome the limitations of statistical approaches to \ac{ddos} detection, machine learning techniques have been explored.

\subsubsection{Machine Learning for DDoS Detection}\label{sec:related-ml}
As identified by Sommer and Paxson in \cite{sommer2010outside}, there has been extensive research on the application of machine learning to network anomaly detection. The 2016 Buczak and Guven survey \cite{buczak2016survey} cites the use of \ac{svm}, k-Nearest Neighbour (k-NN), Random Forest, Na\"ive Bayes etc. achieving success for cyber security intrusion detection. However, due to the challenges particular to network intrusion detection, such as high cost of errors, variability in traffic etc., adoption of these solutions in the ``real-world'' has been limited. Over recent years, there has been a gradual increase in availability of realistic network traffic data sets and an increased engagement between data scientists and network researchers to improve model explainability such that more practical \ac{ml} solutions for network attack detection can be developed.
Some of the first application of machine learning techniques specific to \ac{ddos} detection has been for traffic classification. Specifically, to distinguish between benign and malicious traffic, techniques such as extra-trees and multi-layer perceptrons have been applied \cite{Idhammad2018,Singh2016EntropyBasedAL}. 

In consideration of the realistic operation of \ac{ddos} attacks from virtual machines, He et al. \cite{he2017machine} evaluate nine \ac{ml} algorithms to identify their capability to detect the \ac{ddos} from the source side in the cloud. The results are promising with high accuracy (99.7\%) and low false positives ($<$ 0.07\%) for the best performing algorithm; \ac{svm} linear kernel. Although there is no information provided regarding the detection time or the datasets used for the evaluation, the results illustrate the variability in accuracy and performance across the range of \ac{ml} models. This is reflected across the literature (e.g., \cite{hoon2018critical,primartha2017anomaly}) with the algorithm performance highly dependent on the selected features (and datasets) evaluated. This has motivated the consideration of deep learning for \ac{ddos} detection, which reduces the emphasis on feature engineering.


\subsubsection{Deep Learning for DDoS Detection}\label{sec:related-dl}
There is a small body of work investigating the application of \ac{dl} to \ac{ddos} detection. For example, in \cite{8343104}, the authors address the problem of threshold setting in entropy-based techniques by combining entropy features with \ac{dl}-based classifiers. The evaluation demonstrates improved performance over the threshold-based approach with higher precision and recall. In \cite{yin2017deep}, a \ac{rnn}-\ac{ids} is compared with a series of previously presented \ac{ml} techniques (e.g., J48, \ac{ann}, Random Forest, and \ac{svm}) applied to the NSL-KDD~\cite{5356528} dataset. The \ac{rnn} technique demonstrates a higher accuracy and detection rate. 

\revised{Some \ac{cnn}-based works \cite{tr-ids, 10.1007/978-3-030-00018-9_15,8666588,homayoun}}, as identified in Sec. \ref{sec:related-stats}, use flow-level statistics (total bytes, flow duration, total number of flags, etc.) as input to the proposed \ac{dl}-based architectures. In addition, \cite{tr-ids} and \cite{10.1007/978-3-030-00018-9_15} combine the statistical features with packet payloads to train the proposed \acp{ids}.

In \cite{wu2018novel}, Kehe Wu et al. present an \ac{ids} based on \ac{cnn} for multi-class traffic classification. The proposed neural network model has been validated with flow-level features from the NSL-KDD dataset encoded into 11x11 arrays. Evaluation results show that the proposed model performs well compared to complex models with 20 times more trainable parameters. A similar approach is taken by the authors of \cite{potluri2018convolutional}, where the \ac{cnn}-based \ac{ids} is validated over datasets NSL-KDD and UNSW-NB-15~\cite{7348942}. In \cite{kwon2018empirical}, the authors study the application of \acp{cnn} to \ac{ids} by comparing a series of architectures (shallow, moderate, and deep, to reflect the number of convolution and pooling layers) across 3 traffic datasets; NSL-KDD, Kyoto Honeypot \cite{song2006description}, and MAWILab \cite{callegari2016statistical}. In the results presented, the shallow \ac{cnn} model with a single convolution layer and single max. pooling layer performed best. However, there is significant variance in the detection accuracy results across the datasets, which indicates instability in the model.

More specific to our \ac{ddos} problem, Ghanbari et al. propose a feature extraction algorithm based on the \textit{discrete wavelet transform} and on the \textit{variance fractal dimension trajectory} to maximize the sensitivity of the \ac{cnn} in detecting \ac{ddos} attacks \cite{8482019}. The evaluation results show that the proposed approach recognises \ac{ddos} attacks with 87.35\% accuracy on the CAIDA \ac{ddos} attack dataset \cite{caida2007}. Although the authors state that their method allows real-time detection of \ac{ddos} attacks in a range of environments, no performance measurements are reported to support this claim.

DeepDefense \cite{7946998} combines \acp{cnn} and \acp{rnn} to translate original traffic traces into arrays that contain packet features collected within sliding time windows. The results presented demonstrate high accuracy in \ac{ddos} attack detection within the selected ISCX2012 dataset \cite{ISCXIDS2012}. However, it is not clear if these results were obtained on unseen test data, or are results from the training phase. Furthermore, the number of trainable parameters in the model is extremely large indicating a long and resource-intensive training phase. This would significantly challenge implementation in an online system with constrained resources, as discussed in Sections \ref{sec:lucid-evaluation} and \ref{sec:usecase}.

Although deep learning offers the potential for an effective \ac{ddos} detection method, as described, existing approaches are limited by their suitability for online implementation in resource-constrained environments, as shown in Section \ref{sec:usecase}. 


\section{High-Performance Server-based DDoS Mitigation}\label{sec:mitigation-xdp}

\graphicspath{{ddos-mitigation-xdp/artworks/}{ddos-mitigation-xdp/graphs/}}
\DeclareGraphicsExtensions{.pdf,.jpeg,.png}

\lettrine[findent=2pt]{\textbf{I}}{ }n recent years, the complexity of the network data plane and its requirements in terms of agility has increased significantly, with many network functions now implemented in software and executed directly in data centre servers. To avoid bottlenecks and to keep up with the ever increasing network speeds, recent approaches propose to move the software packet processing to kernel space using technologies such as the \acf{ebpf} and the \acf{xdp}, or to offload (part of) it to specialised hardware, the so called \acp{smartnic}.

In this chapter, we analyse the aforementioned technologies and we study how to exploit them to build an efficient \ac{ddos} attack mitigation pipeline. In particular, we enhance the \ac{ddos} mitigation capabilities of edge servers by offloading a portion of \ac{ddos} mitigation rules to the \ac{smartnic}, achieving a balanced combination of the \ac{ebpf}/\ac{xdp} flexibility in operating traffic sampling and aggregation in the kernel, with the performance of hardware-based filtering.
We demonstrate the benefits of the proposed processing pipeline over \textit{iptables}, a commonly used technology for packet filtering in Linux-based hosts. We also evaluate the performance of different combinations of host and \ac{smartnic}-based mitigation, showing that offloading part of the \ac{ddos} network function to the SmartNIC can indeed improve the packet processing, but only if combined with additional processing in the host kernel space.

This study has been carried out in collaboration with Politecnico di Torino's Department of Computer and Control Engineering. Moreover, the results have been published in the IEEE Access journal \cite{smartnic-ddos}. 

This rest of the chapter is structured as follows. 
Section \ref{sec:xdp-motivation} provides the motivation behind this study. Section~\ref{sec:xdp-background} presents a high-level overview of \ac{ebpf}, \ac{xdp} and \ac{smartnic} technologies.
Section~\ref{sec:mitigation-approaches} analyses the different approaches that can be used to build an efficient host-based \ac{ddos} mitigation solution.
Section~\ref{sec:xdp-architecture} presents the design of an architecture that uses the above mentioned technologies to mitigate \ac{ddos} attacks.
Finally, Section~\ref{sec:xdp-evaluation} provides the necessary evidence to justify the findings, while Section~\ref{sec:xdp-related-work} briefly discusses the related work.

\subsection{Motivation}\label{sec:xdp-motivation}
The ever-growing network capacity installed in data centre and enterprise networks requires a highly flexible low-latency packet processing, which is difficult to achieve with the current mechanisms adopted in software-based network functions.
Common solutions rely on kernel bypass approaches, such as DPDK~\cite{dpdk:webpage} and Netmap~\cite{rizzo2012netmap}, which map the network hardware buffers directly to user space memory, hence bypassing the operating system. Although these technologies bring an unquestionable performance improvement, they also have two major limitations.
First, they take over one (or more) CPU cores, thus permanently stealing precious CPU cycles to other tasks (e.g., other \acp{vnf} deployed on the server).
Second, they require the installation of additional kernel modules or update of the network card driver, operations that are not always possible in production networks.

Recent technologies such as eBPF~\cite{cilium:ebpf-docs, fleming:a-thorough-introduction-to-ebpf} and \ac{xdp}~\cite{hoiland2018express} offer excellent processing capabilities without requiring the permanent allocation of dedicated resources in the host.
Furthermore, eBPF/\ac{xdp} are included in vanilla Linux kernels, hence avoiding the need to install custom kernel modules or additional device drivers.
\subsection{Background}\label{sec:xdp-background}

The \textbf{\acf{ebpf}} is an enhanced version of the BPF virtual machine \cite{ebpf:original}, originally proposed as a kernel packet filtering mechanism and used to implement network utilities such as \textit{tcpdump}.  Compared to the original version, eBPF enables the execution of custom bytecode (\ac{ebpf} programs) at various points of the Linux kernel in a safe manner. \ac{ebpf} programs can be safely injected in various kernel subsystems for tracing (e.g., kprobes, tracepoints, etc.) and networking purposes (through the \ac{xdp} and Traffic Control (TC) hooks). 
\ac{ebpf} programs share information (e.g., network traffic statistics) with user space applications through data structures called \textit{\ac{ebpf} maps}. Conversely, user space applications can use such maps to set configuration parameters for the \ac{ebpf} programs at run time.

The \textbf{\acf{xdp}} is an execution environment residing at the lowest level of the TCP/IP stack in the Linux kernel. In the \ac{xdp} environment, \ac{ebpf} programs can be executed directly upon the receipt of a packet and immediately after the driver RX queues. In the case of volumetric \ac{ddos} attacks, the combination of \ac{xdp} and \ac{ebpf} can be exploited to efficiently process the network traffic and to drop malicious packets before they reach the system TCP/IP stack and the user space applications, with minimal consumption of the host CPU resources.

\textbf{\acp{smartnic}} are intelligent network adapters that can be used to boost the performance of servers by offloading (part of) the network processing workload from the host CPU to the \ac{smartnic} itself~\cite{netronome:smartnics}.
While a traditional \ac{nic} implements a pre-defined set of basic functions (e.g., transmit/receive, segmentation, checksum computation), a \ac{smartnic} is equipped with a fully-programmable system-on-chip (SoC) multi-core processor that is capable of running a fully-fledged operating system, offering more flexibility and hence potentially taking care of any arbitrary network processing tasks.
\acp{smartnic} are usually equipped with a set of specialised hardware functionalities that can be used to accelerate a specific class of functions (e.g., OpenvSwitch data-plane) or to perform generic packet and flow-filtering at line-rate. Compared to a server, a \ac{smartnic} has limited computing and memory capabilities, confining its application space to lightweight tasks.

\subsection{DDoS Mitigation: Approaches} \label{sec:mitigation-approaches}
Efficient packet dropping is a fundamental part of a \ac{ddos} attack mitigation solution.
In a typical \ac{ddos} mitigation pipeline, a set of mitigation rules are deployed in the server's data plane to filter the malicious traffic.
The strategy used to block the malicious sources may be determined by several factors such as the characteristics of the server (e.g., availability of a SmartNIC, its hardware capabilities), the characteristics of the malicious traffic (e.g., number of attackers) or the type and complexity of the rules that are used to classify the illegitimate traffic.

\subsubsection{Host-based Mitigation}
All the host-based \ac{ddos} mitigation techniques and tools used today fall into two macro-categories, depending on whether packets are processed at kernel or user space level.

Focusing on Linux-based systems, kernel-space approaches are based either on \textit{iptables} and its derivatives, such as \textit{nftables}, or on \ac{xdp} programs. iptables is a popular tool for monitoring, manipulating and filtering the network traffic with the support of the kernel's \textit{netfilter} subsystem. As demonstrated in Section~\ref{sec:xdp-evaluation}, the deep level in the networking stack where the packet processing is executed, and the suboptimal matching algorithm used to monitor the traffic, make iptables and netfilter practically unusable for mitigating today's volumetric \ac{ddos} attacks. On the other hand, \ac{xdp} resides at the lowest levels of the network stack and, compared to netfilter, it intercepts the network traffic earlier, right after the \ac{nic} driver, and exploits a more efficient matching algorithm. Although the attention on the \ac{xdp} technology has been growing in recent years, its adoption in \ac{ddos} mitigation solutions is still in its infancy~\cite{bertin2017xdp,suricata-xdp}.  

User space approaches rely on specialised I/O frameworks (Netmap~\cite{rizzo2012netmap}, DPDK~\cite{dpdk:webpage}, PF\_RING ZC~\cite{pfring}, among others) to obtain direct access to the \ac{nic} device memory, by-passing the kernel network system and its overheads. Although these technologies bring an unquestionable performance improvement, they also have two major limitations.
First, these frameworks require the exclusive access to the \ac{nic}, so that all packets received on the interface are processed by the user space monitoring application. In the case of a user space application for \ac{ddos} mitigation, the benign packets must be injected back to the kernel's network stack, requiring further CPU and memory resources for handling the additional packet copies generated by this process.\footnote{It is worth mentioning that Netmap has a better kernel integration compared to DPDK, as Netmap implements a zero-copy approach for injecting the packets back into the kernel. However, it is subjected to a higher CPU consumption compared to \ac{xdp}.}.
Furthermore, such frameworks require a fixed allocation of one (or more) CPU cores, irrespective of the amount of incoming traffic, hence stealing computing resources from other processes running on the host.

\subsubsection{SmartNIC-based Mitigation}
A strategy to save CPU resources on the host is to offload the \ac{ddos} mitigation task to a \ac{smartnic}. With this approach, the malicious packets are dropped by the \ac{smartnic}, whereas only the surviving benign traffic continues its path towards the final destination, such as the processes executed on the host or the host's routing system, before being sent to the next hop. The availability of a \ac{smartnic} enables three different \ac{ddos} mitigation options: (i) hardware packet filtering: line-rate traffic processing by means of hardware tables available on the \ac{smartnic} (if any), (ii) software packet filtering: mitigation program executed on the \ac{smartnic} CPU when no hardware filtering is possible, and (iii) a combination of (i) and (ii) if the hardware tables do not have the capacity to accommodate the complete list of filtering rules.

\acp{smartnic} can execute programs that are statically or dynamically installed from the host, or directly compiled inside the card~\cite{bosshart2013forwarding}. As not all of the above options are supported by all \acp{smartnic}, the implementation of a generic offloading strategy suitable for cards from multiple manufacturers is usually a challenging task.

\subsubsection{Hybrid (SmartNIC + XDP Host)}
As noted in the previous section, executing the complete mitigation task on a \ac{smartnic} avoids any overhead for the host, hence saving CPU cycles for other applications. However, because of the limited capacity of the \ac{smartnic}'s hardware tables, usually in the order of 1K-2K filtering rules, and the limited processing power of the \ac{smartnic}'s CPU, this approach might not be always the optimal solution. Indeed, in the case of \ac{ddos} attacks, a large number of packets (both \ac{ddos} and benign) would be queued in the buffers and then discarded because they are not handled in the allocated time.

One solution is to split the mitigation pipeline between the \ac{smartnic} and the host. With this ``hybrid'' approach, the filtering rules that do not fit into the \ac{smartnic}'s hardware tables are handled in the host, hence leveraging the faster CPU, compared to the \ac{smartnic}, to speed-up the mitigation process. Of course, this approach leads to better performance with respect to a pure \ac{smartnic}-based mitigation solution, but it is more expensive in terms of CPU resources required on the host. A comparison of these two approaches in terms of dropping rate and CPU usage is presented in Section \ref{sec:xdp-evaluation}.

\subsection{Architecture and Implementation} \label{sec:xdp-architecture}

\begin{figure}[h]
	\centering
	\includegraphics[clip=true, width=1\textwidth, trim= 0cm 0cm 0cm 0cm]{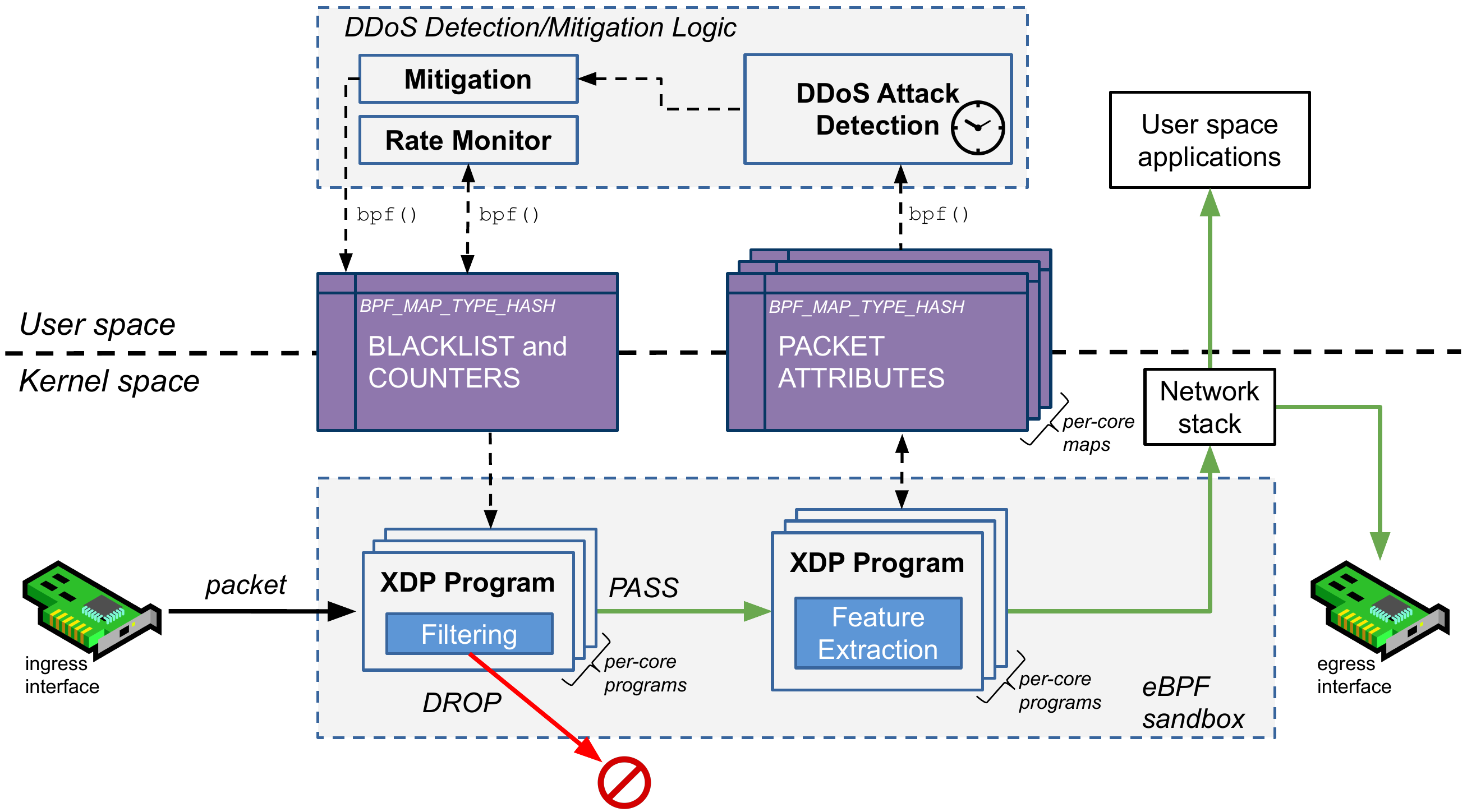}
	\caption{High-level architecture of the system.}
	\label{fig:architecture}
\end{figure}

This section presents the architecture we have designed to evaluate the mitigation approaches discussed in the previous section. The architecture features a data plane composed of a set of \ac{xdp} programs that may run either on the \ac{smartnic}, on the host's kernel, or both, which are in charge of filtering malicious packets and extracting the relevant information from the received traffic.
Extracted features are used by the control plane \ac{ddos} detection algorithm to identify malicious sources and to configure the blacklisted IP addresses in the data plane.
The overall architecture is depicted in Figure~\ref{fig:architecture}; the following sections will present the above components in greater detail.

\subsubsection{Mitigation}
The first program encountered in the pipeline is the \textit{Filtering} module, which matches the incoming packets against the content of a blacklist and drops them if the result is positive; surviving packets are redirected to the next program in the pipeline, the \textit{Feature Extraction}.

At system start-up, the system instantiates the \textit{Filtering} \ac{xdp} program in the host to obtain the necessary traffic information and decide the best mitigation strategy.
If the user space \textit{\ac{ddos} Mitigation} module recognizes the availability of the hardware offload functionality in the \ac{smartnic}, it starts adding the filtering rules into the hardware tables, causing malicious packets to be immediately dropped in hardware.
However, since those tables often have a limited size (typically $\sim$1-2K entries), the strategy is to place the filtering rules of the $k$ most active malicious talkers in the \ac{smartnic}'s hardware tables, where $k$ is the size of those tables. The remaining rules are stored in \ac{ebpf} hash maps (data structures of type \texttt{BPF\_MAP\_} \texttt{TYPE\_HASH}~\cite{bpf-man}) either on the \ac{smartnic} or on the host, depending on the adopted mitigation approach.

\subsubsection{Feature Extraction}
\label{sec:traffic-sampling}
The \textit{Feature Extraction} \ac{xdp} program monitors the incoming traffic and collects relevant packet attributes required by the user space detection algorithm (e.g., IP addresses, protocols, flags, etc.).
Being placed right after the mitigation module, it receives all the (presumed) benign traffic that has not been previously dropped. The benign traffic is then passed to the network stack of the host, which forwards it to the target user space applications or to the next hop in the path towards the final destination.

Packet attributes are stored in \ac{ebpf} hash maps, one for each CPU core, which are periodically read by the user space \ac{ddos} attack detection application.
Since multiple instances of the same \ac{xdp} program are executed in parallel on different CPU cores, each one processing a different packet, the use of dedicated per-core maps guarantees a very fast access to the data, avoiding costly (in terms of CPU cycles) synchronisation operations between the cores.
As a result, each instance of the \textit{feature extraction} works independently, storing the packet attributes in its own private \ac{ebpf} hash map.\\

\subsubsection{Detection}
\label{sec:detection}
The identification of a \ac{ddos} attack is performed by the user space \textit{\ac{ddos} Attack Detection} module, which operates on the traffic statistics collected by the \textit{Feature Extraction} \ac{xdp} program, and exploits the retrieved information to identify the malicious traffic sources. Appropriate filtering rules are then inserted in the \textit{Blacklist} map used by the \textit{Filtering} \ac{xdp} program to drop the traffic. We remind that the challenges related to the detection of \ac{ddos} attacks have been covered in Chapter \ref{sec:detection-cnn}, which also provides a detailed overview of recent solutions in Section \ref{sec:lucid-related}.

Of course, the specific detection algorithm determines the type of traffic information exported by the \textit{Feature Extraction} program. However, the modular design of our architecture and the high level of programmability of \ac{xdp} enable the provisioning of different feature extraction schemes, without impacting on the rest of the architecture.

\subsubsection{Rate Monitor}
The \textit{Blacklist} is also used to keep track of the number of packets dropped for each malicious source stored in it. Such statistics, called \textit{Counters} in Figure~\ref{fig:architecture}, are used by the user space program \textit{Rate Monitor} to remove from the \textit{Blacklist} the sources that are no longer part of a \ac{ddos} attack, or that were erroneously classified as malicious by the \textit{Detection} algorithm. The outcome of this process is twofold: first, the restoration of traffic forwarding from legitimate sources and second, a reduced \textit{Blacklist} size, hence shorter query execution time for the \textit{Filtering} \ac{xdp} program.

The process starts from the global list of blacklisted addresses, sorted according to their traffic volume. The \textit{Rate Monitor} examines the entries that are at the bottom of the list (i.e., the sources sending less traffic), comparing them with a threshold value. If the current transmission rate of the source under consideration is below the threshold, defined as the highest rate of packets with the same source observed under normal network activity, it is removed from the \textit{Blacklist}.
In the case that a malicious source is removed by mistake, the detection algorithm can re-add it to the \textit{Blacklist} during the next iteration.
\subsection{Evaluation}\label{sec:xdp-evaluation}
\pgfplotstableread[header=true]{
SOURCES	HW-XDP-SMARTNIC	IPTABLES	XDP-SMARTNIC	XDP-HOST	HW-XDP-HOST
16		36.996			3.864		14.118			25.594		36.996
64		36.996			2.627		14.037			25.731		36.996
256		36.996			2.017		13.851			25.673		36.996
512		36.996			nan			nan				nan			36.996
1024	23.163			1.004		13.783			25.333		35.163
4096	16.24			0.303		10.384			23.555		28.24
16384	13.757			0.039		9.083			16.926		20.757
65536	7.934			0.003		6.928			15.815		16.934
131072	6.538			0			5.928			10.27		10.538
}\DroppingRate

\pgfplotstableread[header=true]{
	SOURCES	HW-XDP-SMARTNIC	IPTABLES	XDP-SMARTNIC	XDP-HOST	HW-XDP-HOST
	16		1.18			98.59		1.28			95.133		1.28
	64		1.19			98.59		1.43			96.233		1.43
	256		1.49			98.69		2.49			97.361		2.49
	512		1.49			nan			nan				nan			2.49
	1024	2.3				98.79		3.3				97.467		26.3
	4096	3.93			98.89		3.93			97.567		87.13
	16384	4.7				98.99		4.7				98.658		88.7
	65536	4.75			99.19		6.75			98.717		92.75
	131072	5.39			99.69		8.39			99.392		95.39

}\CPUUsage

\pgfplotstableread[header=true]{
	SOURCES	HW-XDP-SMARTNIC	IPTABLES	XDP-SMARTNIC	XDP-HOST	HW-XDP-HOST
	1		60158			32231		66118			60242		66158
	2.5		59623			3149		60624			57572		60623
	5		58342			0			58134			51422		58101
	10		54012			0			55625			44995		55681
	15		52801			0			42834			38424		52801
	20		48245			0			19264			23540		49264
	25		46254			0			2784			7458		49155
	30		40021			0			0				1149		45149
	35		39057			0			0				0			40057
	37		33254			0			0				0			38205	
}\HTTPoneK

\pgfplotstableread[header=true]{
	SOURCES	HW-XDP-SMARTNIC	IPTABLES	XDP-SMARTNIC	XDP-HOST	HW-XDP-HOST
	1		60158			1024		66118			58782		66158
	2.5		59623			0			60624			57623		60623
	5		58342			0			58134			55134		58101
	10		54012			0			55625			51625		55681
	15		52801			0			42834			47494		52801
	20		18245			0			19264			38494		49264
	25		6254			0			2784			8458		49155
	30		4021			0			0				0			10425
	35		0				0			0				0			7214
	37		0				0			0				0			5142
		
}\HTTPfourK

\pgfplotstableread[header=true]{
	SOURCES	HW-XDP-SMARTNIC	IPTABLES	XDP-SMARTNIC	XDP-HOST	HW-XDP-HOST
	1		62118			24			62118			57231		55524
	2.5		56624			0			55624			43688		45245
	5		39234			0			38134			27983		33045
	10		9264			0			9264			19123		21412
	15		3784			0			2784			4009		8254
	20		0				0			0				534			952
	25		0				0			0				0			320
	30		0				0			0				0			0
	35		0				0			0				0			0
	37		0				0			0				0			0
}\HTTPsixtyfourK

This section provides evidence of the benefits of using SmartNICs and \ac{ebpf}/\ac{xdp} in the \ac{ddos} attacks mitigation. For this purpose, the mitigation approaches discussed in Section \ref{sec:mitigation-approaches} are compared against iptables, used as a baseline for the evaluation.

\subsubsection{Experimental Setup}
The testbed comprises two server-class computers, both equipped with a quad-core Intel Xeon~E3-1245@3.50GHz, 8MB of L3 cache, 32GB DDR4-2400 RAM memory, and both running Ubuntu 18.04.2 LTS and the Linux kernel version 4.15. The two servers are directly connected to each other via 25Gbps \acp{smartnic} interfaces.
One server is used to generate volumetric \ac{ddos} attacks against the other server, which runs the \ac{ddos} mitigation pipeline. The attacks are simulated using \textit{Pktgen-DPDK} v3.6.4~\cite{pktgen}, a high-performance testing tool included in the Linux kernel, capable of generating line-rate network traffic. In all the experiments presented below, we use \textit{Pktgen-DPDK} to simulate volumetric \ac{ddos} attacks in the form of streams of 64-byte UDP packets.


\subsubsection{Mitigation Performance}
\label{sec:evaluation-malicious-traffic}

In the first test, we compare the \ac{ddos} mitigation approaches in terms of dropping rate (Mpps, millions of packets per second) and CPU consumption (\%). For this purpose, we generate line-rate traffic at $25$Gbps (corresponding approximately to $37.2$Mpps), with an increasing number of source IPs to simulate botnets of different sizes. 

\begin{figure}[h!]
	\centering
	\captionsetup{justification=centering}
	\begin{tikzpicture}[define rgb/.code={\definecolor{mycolor}{RGB}{#1}}, rgb color/.style={define rgb={#1},mycolor}]
	\begin{semilogxaxis}[  
	legend pos=north east,
	legend columns=1,
	height=6 cm,
	width=0.75\linewidth,
	grid = both,
	xlabel={\# of sources},
	ylabel={Dropping rate (Mpps)},
	scaled y ticks=false,
	scaled x ticks=false,
	xmin=16,
	xmax=131072,
	xtick={16,64,256,1024,4096,16384,65536,131072},
	xticklabels={16,64,256,1K,4K,16K,64K,128K},
	xtick pos=left,
	ymin=0, ymax=40,
	ytick={0,5,10,15,20,25,30,35,40},
	yticklabels={0,5,10,15,20,25,30,35,40},
	ytick pos=left,
	enlargelimits=0.02,
	]
	\addplot [rgb color={45,80,22},style=thick,mark=*,mark size=2.5] table[x index=0,y index=5] {\DroppingRate};
	\addplot [rgb color={68,120,33},style=thick,mark=x,mark size=3.5] table[x index=0,y index=1] {\DroppingRate};
	\addplot [rgb color={90,160,44},style=thick,mark=square*,mark size=2.5] table[x index=0,y index=4] {\DroppingRate};
	\addplot [rgb color={113,200,55},style=thick,mark=triangle*,mark size=3.3] table[x index=0,y index=3] {\DroppingRate};
	\addplot [color=black,style=thick,mark=star,mark size=3.5] table[x index=0,y index=2] {\DroppingRate};
	\node[pin={[pin distance=0cm, pin edge={thick,black, <-,>=stealth'},
		shift={(-2.2cm,-2.5cm)}]
		90:{\sffamily\scriptsize 
			\begin{tabular}{l}
			512 source IPs \\
			(HW tables upper limit) \\
			\end{tabular}
		}
	}] at (axis cs:512,37) {};

	\legend{HW + XDP Host, HW + XDP SmartNIC, XDP Host, XDP SmartNIC, iptables}
	\end{semilogxaxis}
	\end{tikzpicture}
	\caption{Dropping rate with an increasing number of attackers.}
	\label{fig:dropping_rate}
\end{figure}
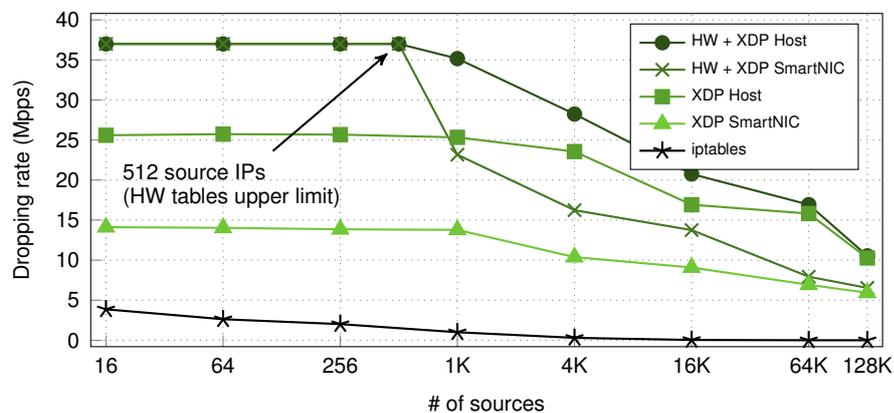

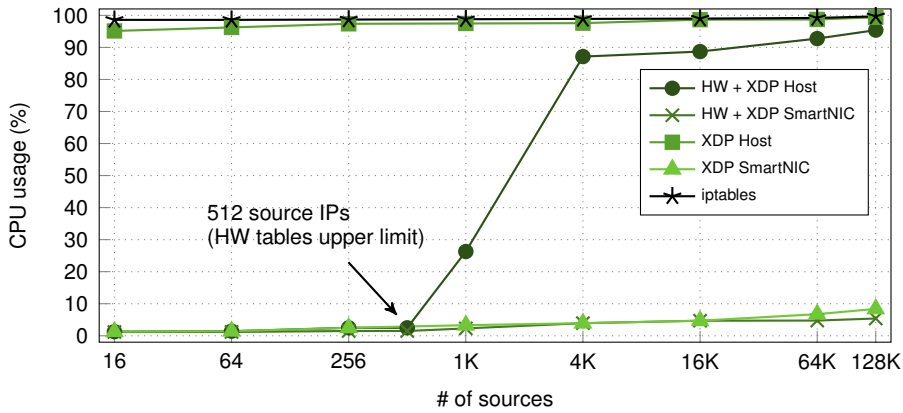
\begin{figure}[h!]
	\centering
	\captionsetup{width=0.75\linewidth}
	\begin{tikzpicture}[define rgb/.code={\definecolor{mycolor}{RGB}{#1}}, rgb color/.style={define rgb={#1},mycolor}]
	\begin{semilogxaxis}[  
	legend style={at={(0.972,0.6)},anchor=east},
	legend columns=1,
	height=6 cm,
	width=0.75\linewidth,
	grid = both,
	xlabel={\# of sources},
	ylabel={CPU usage (\%)},
	scaled y ticks=false,
	scaled x ticks=false,
	xmin=16,
	xmax=131072,
	xtick={16,64,256,1024,4096,16384,65536,131072},
	xticklabels={16,64,256,1K,4K,16K,64K,128K},
	xtick pos=left,
	ymin=0, ymax=100,
	ytick={0,10,20,30,40,50,60,70,80,90,100},
	yticklabels={0,10,20,30,40,50,60,70,80,90,100},
	ytick pos=left,
	enlargelimits=0.02,
	]
	\addplot [rgb color={45,80,22},style=thick,mark=*,mark size=2.5] table[x index=0,y index=5] {\CPUUsage} ;
	\addplot [rgb color={68,120,33},style=thick,mark=x,mark size=3.5] table[x index=0,y index=1] {\CPUUsage};
	\addplot [rgb color={90,160,44},style=thick,mark=square*,mark size=2.5] table[x index=0,y index=4] {\CPUUsage};
	\addplot [rgb color={113,200,55},style=thick,mark=triangle*,mark size=3.3] table[x index=0,y index=3] {\CPUUsage};
	\addplot [color=black,style=thick,mark=star,mark size=3.5] table[x index=0,y index=2] {\CPUUsage};
	\node[pin={[pin distance=0.5cm, pin edge={thick,black, <-,>=stealth'},
		shift={(-1.2cm,0.2cm)}]
		{\sffamily\scriptsize 
			\begin{tabular}{l}
			512 source IPs \\
			(HW tables upper limit) \\
			\end{tabular}
		}
	}] at (axis cs:512,3) {};
	
	\legend{HW + XDP Host, HW + XDP SmartNIC, XDP Host, XDP SmartNIC, iptables}
	\end{semilogxaxis}
	\end{tikzpicture}
	\caption{CPU usage of the different mitigation approaches under a simulated DDoS attack.}
	\label{fig:cpu_usage}
\end{figure}

\textbf{iptables-based mitigation: }
As briefly introduced in Section \ref{sec:mitigation-approaches}, iptables is a user space command line tool for configuring the Linux kernel's netfilter packet filtering framework. To maximize the performance of iptables, in this test we use the netfilter PREROUTING chain, which can access (and drop) the traffic as soon as it enters the network stack, before the kernel's routing operations.

Figure~\ref{fig:dropping_rate} shows that the dropping rate of iptables is limited, around 2.5-4.5Mpps, even with a relatively small number of attack sources. 
This is mainly due to the matching algorithm used by iptables, whose performance degrades linearly with the number of filtering rules inserted in the blacklist, leading to a throughput almost equal to zero with more than 4K rules. 

The CPU consumption (Figure~\ref{fig:cpu_usage}) confirms this limitation: using iptables to mitigate large \ac{ddos} attacks saturates the CPU, leaving almost no computing resources for the other processes running on the host.\\

\textbf{XDP-based mitigation: }
This test runs the entire mitigation pipeline on the host without involving the SmartNIC, which simply forwards the packets to the \ac{xdp} program executed in the host.

The dropping efficiency of XDP is much higher than iptables, being able to process from $\sim$26Mpps in the presence of 1K attackers or less, to $\sim$10Mpps with 128K attackers (\textit{XDP Host} curve in Figure~\ref{fig:dropping_rate}).

In \ac{xdp}, the blacklist is implemented using an \ac{ebpf} hash map (\texttt{BPF\_MAP\_TYPE\_} \texttt{HASH}), whose lookup algorithm is much faster than that of iptables. In fact, although both approaches saturate the host's CPU  (Figure~\ref{fig:cpu_usage}), \ac{xdp} can handle massive \ac{ddos} attacks with more than 4K different sources, whereas iptables stops working.\\

\textbf{SmartNIC-based mitigation: }
In this experiment, the mitigation pipeline is executed entirely on the \ac{smartnic}.
We first perform a test where the attack is mitigated only through an XDP filtering program running on the \ac{smartnic}, without any help from the hardware filter available on the network card. Compared to the host-based mitigation, the slower CPU of the NIC leads to a performance degradation in the dropping rate (Figure~\ref{fig:dropping_rate} (\textit{XDP \ac{smartnic}} curve). On the other hand, no host computing resources are consumed (Figure~\ref{fig:cpu_usage}), as the whole mitigation is executed on the \ac{smartnic}.

A second test exploits a mixture of hardware and XDP-based software filtering on the card. As shown in Figures \ref{fig:dropping_rate} and \ref{fig:cpu_usage}, up to 512 sources, the dropping rate corresponds to the input traffic rate, where 512 is the maximum number of entries (IP addresses) that fit in the \ac{smartnic}'s hardware tables. With larger attacks, part of the blacklist is maintained outside the hardware tables by the \ac{xdp} program running in the \ac{smartnic}, hence leading to a decrease in the overall performance of the mitigation system. \\

\textbf{Hybrid approach (SmartNIC Hardware Tables + XDP Host): }
Here the offloading algorithm splits the mitigation pipeline between the \ac{smartnic}'s hardware tables and the XDP filtering program running in the host. 

Similar to the previous experiment, we can notice a line rate packet dropping up to 512 sources (\textit{HW + XDP Host} in Figure \ref{fig:dropping_rate}). Moreover, the performance of the host CPU leads to a higher dropping rate for large attacks, compared to the approach confined to the \ac{smartnic} discussed above. The price to pay is a consumption of host computing resources, which increases with the number of attackers' IPs exceeding the space available in the hardware tables (Figure~\ref{fig:cpu_usage}).\\

\textbf{Final remarks: }
The benefits of offloading the packet filtering tasks to a \ac{smartnic} are higher when most of the traffic can be handled in hardware, hence maximising the overall performance of the \ac{ddos} mitigation system and saving precious computing resources on the host.

It is worth noticing that the case where a host has to cope with a limited number of malicious sources is rather common, as the incoming traffic in data centres is usually balanced across multiple servers (backends), each one being asked to handle only a portion of the connections and, as a consequence, of the attackers.

\subsubsection{Impact on the Legitimate Traffic}\label{sec:evaluation-legitimate-traffic}

As illustrated in Section \ref{sec:evaluation-malicious-traffic}, mitigating a volumetric \ac{ddos} attack has a cost in terms of CPU usage. Of course, the reduced computing resources can have a negative effect on the other processes running on the victim machine. In this experiment, we evaluate the impact of the \ac{ddos} mitigation approaches under consideration on the web services provided by the server acting as a \ac{ddos} victim. For this test, we generate 1M HTTP requests using 200 concurrent clients towards the \texttt{nginx} web server running on the target server. 
We measure the number of successful HTTP requests per second completed within 5 seconds, as a function of the \ac{ddos} traffic rate.

It is worth recalling that the \ac{smartnic}, similar to traditional NICs, discards incoming packets when its buffer is full. In our case, we can observe such a behaviour when the \ac{ddos} mitigation system is not able to process packets at the rate they arrive at the interface. Of course, the packet rate and the number of attack sources influence the ability of the system to process the traffic in time without forcing the \ac{smartnic} to discard packets. In this experiment, we vary the \ac{ddos} packet rate between 1Mpps and line rate (37.2Mpps), whereas the number of attackers is kept constant at 1K in a first test, and then at 64K in a second test.

\begin{figure}[h!]
	\centering
	\begin{tikzpicture}[define rgb/.code={\definecolor{mycolor}{RGB}{#1}}, rgb color/.style={define rgb={#1},mycolor}]
	\begin{groupplot}[%
	,group style={columns=2},
	legend columns=5,
	legend style={/tikz/every even column/.append style={column sep=0.5cm}},
	height=6 cm,
	width=0.52\linewidth,
	grid = both,
	scaled y ticks=false,
	scaled x ticks=false,
	xmin=0,
	xmax=37,
	xtick={0,1,2.5,5,10,15,20,25,30,35},
	xticklabels={0,,,5,10,15,20,25,30,35},
	xtick pos=left,
	ymin=0, ymax=70000,
	ytick={0,10000,20000,30000,40000,50000,60000,70000},
	yticklabels={0,10K,20K,30K,40K,50K,60K,70K},
	ytick pos=left,
	enlargelimits=0.02,
	]
	\nextgroupplot[%
	ylabel={HTTP req/sec},
	xlabel={(a) DDoS Traffic (Mpps) - 1K attackers},
	legend to name=grouplegend,
	mark=none
	]
	\addplot [rgb color={45,80,22},style=thick,mark=*,mark size=2.5] table[x index=0,y index=5] {\HTTPoneK}; \addlegendentry{HW + XDP Host}
	\addplot [rgb color={68,120,33},style=thick,mark=x,mark size=3.5] table[x index=0,y index=1] {\HTTPoneK}; \addlegendentry{HW + XDP SmartNIC}
	\addplot [rgb color={90,160,44},style=thick,mark=square*,mark size=2.5] table[x index=0,y index=4] {\HTTPoneK}; \addlegendentry{XDP Host}
	\addplot [rgb color={113,200,55},style=thick,mark=triangle*,mark size=3.3] table[x index=0,y index=3] {\HTTPoneK};\addlegendentry{XDP SmartNIC}
	\addplot [color=black,style=thick,mark=star,mark size=3.5] table[x index=0,y index=2] {\HTTPoneK}; \addlegendentry{iptables}
	\nextgroupplot[
	xlabel={(b) DDoS Traffic (Mpps) - 64K attackers},
	mark=none
	]
	\addplot [rgb color={45,80,22},style=thick,mark=*,mark size=2.5] table[x index=0,y index=5] {\HTTPsixtyfourK} ;
	\addplot [rgb color={68,120,33},style=thick,mark=x,mark size=3.5] table[x index=0,y index=1] {\HTTPsixtyfourK};
	\addplot [rgb color={90,160,44},style=thick,mark=square*,mark size=2.5] table[x index=0,y index=4] {\HTTPsixtyfourK};
	\addplot [rgb color={113,200,55},style=thick,mark=triangle*,mark size=3.3] table[x index=0,y index=3] {\HTTPsixtyfourK};
	\addplot [color=black,style=thick,mark=star,mark size=3.5] table[x index=0,y index=2] {\HTTPsixtyfourK};
	\end{groupplot} 
	\node at (group c1r1.east) [anchor=south, yshift=2.3cm, xshift=0.5cm] {\ref{grouplegend}};       
	\end{tikzpicture}
	\caption{Number of successfully completed HTTP requests/s under different \ac{ddos} traffic rates.}
	\label{fig:test-legitimate-traffic}
\end{figure}
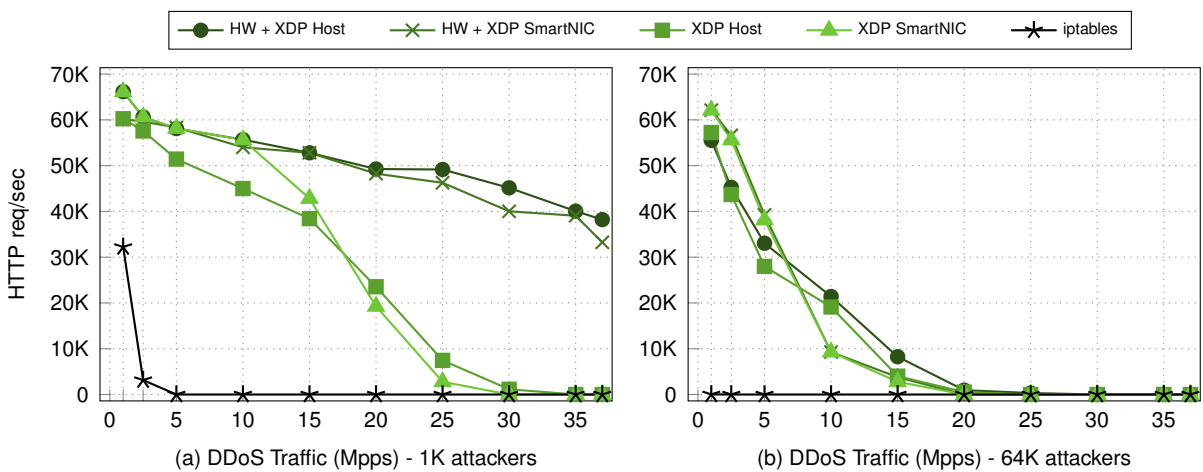

With 1K attackers, the hardware tables process half of the malicious traffic, with noticeable benefits at every \ac{ddos} packet rate (\textit{HW + XDP Host} and \textit{HW + XDP SmartNIC} in Figure~\ref{fig:test-legitimate-traffic}(a)) compared to pure software-based approaches (\textit{XDP Host} and \textit{XDP SmartNIC}). Moreover, at low \ac{ddos} packet rates we can observe a higher number of successful HTTP connections achieved when running \ac{xdp} on the \ac{smartnic} with respect to that obtained with \ac{xdp} executed on the host machine. Indeed, despite a higher computing capacity, the CPU of the server is also busy serving the HTTP requests, whereas the CPU of the \ac{smartnic} is completely devoted to \ac{xdp}. This behaviour disappears at \ac{ddos} packet rates higher than $\backsim$17Mpps, as the impact on the host CPU of the HTTP requests becomes negligible with respect to the load for processing the \ac{ddos} packets. 

Figure~\ref{fig:test-legitimate-traffic}(b) reports the results obtained with 64K attackers. In this case, the performance gain obtained with the hardware tables is less significant, as only one packet in every 128 is processed in hardware, whereas the large majority are handled in software either on the \ac{smartnic} or on the host. More importantly, the long lookup tables require a high search time, leading to high amounts of discarded packets even at low \ac{ddos} packet rates.    

Finally, the \textit{iptables}-based mitigation is infeasible at any rate of \ac{ddos} traffic due to the slow linear search mechanism, causing the failure of the large majority of HTTP requests.
\subsection{Related Work}
\label{sec:xdp-related-work}

The advantages of using XDP to filter packets at high rates have been largely discussed and demonstrated~\cite{brenden:xdp-drop-perf, facebook:xdp-and-lb}; several companies (e.g., Facebook, Cloudflare) have integrated XDP in their data centre networks to protect end hosts from unwanted traffic, given the enormous benefits from both filtering performance and low resource consumption.
In particular, in~\cite{bertin2017xdp} Cloudflare presented a \ac{ddos} mitigation architecture that was initially based on kernel bypass, to overcome the performance limitations of iptables, and classical BPF to filter packets in user space.
However, they subsequently shifted to an XDP-based architecture called \textit{L4Drop}~\cite{cloudflare:l4drop} that performs packet sampling and dropping within an XDP program itself.
Our approach is slightly different; we use an XDP program to extract the relevant packet headers from all the received traffic, instead of sending the entire samples to the user space detection application and we consider simpler filtering rules, which are needed to deal with the SmartNIC hardware limitations. 
Finally, we consider in our architecture the use of SmartNICs to improve the packet processing, which introduces additional complexity (e.g., select rules to offload), which is not needed in a host-based solution.
In this direction, \cite{le2017uno} analysed and proposed a hybrid architecture that use SmartNIC to improve VNFs processing capabilities. However, to the best of our knowledge, this work is the first that analyses and proposes a complete hardware/software architecture for the \ac{ddos} mitigation use case.
\section{Conclusions}\label{sec:conclusions}
	
\lettrine[findent=2pt]{\textbf{I}}{ }n this thesis, we have presented models, algorithms and architectures for the provisioning of security services in ``softwarised'' networks, where the network functions can be implemented as host-based software components running inside virtual machines or containers. 
We have provided a solution for dynamically provisioning security services in a softwarised \ac{tsp} network, where the objective is the minimisation of the consumed computing and network resources.  The proposed approach, called \ac{pess}, takes into account  security and \ac{qos} requirements of user applications and ensures that computing and network resources are accurately utilised. We have discussed the rationale behind our design decisions and presented an \ac{ilp} formulation and a heuristic algorithm that solve the placement problem. The evaluation results demonstrate the benefits of \ac{pess} for both users and telecom operators, with savings in resource utilisation and in end-to-end latency. We have also shown that the heuristic implementation of the proposed application-aware approach produces near-optimal solutions and scales well in large and dense networks, indicating the potential of \ac{pess} in real-world scenarios.
Although \ac{pess} has been designed for \ac{tsp} networks, we are confident that the proposed methods are generic enough to cover different application scenarios. One example in this regard is the \acs{sdbranch}-enabled enterprise~\cite{sd-branch-cisco}, in which branch connectivity, networking and security functions are provided through a centrally managed software-based platform. 
Given the raising availability on the market of \acs{sdbranch} solutions~\cite{sdbranch-fortinet, sdbranch-versa, sdbranch-aruba}, we believe that investigating the applicability of \ac{pess} to softwarised enterprise networks is a promising direction for future work in this area.

We have also proposed host-based software solutions for the detection and mitigation of \ac{ddos} attacks with efficient usage of CPU resources. With respect to the attack detection, we have presented a \ac{dl}-based architecture called \ac{lucid}. Our design has targeted a practical, lightweight implementation with low processing overhead and attack detection time. The benefit of the \ac{dl} model is to remove threshold configuration as required by statistical detection approaches, and reduce feature engineering and the reliance on human experts required by alternative \ac{ml} techniques. This enables practical deployment. In contrast to existing solutions, our unique traffic pre-processing mechanism acknowledges how traffic flows across network devices and is designed to present network traffic to the \ac{dl} model for online \ac{ddos} attack detection. Our evaluation results demonstrate that \ac{lucid} matches the existing state-of-the-art performance. However, distinct from existing work, we have demonstrated consistent detection results across a range of datasets, demonstrating the stability of our solution. Furthermore, our evaluation on a resource-constrained device demonstrates the suitability of our model for deployment in resource-constrained environments. Specifically, we have achieved a 40x improvement in processing time over similar state-of-the-art solutions. We have also presented an activation analysis to explain how \ac{lucid} learns to detect \ac{ddos} traffic, which is lacking in existing works. 

Linked to the activation analysis, the robustness to \ac{aml} attacks is a key consideration for the deployment of \ac{lucid}. As detailed in \cite{CoronaAdversarial:2013:AAA:2479999.2480270}, the two main attacks on \acp{ids} are during training via a poisoning attack (i.e. corruption of the training data), or in testing, when an evasion attack attempts to cause incorrect classification by making small perturbations to observed features. 
Our activation analysis is a first step in the investigation of \ac{lucid} behaviour in adversarial cases with the feature ranking in Table \ref{tab:col-activ} highlighting the features for perturbation for evasion attacks. Of course, the adversary model (goal, knowledge, and capability) dictates the potential for a successful attack. For example, the attacker would require full knowledge of the CNN and kernel activations, and have the ability to forge traffic within the network. The construction of defences robust to adversarial attacks is an open problem \cite{onevaluatingcarlini} and an aspect which we will further explore for \ac{lucid}. 

With respect to the \ac{ddos} attack response, we have conducted an analysis of various approaches for building an efficient and cost-effective \ac{ddos} mitigation pipeline. We have compared the performance of the different mitigation alternatives based on combinations of hardware technologies (a \acf{smartnic}) and recent software technologies (the \acf{ebpf} and the \acf{xdp}). 
According to our experiments, the best approach is a combination of hardware filtering on the \ac{smartnic} and software filtering with \ac{ebpf}/\ac{xdp} on the host, which presents the most efficient results in terms of dropping rate and CPU usage.
However, our findings suggest that current \acp{smartnic} can help mitigating the network load on congested servers, but may not represent a turn-key solution. For instance, an effective \ac{smartnic}-based solution for \ac{ddos} attacks may require the presence of a \ac{ddos}-aware load balancer that distributes incoming datacentre traffic in a way to reduce the amount of attackers landing on each server, whose number should be compatible with the size of the hardware tables of the \ac{smartnic}. Otherwise, the solution may require the software running on the \acp{smartnic} to cooperate with other components running on the host, reducing the effectiveness of the solution in terms of saved resources in the servers.\\ 

Beyond the aspects studied in this thesis, there is a range of open issues for practical and effective implementation of software-based security network functions. As highlighted in \cite{ref_etsinfvsec001}, these include \ac{vsnf} secured boot, \ac{vsnf} secure crash, \ac{vsnf} performance isolation, private keys protection and distribution, back-doors on \ac{vnf} management/test/debug interfaces. One of the most challenging is the \ac{vsnf} performance isolation, which is linked to the characterisation of the CPU usage of the \acp{vsnf} under normal and abnormal conditions. 
In this direction, our current research focuses on implementing a fully-fledged \ac{ddos} defence system built around \ac{lucid} and \ac{ebpf}/\ac{xdp}, with predictable and tunable CPU consumption even under volumetric \ac{ddos} attacks. 
The plan is to design a \ac{vsnf} that works in edge computing environments without compromising the operations of other processes running on the same edge node. 
The problem is complex, as the CPU requirements of \ac{lucid} depend on the number of flows collected in a given time interval, while those of \ac{ebpf}/\ac{xdp} depend on the number of entries in the blacklist. 	
\blankpage
\bibliographystyle{IEEEtran} 
\bibliography{bibliography,my_publications}

\end{document}